\documentclass[reprint, amsmath,amssymb, aip,jcp]{revtex4-2}

\usepackage{graphicx}
\usepackage{dcolumn}
\usepackage{bm}
\usepackage{afterpage}
\usepackage{amssymb}
\usepackage{graphicx}
\usepackage[version=4]{mhchem}
\usepackage[export]{adjustbox}
\usepackage{xcolor}
\usepackage{tabularx}
\usepackage{textgreek}
\usepackage{float}

\begin{document}

\title{Measurement of Coherent Vibrational Dynamics with X-ray Transient Absorption Spectroscopy Simultaneously at the Carbon K- and Chlorine L$_{2,3}$- Edges}

\author{Andrew D. Ross}
\altaffiliation{These authors contributed equally to this work.}
\affiliation{%
 Department of Chemistry, University of California, Berkeley, CA, 94720, USA
}
\affiliation{%
 Chemical Sciences Division, Lawrence Berkeley National Laboratory, Berkeley, CA, 94720, USA
}
\affiliation{%
 Current Address: Toptica Photonics, Inc., Pittsford, NY 14534, USA
}

\author{Diptarka Hait}%
\altaffiliation{These authors contributed equally to this work.}
\affiliation{%
 Department of Chemistry, University of California, Berkeley, CA, 94720, USA
}
\affiliation{%
 Chemical Sciences Division, Lawrence Berkeley National Laboratory, Berkeley, CA, 94720, USA
}
\affiliation{%
 Current Address: Department of Chemistry and PULSE Institute, Stanford University, Stanford, CA 94305, USA
}
\author{Valeriu Scutelnic}%
\affiliation{%
 Department of Chemistry, University of California, Berkeley, CA, 94720, USA
}
\affiliation{%
 Chemical Sciences Division, Lawrence Berkeley National Laboratory, Berkeley, CA, 94720, USA
}

\author{Daniel M. Neumark}%
\affiliation{%
 Department of Chemistry, University of California, Berkeley, CA, 94720, USA
}
\affiliation{%
 Chemical Sciences Division, Lawrence Berkeley National Laboratory, Berkeley, CA, 94720, USA
}
\author{Martin Head-Gordon}%
\affiliation{%
 Department of Chemistry, University of California, Berkeley, CA, 94720, USA
}
\affiliation{%
 Chemical Sciences Division, Lawrence Berkeley National Laboratory, Berkeley, CA, 94720, USA
}

\author{Stephen R. Leone}%
\email{srl@berkeley.edu}
\affiliation{%
 Department of Chemistry, 
 University of California, Berkeley, CA, 94720, USA
}
\affiliation{%
 Chemical Sciences Division, 
 Lawrence Berkeley National Laboratory, Berkeley, CA, 94720, USA
}
\affiliation{%
 Department of Physics, 
 University of California, Berkeley, CA, 94720, USA
}%
\date{\today}

\begin{abstract}
X-ray Transient Absorption Spectroscopy near the carbon K-edge (1s, $\sim$ 285 eV) and chlorine L$_{2,3}$ edges (2p, $\sim$ 200 eV) is used to study the nuclear dynamics of CCl$_4$ vibrationally activated by impulsive stimulated Raman scattering with a few-cycle 800 nm pump pulse. The totally symmetric stretching mode leads to a strong response in the inner-shell spectra, with the concerted elongation (contraction) in bond lengths leading to a red (blue) shift in the X-ray absorption energies associated with core-to-antibonding excitations. The relative slopes of the potential energy surfaces associated with the relevant core-excited states along the symmetric stretching mode are experimentally measured and compared to results from restricted open-shell Kohn-Sham calculations. A combination of experiment and theory indicates that the slope of the core-excited potential energy surface vs totally symmetric bond elongation is $-11.1 \pm 0.8$ eV/{\AA} for the Cl 2p$\to7a_1^*$ excitation, $-9.0\pm0.6$ eV/{\AA} for the Cl 2p$\to8t_2^*$ excitation and $-5.2\pm 0.4$ eV/{\AA} for the C 1s$\to8t_2^*$ excitation, to 95\% confidence. The much larger slopes for the Cl 2p excitations compared to the C 1s state are attributed to greater contributions from Cl to the $7a_1^*$ or $8t_2^*$  antibonding orbitals to which the inner-shell electrons are being excited. No net displacement of the center of the vibrational wavefunction along the other vibrational modes is induced by the pump pulse, leading to absence of transient signal. 
The results highlight the ability of X-ray Transient Absorption Spectroscopy to reveal nuclear dynamics involving tiny ($<0.01$ {\AA}) atomic displacements and also provide direct measurement of forces on core-excited potential energy surfaces.
\end{abstract}

\maketitle

\section{Introduction}
X-ray absorption spectroscopy is a powerful instrumental technique\cite{bhattacherjee2018ultrafast,Geneaux2019ATASReview,bressler2004ultrafast,de2001high} that has been used to measure atomic structure, electronic dynamics, and nuclear motion\cite{bhattacherjee2018ultrafast,ross2022jahn,henderson2014x,wollan1932x,schulke2007electron,hahner2006near,gaffney2007imaging,kraus2018ultrafast,bressler2010molecular,ridente2023femtosecond,zinchenko2021sub,pertotScience2017,haugen2023ultrafast}. The interpretation of experimental X-ray absorption is often done through comparison to quantum chemical calculations, leading to considerable recent theoretical interest in this area\cite{hait2020highly,hait2020accurate,wenzel2014calculating,oosterbaan2018non,vidal2019new,besley2021modeling,norman2018simulating,garner2023spin,yao2022all,carter2022electron}. However, most computational work has focused on predicting X-ray absorption energies at fixed, often equilibrium, geometries, and changes in core-level excitation energies with nuclear displacements have been relatively unexplored, in large part due to the relative lack of high-quality experimental data for comparison. Accurate prediction of such variations is important for estimating spectral line-widths\cite{prince2003near,vaz2019nuclear,hait2023predicting}, probing dynamics on a single electronic state\cite{ross2022jahn,ridente2023femtosecond} and exploring core-excited potential energy surfaces\cite{takahashi2022interpretation,cruzeiro20231b1,barreau2023core} (CEPESs). It would therefore be useful to have experimental measurements that unambiguously report variations in core-excitation energies vs specific nuclear displacements.

\begin{figure}[htb!] 
\centering
\includegraphics[max height=\textheight/2,max width=\columnwidth,keepaspectratio]
{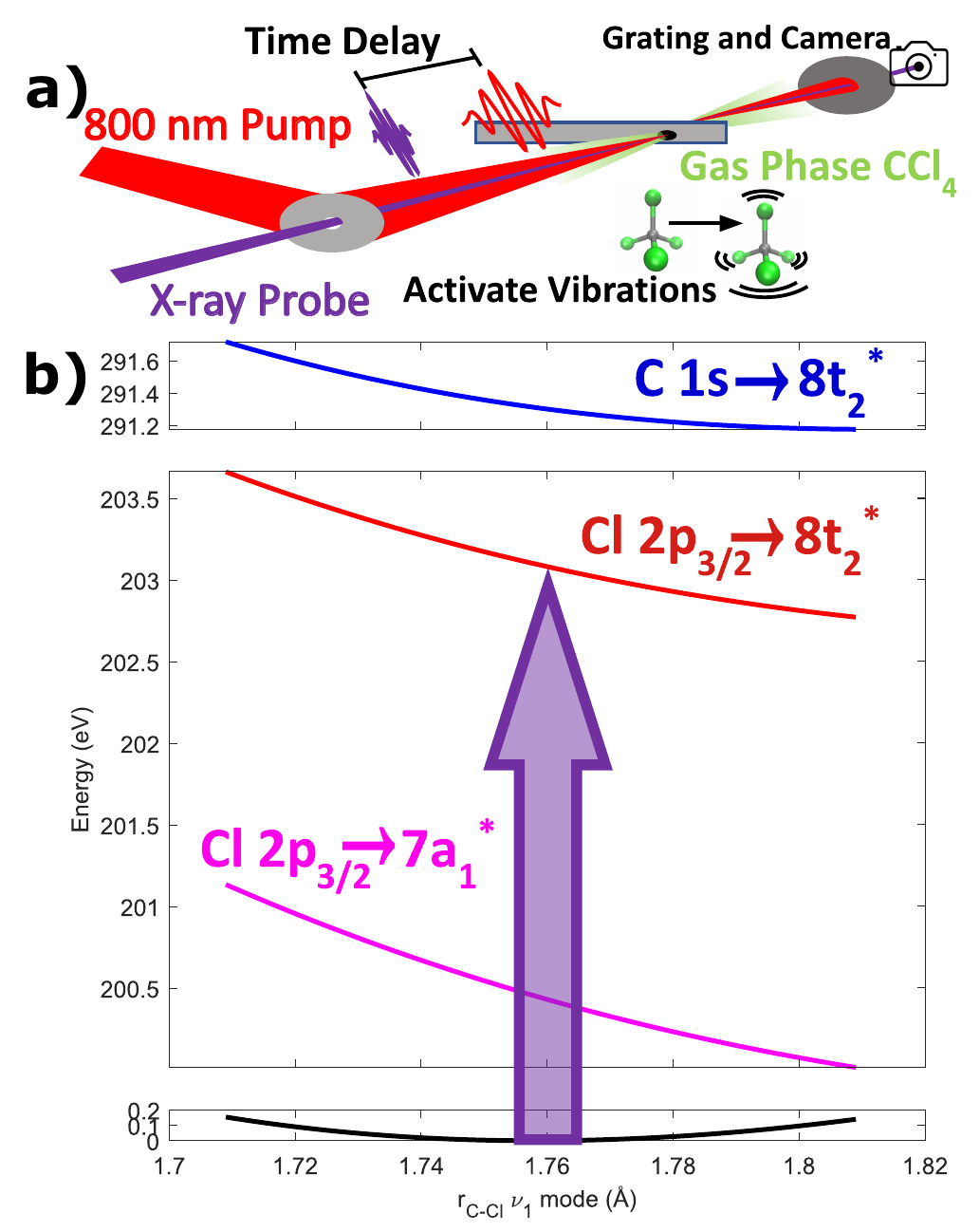}
\vspace*{-20pt}
\caption{\label{fig:Example vibration subraction}
a) Diagram of the experiment. Gas phase \ce{CCl4} is vibrationally excited with a 6 fs duration 800 nm pump and is probed after a variable delay with a few-fs 
X-ray probe pulse, which is dispersed via a grating onto a CCD camera. b) Schematic of the transitions studied along the \ce{CCl4} symmetric stretch mode, with associated PES (y axis energy spacings to scale).
The experimentally measured excitation energy is $\Omega_C(q)=E_C(q)-E_G(q)$. Note that the $E_G(q)$ surface by definition has no slope at the equilibrium bond length. 
}\vspace*{-10pt}
\end{figure}

A relatively straightforward route to obtaining CEPES properties is through X-ray Transient Absorption Spectroscopy (XTAS) of coherent vibrational dynamics. Such dynamics can be experimentally realized by first exciting the sample to a superposition of vibrational states by Impulsive Stimulated Raman Scattering (ISRS), where a pump pulse, such as the short 800 nm pulse shown in Fig. \ref{fig:Example vibration subraction}, launches a wavepacket along the Raman active vibrational modes that have a period more than double the duration of the pump laser pulse\cite{yan1985impulsive,yan1987impulsive}. Following the pump pulse, the nuclear wavepackets oscillate with the associated normal mode frequencies as coherent states of the harmonic oscillator, assuming the harmonic approximation holds and the modes do not couple.  
As the molecule vibrates, changes in the core-level excitation energies can be tracked by the soft X-ray probe at a series of delay times from the initial excitation, revealing features of the CEPES.

The connection between transient core-level excitation energy changes and CEPES features can be understood within a simple Born-Oppenheimer model of classical nuclei moving on electronic potential energy surfaces. Let the molecule be distorted by $\{q_i\}$ along each of the ground state normal modes $\{i\}$ relative to the equilibrium geometry in the electronic ground state. The corresponding core-excitation energy is therefore  $\Omega_C(\{q_i\})=E_C(\{q_i\})-E_G(\{q_i\})$, where $E_C$ and $E_G$ are potential energy surface functions for the core-excited and electronic ground states respectively. $E_G(\{q_i\})$ does not have any terms linear in $\{q_i\}$ as the undistorted geometry is a minimum on the electronic ground state potential energy surfac. This is not true for $E_C(\{q_i\})$ as minima on the CEPES (if they exist at all) are unlikely to be identical with the electronic ground state equilibrium geometry\cite{miron2012imaging,kimberg2014molecular,schreck2016ground,eckert2018one,vaz2019probing}. For small displacements therefore:    
\begin{align}
    E_C(\{q_i\})&\approx E_C(0) + \displaystyle\sum\limits_iq_i\left(\dfrac{\partial E_C}{\partial q_i}\right)_{\{q_i\}=0} 
\end{align}
 where $\left(\dfrac{\partial E_C}{\partial q_i}\right)_{\{q_i\}=0}$ is the slope of the CEPES at the ground state equilibrium geometry along the $i$th ground state mode, and higher order terms in $\{q_i\}$ are neglected. The excitation energies are similarly:
\begin{align}
 \Omega_C(\{q_i\})&\approx E_C(0)-E_G(0)+ \displaystyle\sum\limits_i q_i\left(\dfrac{\partial E_C}{\partial q_i}\right)_{\{q_i\}=0} 
\end{align}

For coherent vibrational motions within the harmonic approximation, the pump stimulated displacements evolve in time as $q_i(t)=A_i\cos(\omega_it+\phi_i)$. A Fourier transform of the time-dependent excitation energies $\Omega_C(\{q_i(t)\})$ would therefore reveal information about the CEPES slopes. Specifically, let us consider two core-excited states $C_1$ and $C_2$, with associated time-dependent core-excitation energies $\Omega_{C_1}(t)$ and $\Omega_{C_2}(t)$ resulting from vibrational dynamics. For a given ground state normal mode $i$ with ground state frequency $\omega_i$, the ratio of the associated Fourier transforms $\mathcal{F}$ then are:  

\begin{align}
    \dfrac{\mathcal{F}\left[\Omega_{C_1}(t)\right](\omega_i)}{\mathcal{F}\left[\Omega_{C_2}(t)\right](\omega_i)}&=\dfrac{\left(\dfrac{\partial E_{C_1}}{\partial q_i}\right)_{\{q_i\}=0}  }{ \left(\dfrac{\partial E_{C_2}}{\partial q_i}\right)_{\{q_i\}=0}  } \label{eq:ratio}
\end{align}
indicating that the ratio of the Fourier transforms of two time-dependent core-excitation energies arising from vibrational dynamics is equivalent to the ratio of the CEPES slopes along that mode at the ground state equilibrium geometry. 
We note that anharmonic effects may lead to deviations from this simplified model, but we do not consider this aspect further in this work.

 XTAS has been used to observe coherent vibrational motion in the past\cite{saito2019real,weisshaupt2017ultrafast,Geneaux2019ATASReview,cammarata2014sequential} and specifically the CEPES for \ce{SF6} at the S L$_{2,3}$ edges recently\cite{barreau2023core,rupprecht2023resolving}. In this work, we experimentally measure the relative CEPES slopes along the symmetric stretching mode of \ce{CCl4}, considering both the C K-edge (1s, $\sim$ 285 eV) and the Cl L$_{2,3}$-edge (2p, $\sim$ 200 eV). We focus on dipole allowed core-excitations to the antibonding $7a_1^*$ and (triply degenerate) $8t_2^*$ levels, which are energetically well separated from each other and readily accessible with the X-ray probe\cite{lu2009core,fock1985shape} (as shown in Fig. \ref{fig:Example vibration subraction}b). Relative CEPES slopes from experiment are compared to the corresponding ratios obtained from absolute CEPES slopes predicted by restricted open-shell Kohn-Sham (ROKS\cite{frank1998molecular,hait2020highly}) calculations. The computed values are found to lie within the experimental error bars, indicating an excellent level of agreement and permitting estimation of absolute experimental CEPES slopes. No signal in the experimental X-ray absorption energy is detected for the other normal modes, as no net displacement along these modes is induced by the pump pulse. 

 \section{Methods}
\subsection{Experiment}
The experimental apparatus has been fully described in Ref \citenum{Barreau2020}. Here, we provide a brief summary of the relevant experimental conditions. The X-ray probe is a nearly femtosecond pulse with photon energies extending up to 370 eV, generated by high harmonic generation (HHG) by a 10 fs, 1300 nm pulse. ISRS is induced by an 800 nm pulse, spectrally broadened, and compressed to about 6 fs in duration with a 1 kHz repetition rate. The maximum power used on the sample was 150 mW, focused to 65 $\mu$m  full width at half maximum (FWHM), which gives an electric field with peak intensity up to $3 \pm 1\times 10^{14}$ W/cm$^2$ as estimated by numerical calculations and by observed relative ionization rates of Ar; details are given in the Supplementary Information (SI). We note that this intensity corresponds to a peak electric field strength of 3.4 $\pm$ 1.1 V/{\AA} (i.e, 0.065 $\pm$ 0.022 a.u.). The X-ray monochromator has 0.2 eV spectral resolution at the C K-edge and 0.1 eV resolution at the Cl L$_{2,3}$-edge. The experimental data are measured as a change in absorbance, or $\Delta$OD. The temporal cross-correlation of the experiment is measured to be $8\pm2$ fs by the autoionization in Ar L$_{2,3}$ lines\cite{Fidler2019FWM}. Experiments are performed over different time ranges to cover a wider range of dynamics; the shortest have a step size of 1 fs and extend to 80 fs and the longest have variable step sizes and extend to 10 ps. Additionally, experiments are run with varying pump power, from $\sim1-3\times10^{14}$ W/cm$^2$, to assess how the power affects the temporal dynamics. Each scan that has a different combination of pump powers, time step sizes, and gas pressures is referred to as a ``data set" in this paper. These data sets are taken as closely in time as possible to make the comparisons between them more consistent. \ce{CCl4} was obtained from Sigma-Aldrich at 99.5$\%$ purity. We note that some of the \ce{CCl4} is strong-field ionized under experimental conditions, leading to dissociative dynamics that have been detailed in Ref \citenum{ross2022jahn}.

The experimental data are noise filtered by two separate noise reduction algorithms, which are fully defined in the SI. In brief, the first is ``edge referencing," which involves using a region with no absorption signal and referencing it to find the correlated noise of the experiment. Then, the correlated noise is subtracted from the data\cite{Geneaux2021XUVNoise}. The second noise reduction is Fourier filtering, wherein a low-pass Fourier filtered spectrum is used as the reference spectrum when calculating $\Delta$OD, instead of a true pump-off reference\cite{ottNature2014}. This method reduces noise associated with fluctuations in X-ray flux from pulse to pulse and allows for better comparison of subsequent time points, critical for observation of vibration signals. Neither of these noise reduction algorithms correlates data between different time points; they only improve the $\Delta$OD signal within a single time point, so no spurious Fourier transform signals should be introduced by these techniques.

\begin{figure*}[htb!] \centering
\includegraphics[max height=\textheight/2,max width=\textwidth]{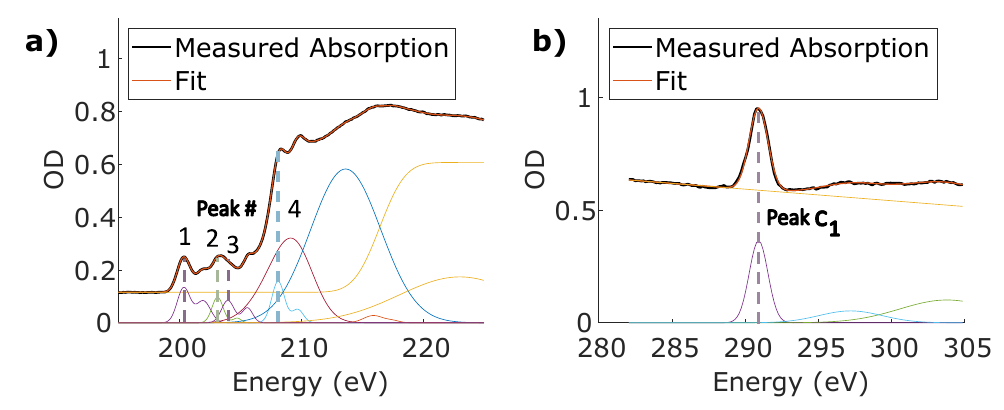}
\vspace*{-12pt}
\caption{\label{fig:CCl4 Absorption Breakdown}
X-ray absorption of \ce{CCl4} at the: a) Cl L$_{2,3}$-edges and b) C K-edge. Additionally, a breakdown of the net absorption into the individual transitions is shown (assignments provided in Table \ref{tab:CCl4 peak assignment}). Cl L$_3$ and L$_2$ excitation peaks have a spin-orbit induced separation of 1.6 eV. Extra Gaussians after the rising edge were used to better reproduce the measured data. 
}
\vspace*{-12pt}
\end{figure*}

The method for obtaining the CEPES slopes from the measured data involves the use of a purpose-built processing software. 
It starts by separating out the static OD data into a sum of Gaussians, as shown in Fig. \ref{fig:CCl4 Absorption Breakdown}. The $\Delta$OD data are simulated by moving the Gaussians in time according to a sine wave, corresponding to the vibration frequency with amplitude as a variable for the slopes in question. The difference between the simulated spectra OD and the measured static ground state OD reproduces the entire $\Delta$OD, across multiple data sets simultaneously. The algorithm is able to concurrently simulate the $\Delta$OD data for dissociation of \ce{CCl4+} and both the edge referencing and Fourier filtering noise reduction techniques, so any errors that might arise from either are minimized. Each fit includes convolution with a Gaussian of 8 fs FWHM to account for the temporal cross-correlation of the experiment and uses 95$\%$ confidence intervals as error bars. A full explanation of the fitting algorithm is provided in the SI.

\begin{table}[htb!]
\begin{tabular}{|l|l|l|l|}
\hline Symmetry & Character     & Experimental $\omega$ & Computed $\omega$                                 \\
 &     & (in cm$^{-1}$) & (in cm$^{-1}$)                                  \\ \hline 
e        & bend          & 218.0 ($\nu_2$)                               & \begin{tabular}[c]{@{}l@{}}224.0\\ 225.0\end{tabular}          \\
\hline
t$_2$    & bend          & 309.7 ($\nu_4$)                              & \begin{tabular}[c]{@{}l@{}}325.6\\ 325.7\\ 326.0\end{tabular} \\
\hline
a$_1$    & sym. stretch  & 456.6 ($\nu_1$)                              & 477.9                                                           \\
\hline
t$_2$    & asym. stretch & 790.6 ($\nu_3$)                              & \begin{tabular}[c]{@{}l@{}}791.9 \\ 794.2\\ 798.6 \end{tabular}\\ \hline 
\end{tabular}
\caption{Comparison of experimental\cite{CHAKRABORTY2002RamanCCl4} and computed (this work) normal mode frequencies for \ce{CCl4}, which are generally in good agreement. The degeneracy of normal modes is broken by the use of a finite grid for computation of DFT exchange-correlation integrals, which breaks rotational invariance. As a result, nine distinct frequencies are obtained from computation, clustered into four groups ($\nu_{1-4}$) corresponding to the experimental frequencies.}
\label{tab:normalmodes}
\end{table}

\subsection{Computational}
Quantum chemical calculations were performed with the Q-Chem software package\cite{epifanovsky2021software}, using the SCAN0 functional\cite{scan0} as it is accurate for the computation of both core-excitation energies\cite{hait2020highly,hait2021orbital} and polarizabilities\cite{hait2018accurate}. The equilibrium geometry and harmonic normal modes were found with the aug-pcseg-2 basis set\cite{jensen2014unifying}. We note that the four lowest energy unoccupied canonical molecular orbitals at this level of theory are antibonding in character and correspond to the 7a$_1^*$ and 8$t_2^*$ levels. ROKS calculations for core-excited states were carried out with the same functional, and a mixed basis approach\cite{hait2020highly} of using the decontracted aug-pcX-2 basis\cite{ambroise2018probing} on the atom type involved in the core-excitation and aug-pcseg-2 on the other atom (i.e. Cl L-edge calculations utilized aug-pcX-2 on Cl and aug-pcseg-2 on C).  Scalar relativistic effects in the core-excitation energy were accounted for with the exact two-component (X2C) method\cite{cunha2022relativistic,saue2011relativistic}. Spin-orbit effects in the Cl L-edge were simulated in the manner described in Ref \citenum{hait2020highly}. Excited state orbital optimization was done with the square gradient minimization (SGM) algorithm\cite{hait2020excited}. Local exchange-correlation integrals were calculated over a radial grid with 99 points and an angular Lebedev grid with 590 points. The CEPES slopes vs symmetric bond stretch were explicitly obtained by fitting the computed core-level excitation energies against the C-Cl bond length for a set of geometries spanning $\pm 0.01$ {\AA} change from the computed equilibrium separation of $1.759$ {\AA}, in increments of $0.001$ {\AA}. Fitting over a larger range of bond lengths ($\pm 0.05$ {\AA} vs equilibrium) yielded very similar slopes (as shown in the SI).

\section{\label{sec:level1c}Results and Discussion}

\subsection{Normal modes of \ce{CCl4}}
The experimental and computed normal modes of \ce{CCl4} are reported in Table \ref{tab:normalmodes}, and they are in reasonable agreement. All normal modes are Raman active\cite{CHAKRABORTY2002RamanCCl4} and have periods ranging from 42 fs (asym. stretch, $\nu_3$) to 153 fs (e bend, $\nu_2$). This indicates that they should all be stimulated by the 6 fs duration 800 nm pump pulse. 
\subsection{General Features of Experimental Spectra}

The ground state static X-ray absorption of \ce{CCl4} is shown in Figure \ref{fig:CCl4 Absorption Breakdown}. It shows the experimentally measured spectra and the separation of electronic transitions as sums of Voigt functions\cite{teodorescu1994approximation,iwamitsu2020spectral} for both the Cl L$_{2,3}$-edge (2p$^{-1}_{3/2}$ and 2p$^{-1}_{1/2}$ starting at 200 eV) and the C K-edge (1s$^{-1}$ starting around 285 eV\cite{lu2009core}). The Voigt functions are determined by fit, and at the Cl L$_{2,3}$-edge we use a second shifted profile with a different amplitude to model the spin-orbit (s-o) splitting of 1.6 eV. The spin-orbit area ratio differs from the expected 2:1 for the L$_3$:L$_2$\cite{briggs2012surface}, which is not unusual for some of these edges\cite{Hudson1993SF6Absorption}. The spectral peak assignments are given in Table \ref{tab:CCl4 peak assignment}\cite{hitchcock1978inner,lu2008state}. 

\begin{table}[htb!]
\begin{tabular}{|l|l|l|l|}
\hline
\multicolumn{4}{|c|}{\textbf{Peak Assignments}}\\ \hline
Peak & Exp. Energy (eV) & Calc. Energy (eV) & Assignment\\ \hline
1   & 200.35  & 200.44      & Cl 2p$_{3/2}\to 7a_1^*$\\ \hline
1$_{s-o}$    & 201.95  & 202.05    & Cl 2p$_{1/2}\to7a_1^*$\\ \hline
2    & 203.15   &  203.09   & Cl 2p$_{3/2}\to8t_2^*$\\ \hline
2$_{s-o}$    & 204.75   & 204.70   & Cl 2p$_{1/2}\to8t_2^*$ \\ \hline
3    & 203.98   &  - & Cl Rydberg\\ \hline
3$_{s-o}$    & 205.58  &  -  & Cl Rydberg\\ \hline
4    & 208.1    &  - & Shake-up State\\ \hline
4$_{s-o}$    & 209.7    & -  & Shake-up State\\ \hline
C$_1$  & 290.91  & 291.30    & C 1s$\to8t_2^*$\\ \hline
\end{tabular}

\caption{\label{tab:CCl4 peak assignment}
The assignments of the fitted absorption components, based on previous literature\cite{hitchcock1978inner,lu2008state}, compared to ROKS calculations. The Cl L$_2$-edge (s-o peaks) are 1.6 eV higher in energy than the corresponding  the L$_3$-edge peaks due to spin-orbit effects. 
}
\end{table}

The main features of interest in the Cl L-edge are Peaks 1 and 2 at 200.35 and 203.15 eV, which correspond to excitations to the $7a_1^*$ and $8t_2^*$ orbitals, respectively. Peak C$_1$ is the only feature of interest in the C K-edge, and it arises from the C 1s$\to 8t_2^*$ excitation. 
Unfortunately, the C 1s$\to 7a_1^*$ excitation is not a dipole allowed transition and so it cannot be probed by our experiment\cite{lu2009core}. In the Cl absorption, there are additional peaks, Peaks 3 and 4, that arise due to Rydberg transitions and shape resonances\cite{lu2009core}; although, the corresponding transient spectral features do not show prominent vibrational oscillations comparable to Peaks 1 and 2. The Cl L$_1$-edge ($\sim$ 260 eV) involving the 2s orbitals is also accessible with the probe, 
but that edge is too broad to extract useful information, especially for vibrational dynamics, where the transient signal strength resulting from coherent vibrational motion is proportional to the sharpness of the static absorption. In the static absorption, a downside of measuring simultaneous element absorptions becomes apparent; the absorption from the Cl edge extends all the way to the C K-edge, where it represents $\sim$50$\%$ of the absorption, even at the point of highest carbon absorption. This leads to smaller effective photon flux available at the C K-edge and increases the relative noise.

\begin{figure*}[!t] \centering
\includegraphics[max height=\textheight/2,max width=\textwidth]{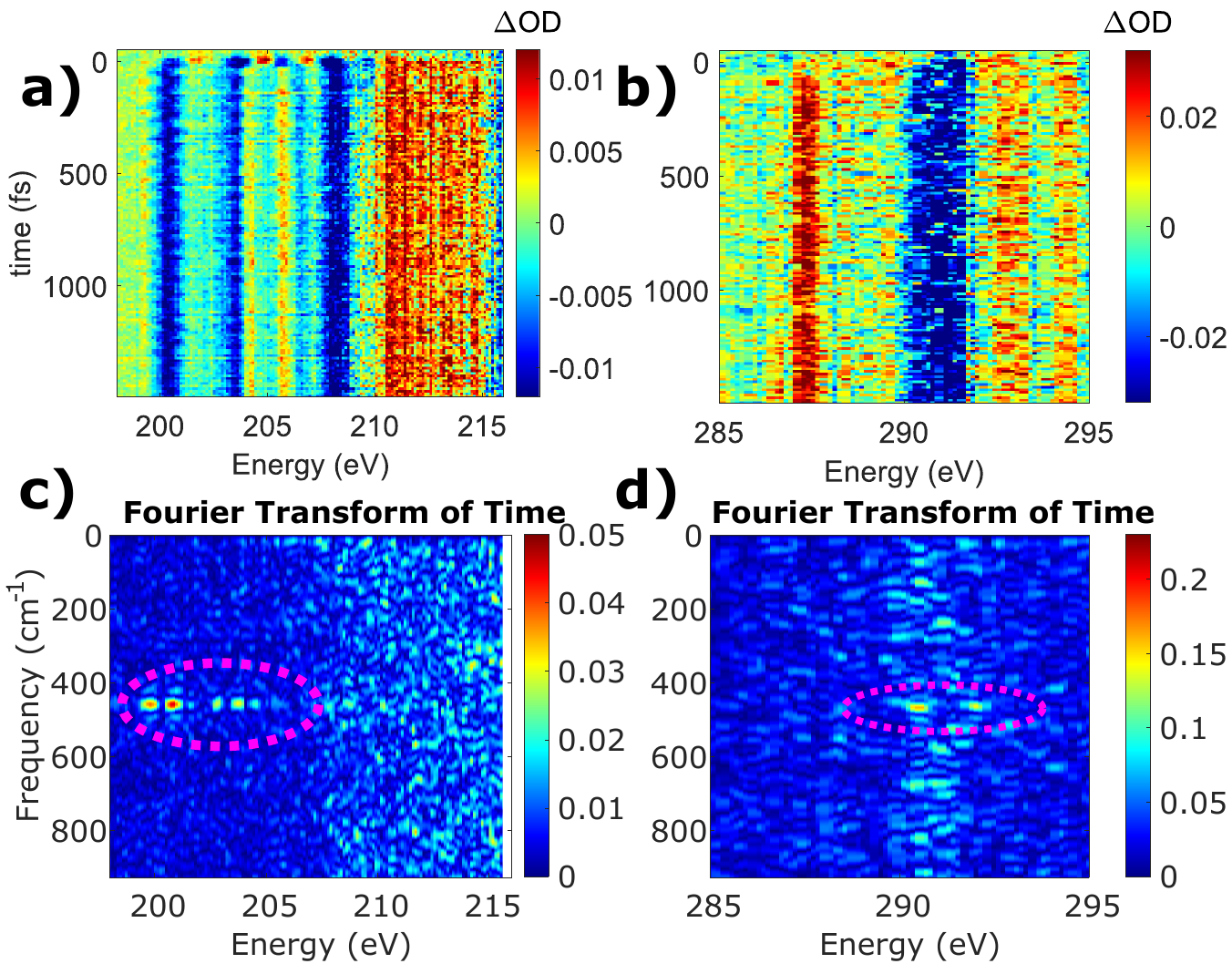}
\caption{\label{fig:CCl4 vibration dOD datastuffs}
a) and b) Example $\Delta$OD data at the Cl L$_{2,3}$-edges and C K-edge, respectively. Positive time indicates pump arriving before probe. Positive (red) $\Delta$OD signal indicates increased absorption, while negative (blue) indicates less absorption. The long-lived positive signals arise from dissociation products of \ce{CCl4+}. Vibrational dynamics correspond to the oscillatory signals that are most clear around 200-205 eV. c) and d) The above data is Fourier transformed along the time axis at each energy. Only the $\nu_1$ mode frequency is visible at 450 cm$^{-1}$ (with the pink dotted line aiding the eye). Other modes are not visible above the noise at or around 218 cm$^{-1}$, 791 cm$^{-1}$, nor 310 cm$^{-1}$ ($\nu_{2-4}$, respectively), nor at the double of those frequencies. The positions of the vibrating electronic states are shown in Fig. \ref{fig:CCl4 Absorption Breakdown} as peaks 1, 2, and C$_1$.}
\end{figure*}

The time-resolved changes to the static absorption spectrum, or $\Delta$OD, after excitation with the few-fs 800 nm pump pulse are shown in Figure \ref{fig:CCl4 vibration dOD datastuffs}. The blue areas show where the absorption decreases, and the red are where absorption increases. Several transient spectral features that are relatively constant in time after $t=0$ are apparent in both the Cl and C edges. These features arise from \ce{CCl4+} formed by pump induced strong-field ionization, whose Jahn-Teller and dissociative dynamics have already been studied in detail in a previous work\cite{ross2022jahn}.

The $\Delta$OD signals that are of interest to this paper are the oscillatory signals that occur surrounding the electronic transition peaks and their spin-orbit splittings at 200.35 eV, 203.15 eV, and 290.91 eV from Peaks 1, 2, and C$_1$, respectively. The positive-negative-etc. oscillations in time that are typical of vibrational features in XTAS\cite{barreau2023core} are clear at 200.35 eV in Fig. \ref{fig:CCl4 vibration dOD datastuffs} a) but are not as clear at some of the other energies, especially at the C K-edge. This is because of obscuring transient absorption from ionized species. We were not able to obtain high quality pure vibrational signals without also inducing substantial strong-field ionization. The effect of the cation signals are however accounted for in the fitting program, and the existence of the oscillations can be confirmed by the lower panels in Fig. \ref{fig:CCl4 vibration dOD datastuffs}, which show the Fourier transform in time at the different probe energies. 
Signals above the noise are apparent at 456 cm$^{-1}$ (73.05 fs), corresponding to the totally symmetric $a_1$ symmetry mode\cite{CHAKRABORTY2002RamanCCl4}, with the pink dotted oval aiding the eye. The Fourier transform signals appear as doublets to either side of the static absorption peak, but not at the peak center, as is most clear at 200.35 eV. This is due to the nature of these signals, which occur by essentially shifting the absorption peak center energy to higher and lower energies as the molecule vibrates. The edges of the absorption peak, where the derivative of absorption vs energy is the largest, produce the greatest signal, while the middle of the absorption peaks produce the smallest signal, as the derivative goes to zero. 

At longer times than those shown in Fig. \ref{fig:CCl4 vibration dOD datastuffs}, $\sim$3 ps, the different isotopic masses of Cl lead to dephasing and an inability to distinguish vibrations. At 10.5 ps, the various frequencies of different isotopes rephase as expected\cite{chakraborty2006comparative,gaynor2015vibrational} and the signal can be seen again, as shown in the SI. However, most data taken focus on the $<$1 ps range to circumvent this dephasing effect.

Notably, the Fourier transform does not reveal signals at frequencies corresponding to the $e$ and $t_2$ symmetry vibrations, for reasons that are discussed later. Fourier transformation of XTAS data often amplifies noise, so some regions may appear to show additional signals; however, extensive repetition of the experiment only shows reproducibility in the signals reported, and some additional data sets are shown in the SI.

\subsection{Core-Excited Potential Slopes vs Symmetric Stretch}

The experimentally measured signals were analyzed by the aforementioned software where the transient signal was recreated by simulating the vibrations and ionization signals simultaneously. The value that can be extracted from the data most directly is the change in central energy of each of the X-ray transition peaks as a result of vibrational dynamics. These shifts are on the order of 0.01 eV, which may seem quite small relative to the 0.1-0.2 eV resolution of the experiment but they induce quite large changes in $\Delta$OD. 

However, the peak center energy shifts alone do not define the desired CEPES slopes. It is also necessary to know the displacement along the vibrational coordinate or, in the case of the symmetric stretch, the C-Cl bond distance change induced by the pump pulse. That bond distance cannot be obtained directly from the XTAS data without prior knowledge of the CEPES, and so it needs to be estimated by other means. Typically, this stretch can be estimated from numerical solutions of the time-dependent Schr{\"o}dinger equation if the time-dependent intensity profile of the pump pulse is very well known\cite{schmidt2018wavepacket}, but this estimation introduces far more uncertainty than the XTAS under the present setup. 
Precise knowledge of the pulse intensity is challenging on account of losses on the annular mirror used for pump-probe co-linearity, non-Gaussian focal spots due to the annular mirror, losses on the \ce{CCl4} gas cell aperture, and perhaps most importantly, non-Gaussian temporal pulses. Because these experiments were undertaken with pump pulses close to a single cycle, third-order and higher dispersions have a large effect on the exact pulse envelope shape\cite{hong2004adaptive,meng2010enhanced}, and the tools to measure that envelope are not consistent enough in reconstruction to provide an exact pulse shape without error\cite{miranda2012characterization}. With these reservations, we numerically estimate the C-Cl bond length changes to be 0.0075 $\pm$ 0.0025 {\AA} (i.e. $\sim$ 33\% error) at the 3$\pm 1\times$10$^{14}$ W/cm$^2$ pump intensity, utilizing Ehrenfest's theorem\cite{ehrenfest1927bemerkung,shankar2012principles}. This is consistent with previously derived analytic results for Gaussian pump pulse envelopes\cite{yan1985impulsive}.

\begin{figure}[h] \centering
\includegraphics[max height=\textheight/2,max width=\columnwidth]{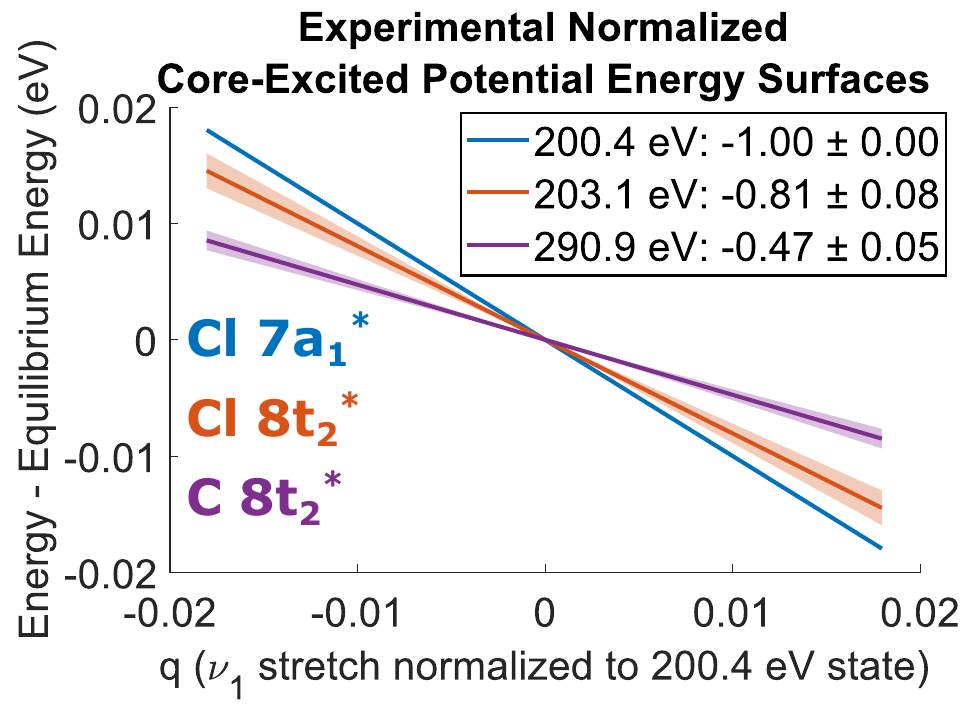}
\caption{\label{fig:CCl4 CEPES}
Extracted values for shifts in the excitation energies (vs the equilibrium geometry) for each core-excited state (blue: Cl 2p$_{3/2} \to 7a_1^*$, red: Cl 2p$_{3/2} \to 8t_2^*$, purple: C 1s$ \to 8t_2^*$), via linear fits. The shading shows 95$\%$ confidence intervals. The vibrational coordinate, q, is scaled such that the Cl 2p$_{3/2} \to 7a_1^*$ slope is definitionally -1.}
\end{figure}

It is possible to bypass errors in bond stretch estimation by comparing different core excited states relative to each other for a given data set, as the signal arises from the same vibrational motion. 
By normalizing to one of the CEPES as discussed in the introduction (Eqn. \ref{eq:ratio}),  
the bond length stretch parameter can be eliminated. This allows comparison of the CEPES slopes purely based on the XTAS data. For the purposes of this paper, this means we define the slope of the first state, Cl 2p$_{3/2} \to 7a_1^*$, to -1 eV/q, where q is an arbitrary bond stretch lengthscale. The negative sign is explicitly chosen as the core-level excitation energy to an antibonding orbital is expected to monotonically decrease with bond length increase\cite{barreau2023core,vaz2019probing,schreck2016ground,miron2012imaging,ross2022jahn}, as the energy of the associated antibonding levels is lowered. 
The experimental relative slopes subsequently are: Cl 2p$_{3/2} \to 7a_1^*$: -1, Cl 2p$_{3/2} \to 8t_2^*$: -0.81 $\pm$ 0.08, and C 1s$ \to 8t_2^*$: -0.47 $\pm$ 0.05 (to 95\% confidence), which are also shown in Fig. \ref{fig:CCl4 CEPES} and Table \ref{table:CCl4 coefficient table}. The overlap of absorption features in the Cl L$_{2,3}$-edge static spectrum and the lower signal-to-noise, especially at the C K-edge, leads to somewhat sizeable error bars. 
These slopes show a significantly larger change for the Cl L-edge excitation energies along the normal coordinate compared to the C K-edge excitation.

\begin{table}[h]
\begin{tabular}{|l|l|l|l|}
\hline
\multicolumn{4}{|c|}{\textbf{Core-Excited Potential Energy }}\\
\multicolumn{4}{|c|}{\textbf{Surfaces Slopes for $\nu_1$}}\\ \hline
\begin{tabular}[c]{@{}l@{}}Slope\\ \end{tabular} & \begin{tabular}[c]{@{}l@{}}Experimental\\ Relative\end{tabular} & \begin{tabular}[c]{@{}l@{}}Calculation\\ Relative\end{tabular}& \begin{tabular}[c]{@{}l@{}}Calculation\\ Absolute (in eV/{\AA})\end{tabular} \\ \hline
Cl 2p$_{3/2} \to 7a_1^*$                                                     & -1                                                            & -1    & -11.13                                                     \\ \hline
Cl 2p$_{3/2} \to 8t_2^*$                                                     & -0.81$\pm$0.08                                                & -0.79                              & -8.81                        \\ \hline
C 1s$ \to 8t_2^*$                                                            & -0.47$\pm$0.05                                                & -0.48                               & -5.30                       \\ \hline
\end{tabular}
\caption{\label{table:CCl4 coefficient table}
The experimentally extracted slopes for \ce{CCl4}, as compared to ROKS calculations. 
}
\end{table}

These experimental extracted relative slopes can be compared to absolute slopes from ROKS calculations. The computed absolute slopes for core-excitation energies vs C-Cl symmetric bond stretch are Cl 2p$_{3/2} \to 7a_1^*$: -11.13 eV/{\AA}, Cl 2p$_{3/2} \to 8t_2^*$: -8.81 eV/{\AA}, and C 1s$ \to 8t_2^*$: -5.30 eV/{\AA}. The computed relative slopes for these states therefore are: Cl 2p$_{3/2} \to 7a_1^*$: -1, Cl 2p$_{3/2} \to 8t_2^*$: -0.79, and C 1s$ \to 8t_2^*$: -0.48, which agree very well with the experimental values (as shown in Table  \ref{table:CCl4 coefficient table}).  

Encouraged by the strong agreement between experiment and theory for relative slopes, we apply the theoretical absolute values of CEPES slopes to the experimentally extracted relative values, which yields estimated experimental absolute slopes of Cl 2p$_{3/2} \to 7a_1^*$: -11.13 eV/{\AA}, Cl 2p$_{3/2} \to 8t_2^*$: -9.0 eV/{\AA}, and C 1s$ \to 8t_2^*$: -5.2 eV/{\AA}. Both experimental relative slopes have $\sim 10\%$ uncertainty, and if we assume that this arises from equal percentage uncertainties for both the numerator and denominator, we obtain the 95\% error bars to be $\pm$ 0.8 eV/{\AA}, $\pm$ 0.6 eV/{\AA}, and $\pm$ 0.4 eV/{\AA}, respectively for the slopes of the three states. We note that the assumption that all three absolute slopes have the same relative uncertainty is not necessarily correct (especially considering the lower signal to noise in the C K-edge region), but the relative errors in the individual CEPES slopes are less than the $\sim$10\% error in the relative slopes, if the errors between slopes of different states are uncorrelated. 

From the estimates for the absolute slopes, the amount of nuclear movement can also be estimated for each experimental data set. The smallest symmetric bond stretch amplitude across all data sets was 0.0018 {\AA} and the largest was 0.0045 {\AA}. These numbers seem very small, but they are close to the numeric solution of the Schr{\"o}dinger equation, which yielded 0.0075 {\AA}. The difference likely arises from errors in the expected intensity or deviations in the pump pulse shape from a perfect Gaussian envelope or, from an experimental perspective, the use of an incorrect pump pulse temporal broadening in the extraction. These bond length changes result in only $\sim$0.05 eV changes in the absorption peak centers from the ground state, but the resulting transient signal is measurable by the experiment. The $<0.01$ {\AA} pump induced bond stretch also indicates that the first-order Taylor series approximation utilized in Eqn. \ref{eq:ratio} is likely valid, and higher order terms in bond stretch likely have a negligible impact on the transient X-ray absorption signal arising from vibrational dynamics of neutral \ce{CCl4}. 

It is also interesting to consider why the CEPES slopes for the Cl 2p$_{3/2} \to 7a_1^*$ and 2p$_{3/2} \to 8t_2^*$ states are much larger than the corresponding slope for the C 1s$ \to 8t_2^*$ excitation. This can be possibly rationalized in terms of the composition of the antibonding orbital to which the core-electron is being excited, as the core-excitation energy is expected to be sensitive to this orbital. Subshell L{\"o}wdin population analysis\cite{lowdin1950non} on the relaxed orbitals obtained from ROKS indicate that the $7a_1^*$ orbital of the Cl 2p$_{3/2} \to 7a_1^*$ excited state and the $8t_2^*$ orbital of both the Cl 2p$_{3/2} \to 8t_2^*$ and C 1s$ \to 8t_2^*$ excited states are of $\sim 33-37\%$ carbon character and $\sim 63-67\%$ Cl character. The greater contribution from Cl to these antibonding orbitals likely leads to the Cl 2p hole more strongly affecting the core-excitation potential than a C 1s hole, leading to larger slopes for Cl L-edge states than for the C K-edge states. We note that the Cl contribution to the antibonding levels is not very different for the Cl 2p$_{3/2} \to 7a_1^*$ and 2p$_{3/2} \to 8t_2^*$ excited states, and therefore this heuristic thus does not explain the 20\% difference between the two Cl L-edge slopes. However, we believe that the large differences in behavior between CEPES slopes arising from different atom types in a molecule can be rationalized by an analysis of the composition of the orbital to which the core-electrons are being excited, post orbital relaxation in response to the formation of the core-hole.

\subsection{The lack of signal from other vibrational modes}
The Fourier transform of the experimental XTAS does not show any perceptible signal associated with the normal modes other than the symmetric stretch, despite all the modes being Raman active\cite{chakraborty2006comparative}. This behavior can arise either from  the center of the vibrational wavepacket not coherently oscillating along asymmetric modes, or from very small CEPES slopes along these modes. We find both factors to be relevant for the present study.  

\ce{CCl4} has no permanent dipole and possesses an experimental\cite{yamamoto1984diode} rotational constant of 0.057 cm$^{-1}$. The molecule consequently cannot rotationally reorient within the duration of the 6 fs FWHM 800 nm pump pulse and thus experiences a spatially isotropic electric field $\mathcal{E}(t)$ that lowers the energy by $\dfrac{1}{2}\alpha_{iso}(\{q_i\})|\mathcal{E}(t)|^2$ (where $\alpha_{iso}(\{q_i\})$ is the isotropic polarizability). 
Small bond stretches are induced by $\mathcal{E}(t)$ as the resulting increase in polarizability\cite{hait2023bond} can provide stabilization. 
It is however not possible for an isotropic electric field to spontaneously break the $T_d$ symmetry of \ce{CCl4}, implying that $\left(\dfrac{d \alpha_{iso}}{d q_i}\right)_{\{q_i\}=0}=0$ for the asymmetric modes. 
Ehrenfest's theorem\cite{ehrenfest1927bemerkung,shankar2012principles} therefore indicates that there is no field-induced force to displace the center of the vibrational wavefunction along the asymmetric modes, and ISRS cannot produce coherent oscillation of the wavefunction center\cite{yan1985impulsive} (see discussion in the SI). This is confirmed by time-dependent Schr{\"o}dinger equation simulations utilizing the full polarizability tensor (shown in SI), which demonstrate that the wavepacket center  position $\langle \{q_i\}\rangle(t)$ only oscillates along the symmetric stretch.  ISRS can however produce oscillations in the spatial width of the wavefunction along all normal modes, which occur at twice the normal mode frequency.

In addition, we note that small displacements in the wavepacket center along the asymmetric modes would nonetheless not lead to a strong response in the XTAS for \ce{CCl4} because of degeneracies between core-excited states that contribute to the spectral features. The C 1s$\to 8t_2^*$ excitation is triply degenerate and the individual constituent states should therefore possess CEPES forces along some  asymmetric modes from the Jahn-Teller effect\cite{jahn1937stability}. However, the observed C$_1$ peak in Fig. \ref{fig:CCl4 Absorption Breakdown} would not separate into three distinct, non-overlapping entities in the limit of small asymmetric mode displacements and any observed shift in the peak center position would represent an average over all three core-excited states. This scenario also applies to the Cl L-edge, where the presence of 4 Cl atoms similarly leads to a number of degenerate or nearly degenerate states that contribute to the observed spectral features like Peaks 1 and 2 in Fig. \ref{fig:CCl4 Absorption Breakdown}. It is possible for the individual states to have CEPES slopes of opposite signs along asymmetric modes (e.g. an asymmetric stretching mode elevating the energy of one C 1s$\to 8t_2^*$ state while reducing the energy of the other two), leading to small \textit{average} slopes and consequently small shifts in the peak center energy. 

To investigate effects of averaging CEPES slopes over (nearly) degenerate core-excited states, we have performed nonrelativistic CEPES gradient calculations with ROKS\cite{hait2020highly,kowalczyk2013excitation} for all three C 1s$\to 8t_2^*$ excitations and for all twelve possible Cl 2p$ \to 7a_1^*$ excitations. Projecting these gradients onto unit vectors along each normal mode direction shows that CEPES slopes averaged over all C 1s$\to 8t_2^*$ or Cl 2p$ \to 7a_1^*$ excitations are essentially zero ($\sim 0.03$ eV/{\AA} or smaller, as shown in the SI) along the asymmetric modes.
We do note that even though the average slope along any asymmetric mode is quite small, individual C 1s and Cl 2p excited states may have substantial slopes along some modes. This could lead to time-dependent changes in the width of XTAS features as the energies of the constituent states move further apart (closer together) with displacement from (towards) T$_d$ symmetry. Such effects should lead to signal at twice the normal mode frequency, but were not experimentally detected above the noise.

These issues raise an important question as to whether all vibrational modes can be reliably measured with XTAS in general molecular systems. In our view, a system with reduced symmetry where core-excitation energies are well-separated ought not to show either of the previously discussed issues. For instance, all the vibrational modes of a molecule like CFClBrI will be totally symmetric ($a$ symmetry) and the C 1s absorption features will arise from nondegenerate core-level excitations. Such a molecule would therefore be ideal for CEPES slope measurements with XTAS subsequent to vibrational activation. It would nonetheless be quite helpful to have a significant force/slope on the CEPES at the equilibrium geometry, suggesting that core-level excitations involving antibonding orbitals would be ideal targets.

\section{Conclusions}

In this work, the impulsive stimulated Raman scattering induced coherent vibrations of \ce{CCl4} have been studied with X-ray transient absorption spectroscopy, and linear slopes of the core-excited potential energy surfaces along the totally symmetric bond stretching mode have been extracted. Normalizing to the first state, these are Cl 2p$_{3/2} \to 7a_1^*$: -1, Cl 2p$_{3/2} \to 8t_2^*$: -0.81 $\pm$ 0.08, and C 1s$ \to 8t_2^*$: -0.47 $\pm$ 0.05, which compares well with theoretical values of -1, -0.79, and -0.48, respectively. Scaling the experimental relative slopes to the theoretical absolute slopes give $-11.1\pm 0.8$ eV/{\AA}, $-9.0\pm0.6$ eV/{\AA}, and -$5.2\pm 0.4$ eV/{\AA}, respectively. These slopes indicate a small ISRS induced totally symmetric bond elongation of $<0.01$ {\AA}, in conjunction with the experimentally observed shifts in peak center energies ($\sim$ 0.05 eV). The ability of XTAS to report on the dynamics arising from such small displacements indicate the sensitivity of this approach for studying the X-ray potential response to the dynamics of nuclear motion. 

The larger slopes for Cl L-edge excitations relative to the C K-edge excitation is likely a result of the excited electron occupancy of the  $7a_1^*/8t_2^*$ levels that ROKS calculations indicate have a $\sim 33\%$ contribution from C and $\sim 66\%$ contribution from Cl and thereby are more sensitive to the presence of a Cl core hole. It would be interesting to explore whether similar relationships between CEPES slopes can be observed in other molecules where multiple different inner-shell edges can be probed. 

We note that no XTAS signal from vibrational dynamics from other normal modes of \ce{CCl4} is experimentally detected. This appears to be a consequence of the ISRS process not displacing the center of the vibrational wavefunction along these modes. It also appears that there is no or little CEPES slope along the asymmetric modes after averaging over all (near) degenerate core-level excited states that contribute to the observed spectral features, leading to lack of transient X-ray absorption under current experimental conditions. Less symmetric molecules with well-resolved core-excitation energies at the equilibrium geometry would be unlikely to have such issues and therefore are likely to reveal XTAS signals corresponding to vibrational dynamics of many more modes. 

\section*{Acknowledgments}
This work is funded by the DOE Office of Science, Basic Energy Science (BES) Program, Chemical Sciences,
Geosciences and Biosciences Division under Contract no. DE-AC02-05CH11231, through the Gas Phase Chemical
Physics program (ADR, VS, DMN, SRL) and Atomic, Molecular, and Optical Sciences program (DH and MHG).
The instrument was built with funds from the National Science Foundation through NSF MRI 1624322 and matching
funds from the Lawrence Berkeley National Laboratory, the College of Chemistry, the Department of Physics, and the
Vice Chancellor for Research at UC Berkeley. This research used resources of the National Energy Research Scientific
Computing Center, a DOE Office of Science User Facility supported by the Office of Science of the U.S. Department of
Energy under Contract No. DE-AC02-05CH11231 using NERSC award BES-ERCAP0027716. ADR was additionally
funded by the U.S. Department of Energy, Office of Science, Office of Basic Energy Sciences, Materials Sciences
and Engineering Division, under Contract No. DE-AC02-05-CH11231 within the Physical Chemistry of Inorganic
Nanostructures Program (KC3103), by the W. M. Keck Foundation Grant No. 042982, and by the U.S. Army
Research Office (ARO) under Grant No. W911NF-20-1-0127. SRL acknowledges related support of NSF grant CHE-2243756.

\section*{Data Availability}
The Supplementary Information contains additional information about the experiment and the time-dependent Schr{\"o}dinger equation simulations, as well as the results from these simulations (PDF). The computed X-ray absorption energies at various (symmetric) bond distances that were used to obtain the computed slopes are also provided (XLXS). 
The code for the time-dependent Schr{\"o}dinger equation simulations and the molecular orbitals optimized by ROKS at the ground state equilibrium geometry have been made publicly available via Zenodo\cite{ross_2024_11153002}.
\bibliography{biblio}

\begin{thebibliography}{80}%
\makeatletter
\providecommand \@ifxundefined [1]{%
 \@ifx{#1\undefined}
}%
\providecommand \@ifnum [1]{%
 \ifnum #1\expandafter \@firstoftwo
 \else \expandafter \@secondoftwo
 \fi
}%
\providecommand \@ifx [1]{%
 \ifx #1\expandafter \@firstoftwo
 \else \expandafter \@secondoftwo
 \fi
}%
\providecommand \natexlab [1]{#1}%
\providecommand \enquote  [1]{``#1''}%
\providecommand \bibnamefont  [1]{#1}%
\providecommand \bibfnamefont [1]{#1}%
\providecommand \citenamefont [1]{#1}%
\providecommand \href@noop [0]{\@secondoftwo}%
\providecommand \href [0]{\begingroup \@sanitize@url \@href}%
\providecommand \@href[1]{\@@startlink{#1}\@@href}%
\providecommand \@@href[1]{\endgroup#1\@@endlink}%
\providecommand \@sanitize@url [0]{\catcode `\\12\catcode `\$12\catcode `\&12\catcode `\#12\catcode `\^12\catcode `\_12\catcode `\%12\relax}%
\providecommand \@@startlink[1]{}%
\providecommand \@@endlink[0]{}%
\providecommand \url  [0]{\begingroup\@sanitize@url \@url }%
\providecommand \@url [1]{\endgroup\@href {#1}{\urlprefix }}%
\providecommand \urlprefix  [0]{URL }%
\providecommand \Eprint [0]{\href }%
\providecommand \doibase [0]{https://doi.org/}%
\providecommand \selectlanguage [0]{\@gobble}%
\providecommand \bibinfo  [0]{\@secondoftwo}%
\providecommand \bibfield  [0]{\@secondoftwo}%
\providecommand \translation [1]{[#1]}%
\providecommand \BibitemOpen [0]{}%
\providecommand \bibitemStop [0]{}%
\providecommand \bibitemNoStop [0]{.\EOS\space}%
\providecommand \EOS [0]{\spacefactor3000\relax}%
\providecommand \BibitemShut  [1]{\csname bibitem#1\endcsname}%
\let\auto@bib@innerbib\@empty
\bibitem [{\citenamefont {Bhattacherjee}\ and\ \citenamefont {Leone}(2018)}]{bhattacherjee2018ultrafast}%
  \BibitemOpen
  \bibfield  {author} {\bibinfo {author} {\bibfnamefont {A.}~\bibnamefont {Bhattacherjee}}\ and\ \bibinfo {author} {\bibfnamefont {S.~R.}\ \bibnamefont {Leone}},\ }\bibfield  {title} {\enquote {\bibinfo {title} {Ultrafast x-ray transient absorption spectroscopy of gas-phase photochemical reactions: A new universal probe of photoinduced molecular dynamics},}\ }\href@noop {} {\bibfield  {journal} {\bibinfo  {journal} {Accounts of chemical research}\ }\textbf {\bibinfo {volume} {51}},\ \bibinfo {pages} {3203--3211} (\bibinfo {year} {2018})}\BibitemShut {NoStop}%
\bibitem [{\citenamefont {Geneaux}\ \emph {et~al.}(2019)\citenamefont {Geneaux}, \citenamefont {Marroux}, \citenamefont {Guggenmos}, \citenamefont {Neumark},\ and\ \citenamefont {Leone}}]{Geneaux2019ATASReview}%
  \BibitemOpen
  \bibfield  {author} {\bibinfo {author} {\bibfnamefont {R.}~\bibnamefont {Geneaux}}, \bibinfo {author} {\bibfnamefont {H.~J.~B.}\ \bibnamefont {Marroux}}, \bibinfo {author} {\bibfnamefont {A.}~\bibnamefont {Guggenmos}}, \bibinfo {author} {\bibfnamefont {D.~M.}\ \bibnamefont {Neumark}},\ and\ \bibinfo {author} {\bibfnamefont {S.~R.}\ \bibnamefont {Leone}},\ }\bibfield  {title} {\enquote {\bibinfo {title} {Transient absorption spectroscopy using high harmonic generation: a review of ultrafast x-ray dynamics in molecules and solids},}\ }\href {https://doi.org/10.1098/rsta.2017.0463} {\bibfield  {journal} {\bibinfo  {journal} {Philosophical Transactions of the Royal Society A: Mathematical, Physical and Engineering Sciences}\ }\textbf {\bibinfo {volume} {377}},\ \bibinfo {pages} {20170463} (\bibinfo {year} {2019})}\BibitemShut {NoStop}%
\bibitem [{\citenamefont {Bressler}\ and\ \citenamefont {Chergui}(2004)}]{bressler2004ultrafast}%
  \BibitemOpen
  \bibfield  {author} {\bibinfo {author} {\bibfnamefont {C.}~\bibnamefont {Bressler}}\ and\ \bibinfo {author} {\bibfnamefont {M.}~\bibnamefont {Chergui}},\ }\bibfield  {title} {\enquote {\bibinfo {title} {Ultrafast x-ray absorption spectroscopy},}\ }\href@noop {} {\bibfield  {journal} {\bibinfo  {journal} {Chemical reviews}\ }\textbf {\bibinfo {volume} {104}},\ \bibinfo {pages} {1781--1812} (\bibinfo {year} {2004})}\BibitemShut {NoStop}%
\bibitem [{\citenamefont {De~Groot}(2001)}]{de2001high}%
  \BibitemOpen
  \bibfield  {author} {\bibinfo {author} {\bibfnamefont {F.}~\bibnamefont {De~Groot}},\ }\bibfield  {title} {\enquote {\bibinfo {title} {High-resolution x-ray emission and x-ray absorption spectroscopy},}\ }\href@noop {} {\bibfield  {journal} {\bibinfo  {journal} {Chemical Reviews}\ }\textbf {\bibinfo {volume} {101}},\ \bibinfo {pages} {1779--1808} (\bibinfo {year} {2001})}\BibitemShut {NoStop}%
\bibitem [{\citenamefont {Ross}\ \emph {et~al.}(2022)\citenamefont {Ross}, \citenamefont {Hait}, \citenamefont {Scutelnic}, \citenamefont {Haugen}, \citenamefont {Ridente}, \citenamefont {Balkew}, \citenamefont {Neumark}, \citenamefont {Head-Gordon},\ and\ \citenamefont {Leone}}]{ross2022jahn}%
  \BibitemOpen
  \bibfield  {author} {\bibinfo {author} {\bibfnamefont {A.~D.}\ \bibnamefont {Ross}}, \bibinfo {author} {\bibfnamefont {D.}~\bibnamefont {Hait}}, \bibinfo {author} {\bibfnamefont {V.}~\bibnamefont {Scutelnic}}, \bibinfo {author} {\bibfnamefont {E.~A.}\ \bibnamefont {Haugen}}, \bibinfo {author} {\bibfnamefont {E.}~\bibnamefont {Ridente}}, \bibinfo {author} {\bibfnamefont {M.~B.}\ \bibnamefont {Balkew}}, \bibinfo {author} {\bibfnamefont {D.~M.}\ \bibnamefont {Neumark}}, \bibinfo {author} {\bibfnamefont {M.}~\bibnamefont {Head-Gordon}},\ and\ \bibinfo {author} {\bibfnamefont {S.~R.}\ \bibnamefont {Leone}},\ }\bibfield  {title} {\enquote {\bibinfo {title} {Jahn-teller distortion and dissociation of ccl 4+ by transient x-ray spectroscopy simultaneously at the carbon k-and chlorine l-edge},}\ }\href@noop {} {\bibfield  {journal} {\bibinfo  {journal} {Chemical science}\ }\textbf {\bibinfo {volume} {13}},\ \bibinfo {pages} {9310--9320} (\bibinfo {year} {2022})}\BibitemShut {NoStop}%
\bibitem [{\citenamefont {Henderson}, \citenamefont {De~Groot},\ and\ \citenamefont {Moulton}(2014)}]{henderson2014x}%
  \BibitemOpen
  \bibfield  {author} {\bibinfo {author} {\bibfnamefont {G.~S.}\ \bibnamefont {Henderson}}, \bibinfo {author} {\bibfnamefont {F.~M.}\ \bibnamefont {De~Groot}},\ and\ \bibinfo {author} {\bibfnamefont {B.~J.}\ \bibnamefont {Moulton}},\ }\bibfield  {title} {\enquote {\bibinfo {title} {X-ray absorption near-edge structure (xanes) spectroscopy},}\ }\href@noop {} {\bibfield  {journal} {\bibinfo  {journal} {Reviews in Mineralogy and Geochemistry}\ }\textbf {\bibinfo {volume} {78}},\ \bibinfo {pages} {75--138} (\bibinfo {year} {2014})}\BibitemShut {NoStop}%
\bibitem [{\citenamefont {Wollan}(1932)}]{wollan1932x}%
  \BibitemOpen
  \bibfield  {author} {\bibinfo {author} {\bibfnamefont {E.~O.}\ \bibnamefont {Wollan}},\ }\bibfield  {title} {\enquote {\bibinfo {title} {X-ray scattering and atomic structure},}\ }\href@noop {} {\bibfield  {journal} {\bibinfo  {journal} {Reviews of Modern Physics}\ }\textbf {\bibinfo {volume} {4}},\ \bibinfo {pages} {205} (\bibinfo {year} {1932})}\BibitemShut {NoStop}%
\bibitem [{\citenamefont {Sch{\"u}lke}(2007)}]{schulke2007electron}%
  \BibitemOpen
  \bibfield  {author} {\bibinfo {author} {\bibfnamefont {W.}~\bibnamefont {Sch{\"u}lke}},\ }\href@noop {} {\emph {\bibinfo {title} {Electron dynamics by inelastic X-ray scattering}}},\ Vol.~\bibinfo {volume} {7}\ (\bibinfo  {publisher} {OUP Oxford},\ \bibinfo {year} {2007})\BibitemShut {NoStop}%
\bibitem [{\citenamefont {H{\"a}hner}(2006)}]{hahner2006near}%
  \BibitemOpen
  \bibfield  {author} {\bibinfo {author} {\bibfnamefont {G.}~\bibnamefont {H{\"a}hner}},\ }\bibfield  {title} {\enquote {\bibinfo {title} {Near edge x-ray absorption fine structure spectroscopy as a tool to probe electronic and structural properties of thin organic films and liquids},}\ }\href@noop {} {\bibfield  {journal} {\bibinfo  {journal} {Chemical Society Reviews}\ }\textbf {\bibinfo {volume} {35}},\ \bibinfo {pages} {1244--1255} (\bibinfo {year} {2006})}\BibitemShut {NoStop}%
\bibitem [{\citenamefont {Gaffney}\ and\ \citenamefont {Chapman}(2007)}]{gaffney2007imaging}%
  \BibitemOpen
  \bibfield  {author} {\bibinfo {author} {\bibfnamefont {K.}~\bibnamefont {Gaffney}}\ and\ \bibinfo {author} {\bibfnamefont {H.~N.}\ \bibnamefont {Chapman}},\ }\bibfield  {title} {\enquote {\bibinfo {title} {Imaging atomic structure and dynamics with ultrafast x-ray scattering},}\ }\href@noop {} {\bibfield  {journal} {\bibinfo  {journal} {science}\ }\textbf {\bibinfo {volume} {316}},\ \bibinfo {pages} {1444--1448} (\bibinfo {year} {2007})}\BibitemShut {NoStop}%
\bibitem [{\citenamefont {Kraus}\ \emph {et~al.}(2018)\citenamefont {Kraus}, \citenamefont {Z{\"u}rch}, \citenamefont {Cushing}, \citenamefont {Neumark},\ and\ \citenamefont {Leone}}]{kraus2018ultrafast}%
  \BibitemOpen
  \bibfield  {author} {\bibinfo {author} {\bibfnamefont {P.~M.}\ \bibnamefont {Kraus}}, \bibinfo {author} {\bibfnamefont {M.}~\bibnamefont {Z{\"u}rch}}, \bibinfo {author} {\bibfnamefont {S.~K.}\ \bibnamefont {Cushing}}, \bibinfo {author} {\bibfnamefont {D.~M.}\ \bibnamefont {Neumark}},\ and\ \bibinfo {author} {\bibfnamefont {S.~R.}\ \bibnamefont {Leone}},\ }\bibfield  {title} {\enquote {\bibinfo {title} {The ultrafast x-ray spectroscopic revolution in chemical dynamics},}\ }\href@noop {} {\bibfield  {journal} {\bibinfo  {journal} {Nature Reviews Chemistry}\ }\textbf {\bibinfo {volume} {2}},\ \bibinfo {pages} {82--94} (\bibinfo {year} {2018})}\BibitemShut {NoStop}%
\bibitem [{\citenamefont {Bressler}\ and\ \citenamefont {Chergui}(2010)}]{bressler2010molecular}%
  \BibitemOpen
  \bibfield  {author} {\bibinfo {author} {\bibfnamefont {C.}~\bibnamefont {Bressler}}\ and\ \bibinfo {author} {\bibfnamefont {M.}~\bibnamefont {Chergui}},\ }\bibfield  {title} {\enquote {\bibinfo {title} {Molecular structural dynamics probed by ultrafast x-ray absorption spectroscopy},}\ }\href@noop {} {\bibfield  {journal} {\bibinfo  {journal} {Annual review of physical chemistry}\ }\textbf {\bibinfo {volume} {61}},\ \bibinfo {pages} {263--282} (\bibinfo {year} {2010})}\BibitemShut {NoStop}%
\bibitem [{\citenamefont {Ridente}\ \emph {et~al.}(2023)\citenamefont {Ridente}, \citenamefont {Hait}, \citenamefont {Haugen}, \citenamefont {Ross}, \citenamefont {Neumark}, \citenamefont {Head-Gordon},\ and\ \citenamefont {Leone}}]{ridente2023femtosecond}%
  \BibitemOpen
  \bibfield  {author} {\bibinfo {author} {\bibfnamefont {E.}~\bibnamefont {Ridente}}, \bibinfo {author} {\bibfnamefont {D.}~\bibnamefont {Hait}}, \bibinfo {author} {\bibfnamefont {E.~A.}\ \bibnamefont {Haugen}}, \bibinfo {author} {\bibfnamefont {A.~D.}\ \bibnamefont {Ross}}, \bibinfo {author} {\bibfnamefont {D.~M.}\ \bibnamefont {Neumark}}, \bibinfo {author} {\bibfnamefont {M.}~\bibnamefont {Head-Gordon}},\ and\ \bibinfo {author} {\bibfnamefont {S.~R.}\ \bibnamefont {Leone}},\ }\bibfield  {title} {\enquote {\bibinfo {title} {Femtosecond symmetry breaking and coherent relaxation of methane cations via x-ray spectroscopy},}\ }\href@noop {} {\bibfield  {journal} {\bibinfo  {journal} {Science}\ }\textbf {\bibinfo {volume} {380}},\ \bibinfo {pages} {713--717} (\bibinfo {year} {2023})}\BibitemShut {NoStop}%
\bibitem [{\citenamefont {Zinchenko}\ \emph {et~al.}(2021)\citenamefont {Zinchenko}, \citenamefont {Ardana-Lamas}, \citenamefont {Seidu}, \citenamefont {Neville}, \citenamefont {van~der Veen}, \citenamefont {Lanfaloni}, \citenamefont {Schuurman},\ and\ \citenamefont {W{\"o}rner}}]{zinchenko2021sub}%
  \BibitemOpen
  \bibfield  {author} {\bibinfo {author} {\bibfnamefont {K.~S.}\ \bibnamefont {Zinchenko}}, \bibinfo {author} {\bibfnamefont {F.}~\bibnamefont {Ardana-Lamas}}, \bibinfo {author} {\bibfnamefont {I.}~\bibnamefont {Seidu}}, \bibinfo {author} {\bibfnamefont {S.~P.}\ \bibnamefont {Neville}}, \bibinfo {author} {\bibfnamefont {J.}~\bibnamefont {van~der Veen}}, \bibinfo {author} {\bibfnamefont {V.~U.}\ \bibnamefont {Lanfaloni}}, \bibinfo {author} {\bibfnamefont {M.~S.}\ \bibnamefont {Schuurman}},\ and\ \bibinfo {author} {\bibfnamefont {H.~J.}\ \bibnamefont {W{\"o}rner}},\ }\bibfield  {title} {\enquote {\bibinfo {title} {Sub-7-femtosecond conical-intersection dynamics probed at the carbon k-edge},}\ }\href@noop {} {\bibfield  {journal} {\bibinfo  {journal} {Science}\ }\textbf {\bibinfo {volume} {371}},\ \bibinfo {pages} {489--494} (\bibinfo {year} {2021})}\BibitemShut {NoStop}%
\bibitem [{\citenamefont {Pertot}\ \emph {et~al.}(2017)\citenamefont {Pertot}, \citenamefont {Schmidt}, \citenamefont {Matthews}, \citenamefont {Chauvet}, \citenamefont {Huppert}, \citenamefont {Svoboda}, \citenamefont {von Conta}, \citenamefont {Tehlar}, \citenamefont {Baykusheva}, \citenamefont {Wolf} \emph {et~al.}}]{pertotScience2017}%
  \BibitemOpen
  \bibfield  {author} {\bibinfo {author} {\bibfnamefont {Y.}~\bibnamefont {Pertot}}, \bibinfo {author} {\bibfnamefont {C.}~\bibnamefont {Schmidt}}, \bibinfo {author} {\bibfnamefont {M.}~\bibnamefont {Matthews}}, \bibinfo {author} {\bibfnamefont {A.}~\bibnamefont {Chauvet}}, \bibinfo {author} {\bibfnamefont {M.}~\bibnamefont {Huppert}}, \bibinfo {author} {\bibfnamefont {V.}~\bibnamefont {Svoboda}}, \bibinfo {author} {\bibfnamefont {A.}~\bibnamefont {von Conta}}, \bibinfo {author} {\bibfnamefont {A.}~\bibnamefont {Tehlar}}, \bibinfo {author} {\bibfnamefont {D.}~\bibnamefont {Baykusheva}}, \bibinfo {author} {\bibfnamefont {J.-P.}\ \bibnamefont {Wolf}}, \emph {et~al.},\ }\bibfield  {title} {\enquote {\bibinfo {title} {Time-resolved x-ray absorption spectroscopy with a water window high-harmonic source},}\ }\href {https://doi.org/https://doi.org/10.1126/science.aah6114} {\bibfield  {journal} {\bibinfo  {journal} {Science}\ }\textbf {\bibinfo {volume} {355}},\ \bibinfo {pages} {264--267} (\bibinfo {year}
  {2017})}\BibitemShut {NoStop}%
\bibitem [{\citenamefont {Haugen}\ \emph {et~al.}(2023)\citenamefont {Haugen}, \citenamefont {Hait}, \citenamefont {Scutelnic}, \citenamefont {Xue}, \citenamefont {Head-Gordon},\ and\ \citenamefont {Leone}}]{haugen2023ultrafast}%
  \BibitemOpen
  \bibfield  {author} {\bibinfo {author} {\bibfnamefont {E.~A.}\ \bibnamefont {Haugen}}, \bibinfo {author} {\bibfnamefont {D.}~\bibnamefont {Hait}}, \bibinfo {author} {\bibfnamefont {V.}~\bibnamefont {Scutelnic}}, \bibinfo {author} {\bibfnamefont {T.}~\bibnamefont {Xue}}, \bibinfo {author} {\bibfnamefont {M.}~\bibnamefont {Head-Gordon}},\ and\ \bibinfo {author} {\bibfnamefont {S.~R.}\ \bibnamefont {Leone}},\ }\bibfield  {title} {\enquote {\bibinfo {title} {Ultrafast x-ray spectroscopy of intersystem crossing in hexafluoroacetylacetone: Chromophore photophysics and spectral changes in the face of electron-withdrawing groups},}\ }\href@noop {} {\bibfield  {journal} {\bibinfo  {journal} {The Journal of Physical Chemistry A}\ }\textbf {\bibinfo {volume} {127}},\ \bibinfo {pages} {634--644} (\bibinfo {year} {2023})}\BibitemShut {NoStop}%
\bibitem [{\citenamefont {Hait}\ and\ \citenamefont {Head-Gordon}(2020{\natexlab{a}})}]{hait2020highly}%
  \BibitemOpen
  \bibfield  {author} {\bibinfo {author} {\bibfnamefont {D.}~\bibnamefont {Hait}}\ and\ \bibinfo {author} {\bibfnamefont {M.}~\bibnamefont {Head-Gordon}},\ }\bibfield  {title} {\enquote {\bibinfo {title} {Highly accurate prediction of core spectra of molecules at density functional theory cost: Attaining sub-electronvolt error from a restricted open-shell kohn--sham approach},}\ }\href@noop {} {\bibfield  {journal} {\bibinfo  {journal} {J. Phys. Chem. Lett.}\ }\textbf {\bibinfo {volume} {11}},\ \bibinfo {pages} {775--786} (\bibinfo {year} {2020}{\natexlab{a}})}\BibitemShut {NoStop}%
\bibitem [{\citenamefont {Hait}\ \emph {et~al.}(2020)\citenamefont {Hait}, \citenamefont {Haugen}, \citenamefont {Yang}, \citenamefont {Oosterbaan}, \citenamefont {Leone},\ and\ \citenamefont {Head-Gordon}}]{hait2020accurate}%
  \BibitemOpen
  \bibfield  {author} {\bibinfo {author} {\bibfnamefont {D.}~\bibnamefont {Hait}}, \bibinfo {author} {\bibfnamefont {E.~A.}\ \bibnamefont {Haugen}}, \bibinfo {author} {\bibfnamefont {Z.}~\bibnamefont {Yang}}, \bibinfo {author} {\bibfnamefont {K.~J.}\ \bibnamefont {Oosterbaan}}, \bibinfo {author} {\bibfnamefont {S.~R.}\ \bibnamefont {Leone}},\ and\ \bibinfo {author} {\bibfnamefont {M.}~\bibnamefont {Head-Gordon}},\ }\bibfield  {title} {\enquote {\bibinfo {title} {Accurate prediction of core-level spectra of radicals at density functional theory cost via square gradient minimization and recoupling of mixed configurations},}\ }\href@noop {} {\bibfield  {journal} {\bibinfo  {journal} {J. Chem. Phys.}\ }\textbf {\bibinfo {volume} {153}},\ \bibinfo {pages} {134108} (\bibinfo {year} {2020})}\BibitemShut {NoStop}%
\bibitem [{\citenamefont {Wenzel}, \citenamefont {Wormit},\ and\ \citenamefont {Dreuw}(2014)}]{wenzel2014calculating}%
  \BibitemOpen
  \bibfield  {author} {\bibinfo {author} {\bibfnamefont {J.}~\bibnamefont {Wenzel}}, \bibinfo {author} {\bibfnamefont {M.}~\bibnamefont {Wormit}},\ and\ \bibinfo {author} {\bibfnamefont {A.}~\bibnamefont {Dreuw}},\ }\bibfield  {title} {\enquote {\bibinfo {title} {Calculating core-level excitations and x-ray absorption spectra of medium-sized closed-shell molecules with the algebraic-diagrammatic construction scheme for the polarization propagator},}\ }\href@noop {} {\bibfield  {journal} {\bibinfo  {journal} {J. Comput. Chem.}\ }\textbf {\bibinfo {volume} {35}},\ \bibinfo {pages} {1900--1915} (\bibinfo {year} {2014})}\BibitemShut {NoStop}%
\bibitem [{\citenamefont {Oosterbaan}, \citenamefont {White},\ and\ \citenamefont {Head-Gordon}(2018)}]{oosterbaan2018non}%
  \BibitemOpen
  \bibfield  {author} {\bibinfo {author} {\bibfnamefont {K.~J.}\ \bibnamefont {Oosterbaan}}, \bibinfo {author} {\bibfnamefont {A.~F.}\ \bibnamefont {White}},\ and\ \bibinfo {author} {\bibfnamefont {M.}~\bibnamefont {Head-Gordon}},\ }\bibfield  {title} {\enquote {\bibinfo {title} {Non-orthogonal configuration interaction with single substitutions for the calculation of core-excited states},}\ }\href@noop {} {\bibfield  {journal} {\bibinfo  {journal} {The Journal of chemical physics}\ }\textbf {\bibinfo {volume} {149}} (\bibinfo {year} {2018})}\BibitemShut {NoStop}%
\bibitem [{\citenamefont {Vidal}\ \emph {et~al.}(2019)\citenamefont {Vidal}, \citenamefont {Feng}, \citenamefont {Epifanovsky}, \citenamefont {Krylov},\ and\ \citenamefont {Coriani}}]{vidal2019new}%
  \BibitemOpen
  \bibfield  {author} {\bibinfo {author} {\bibfnamefont {M.~L.}\ \bibnamefont {Vidal}}, \bibinfo {author} {\bibfnamefont {X.}~\bibnamefont {Feng}}, \bibinfo {author} {\bibfnamefont {E.}~\bibnamefont {Epifanovsky}}, \bibinfo {author} {\bibfnamefont {A.~I.}\ \bibnamefont {Krylov}},\ and\ \bibinfo {author} {\bibfnamefont {S.}~\bibnamefont {Coriani}},\ }\bibfield  {title} {\enquote {\bibinfo {title} {New and efficient equation-of-motion coupled-cluster framework for core-excited and core-ionized states},}\ }\href@noop {} {\bibfield  {journal} {\bibinfo  {journal} {J. Chem. Theory Comput.}\ }\textbf {\bibinfo {volume} {15}},\ \bibinfo {pages} {3117--3133} (\bibinfo {year} {2019})}\BibitemShut {NoStop}%
\bibitem [{\citenamefont {Besley}(2021)}]{besley2021modeling}%
  \BibitemOpen
  \bibfield  {author} {\bibinfo {author} {\bibfnamefont {N.~A.}\ \bibnamefont {Besley}},\ }\bibfield  {title} {\enquote {\bibinfo {title} {Modeling of the spectroscopy of core electrons with density functional theory},}\ }\href@noop {} {\bibfield  {journal} {\bibinfo  {journal} {Wiley Interdiscip. Rev. Comput. Mol. Sci.}\ }\textbf {\bibinfo {volume} {11}},\ \bibinfo {pages} {e1527} (\bibinfo {year} {2021})}\BibitemShut {NoStop}%
\bibitem [{\citenamefont {Norman}\ and\ \citenamefont {Dreuw}(2018)}]{norman2018simulating}%
  \BibitemOpen
  \bibfield  {author} {\bibinfo {author} {\bibfnamefont {P.}~\bibnamefont {Norman}}\ and\ \bibinfo {author} {\bibfnamefont {A.}~\bibnamefont {Dreuw}},\ }\bibfield  {title} {\enquote {\bibinfo {title} {Simulating x-ray spectroscopies and calculating core-excited states of molecules},}\ }\href@noop {} {\bibfield  {journal} {\bibinfo  {journal} {Chemical reviews}\ }\textbf {\bibinfo {volume} {118}},\ \bibinfo {pages} {7208--7248} (\bibinfo {year} {2018})}\BibitemShut {NoStop}%
\bibitem [{\citenamefont {Garner}\ \emph {et~al.}(2023)\citenamefont {Garner}, \citenamefont {Haugen}, \citenamefont {Leone},\ and\ \citenamefont {Neuscamman}}]{garner2023spin}%
  \BibitemOpen
  \bibfield  {author} {\bibinfo {author} {\bibfnamefont {S.~M.}\ \bibnamefont {Garner}}, \bibinfo {author} {\bibfnamefont {E.~A.}\ \bibnamefont {Haugen}}, \bibinfo {author} {\bibfnamefont {S.~R.}\ \bibnamefont {Leone}},\ and\ \bibinfo {author} {\bibfnamefont {E.}~\bibnamefont {Neuscamman}},\ }\bibfield  {title} {\enquote {\bibinfo {title} {Spin coupling effect on geometry-dependent x-ray absorption of diradicals},}\ }\href@noop {} {\bibfield  {journal} {\bibinfo  {journal} {arXiv preprint arXiv:2307.15207}\ } (\bibinfo {year} {2023})}\BibitemShut {NoStop}%
\bibitem [{\citenamefont {Yao}\ \emph {et~al.}(2022)\citenamefont {Yao}, \citenamefont {Golze}, \citenamefont {Rinke}, \citenamefont {Blum},\ and\ \citenamefont {Kanai}}]{yao2022all}%
  \BibitemOpen
  \bibfield  {author} {\bibinfo {author} {\bibfnamefont {Y.}~\bibnamefont {Yao}}, \bibinfo {author} {\bibfnamefont {D.}~\bibnamefont {Golze}}, \bibinfo {author} {\bibfnamefont {P.}~\bibnamefont {Rinke}}, \bibinfo {author} {\bibfnamefont {V.}~\bibnamefont {Blum}},\ and\ \bibinfo {author} {\bibfnamefont {Y.}~\bibnamefont {Kanai}},\ }\bibfield  {title} {\enquote {\bibinfo {title} {All-electron bse@ gw method for k-edge core electron excitation energies},}\ }\href@noop {} {\bibfield  {journal} {\bibinfo  {journal} {Journal of Chemical Theory and Computation}\ }\textbf {\bibinfo {volume} {18}},\ \bibinfo {pages} {1569--1583} (\bibinfo {year} {2022})}\BibitemShut {NoStop}%
\bibitem [{\citenamefont {Carter-Fenk}\ \emph {et~al.}(2022)\citenamefont {Carter-Fenk}, \citenamefont {Cunha}, \citenamefont {Arias-Martinez},\ and\ \citenamefont {Head-Gordon}}]{carter2022electron}%
  \BibitemOpen
  \bibfield  {author} {\bibinfo {author} {\bibfnamefont {K.}~\bibnamefont {Carter-Fenk}}, \bibinfo {author} {\bibfnamefont {L.~A.}\ \bibnamefont {Cunha}}, \bibinfo {author} {\bibfnamefont {J.~E.}\ \bibnamefont {Arias-Martinez}},\ and\ \bibinfo {author} {\bibfnamefont {M.}~\bibnamefont {Head-Gordon}},\ }\bibfield  {title} {\enquote {\bibinfo {title} {Electron-affinity time-dependent density functional theory: Formalism and applications to core-excited states},}\ }\href@noop {} {\bibfield  {journal} {\bibinfo  {journal} {The Journal of Physical Chemistry Letters}\ }\textbf {\bibinfo {volume} {13}},\ \bibinfo {pages} {9664--9672} (\bibinfo {year} {2022})}\BibitemShut {NoStop}%
\bibitem [{\citenamefont {Prince}\ \emph {et~al.}(2003)\citenamefont {Prince}, \citenamefont {Richter}, \citenamefont {de~Simone}, \citenamefont {Alagia},\ and\ \citenamefont {Coreno}}]{prince2003near}%
  \BibitemOpen
  \bibfield  {author} {\bibinfo {author} {\bibfnamefont {K.~C.}\ \bibnamefont {Prince}}, \bibinfo {author} {\bibfnamefont {R.}~\bibnamefont {Richter}}, \bibinfo {author} {\bibfnamefont {M.}~\bibnamefont {de~Simone}}, \bibinfo {author} {\bibfnamefont {M.}~\bibnamefont {Alagia}},\ and\ \bibinfo {author} {\bibfnamefont {M.}~\bibnamefont {Coreno}},\ }\bibfield  {title} {\enquote {\bibinfo {title} {Near edge x-ray absorption spectra of some small polyatomic molecules},}\ }\href@noop {} {\bibfield  {journal} {\bibinfo  {journal} {The Journal of Physical Chemistry A}\ }\textbf {\bibinfo {volume} {107}},\ \bibinfo {pages} {1955--1963} (\bibinfo {year} {2003})}\BibitemShut {NoStop}%
\bibitem [{\citenamefont {Vaz~da Cruz}\ \emph {et~al.}(2019{\natexlab{a}})\citenamefont {Vaz~da Cruz}, \citenamefont {Ignatova}, \citenamefont {Couto}, \citenamefont {Fedotov}, \citenamefont {Rehn}, \citenamefont {Savchenko}, \citenamefont {Norman}, \citenamefont {{\AA}gren}, \citenamefont {Polyutov}, \citenamefont {Niskanen} \emph {et~al.}}]{vaz2019nuclear}%
  \BibitemOpen
  \bibfield  {author} {\bibinfo {author} {\bibfnamefont {V.}~\bibnamefont {Vaz~da Cruz}}, \bibinfo {author} {\bibfnamefont {N.}~\bibnamefont {Ignatova}}, \bibinfo {author} {\bibfnamefont {R.~C.}\ \bibnamefont {Couto}}, \bibinfo {author} {\bibfnamefont {D.~A.}\ \bibnamefont {Fedotov}}, \bibinfo {author} {\bibfnamefont {D.~R.}\ \bibnamefont {Rehn}}, \bibinfo {author} {\bibfnamefont {V.}~\bibnamefont {Savchenko}}, \bibinfo {author} {\bibfnamefont {P.}~\bibnamefont {Norman}}, \bibinfo {author} {\bibfnamefont {H.}~\bibnamefont {{\AA}gren}}, \bibinfo {author} {\bibfnamefont {S.}~\bibnamefont {Polyutov}}, \bibinfo {author} {\bibfnamefont {J.}~\bibnamefont {Niskanen}}, \emph {et~al.},\ }\bibfield  {title} {\enquote {\bibinfo {title} {Nuclear dynamics in resonant inelastic x-ray scattering and x-ray absorption of methanol},}\ }\href@noop {} {\bibfield  {journal} {\bibinfo  {journal} {The Journal of chemical physics}\ }\textbf {\bibinfo {volume} {150}},\ \bibinfo {pages} {234301} (\bibinfo {year}
  {2019}{\natexlab{a}})}\BibitemShut {NoStop}%
\bibitem [{\citenamefont {Hait}\ and\ \citenamefont {Mart{\'\i}nez}(2023)}]{hait2023predicting}%
  \BibitemOpen
  \bibfield  {author} {\bibinfo {author} {\bibfnamefont {D.}~\bibnamefont {Hait}}\ and\ \bibinfo {author} {\bibfnamefont {T.~J.}\ \bibnamefont {Mart{\'\i}nez}},\ }\bibfield  {title} {\enquote {\bibinfo {title} {Predicting the x-ray absorption spectrum of ozone with single configuration state functions},}\ }\href@noop {} {\bibfield  {journal} {\bibinfo  {journal} {Journal of Chemical Theory and Computation}\ } (\bibinfo {year} {2023})}\BibitemShut {NoStop}%
\bibitem [{\citenamefont {Takahashi}\ \emph {et~al.}(2022)\citenamefont {Takahashi}, \citenamefont {Yamamura}, \citenamefont {Tokushima},\ and\ \citenamefont {Harada}}]{takahashi2022interpretation}%
  \BibitemOpen
  \bibfield  {author} {\bibinfo {author} {\bibfnamefont {O.}~\bibnamefont {Takahashi}}, \bibinfo {author} {\bibfnamefont {R.}~\bibnamefont {Yamamura}}, \bibinfo {author} {\bibfnamefont {T.}~\bibnamefont {Tokushima}},\ and\ \bibinfo {author} {\bibfnamefont {Y.}~\bibnamefont {Harada}},\ }\bibfield  {title} {\enquote {\bibinfo {title} {Interpretation of the x-ray emission spectra of liquid water through temperature and isotope dependence},}\ }\href@noop {} {\bibfield  {journal} {\bibinfo  {journal} {Physical Review Letters}\ }\textbf {\bibinfo {volume} {128}},\ \bibinfo {pages} {086002} (\bibinfo {year} {2022})}\BibitemShut {NoStop}%
\bibitem [{\citenamefont {Cruzeiro}\ \emph {et~al.}(2023)\citenamefont {Cruzeiro}, \citenamefont {Hait}, \citenamefont {Shea}, \citenamefont {Hohenstein},\ and\ \citenamefont {Martinez}}]{cruzeiro20231b1}%
  \BibitemOpen
  \bibfield  {author} {\bibinfo {author} {\bibfnamefont {V.~W.~D.}\ \bibnamefont {Cruzeiro}}, \bibinfo {author} {\bibfnamefont {D.}~\bibnamefont {Hait}}, \bibinfo {author} {\bibfnamefont {J.}~\bibnamefont {Shea}}, \bibinfo {author} {\bibfnamefont {E.~G.}\ \bibnamefont {Hohenstein}},\ and\ \bibinfo {author} {\bibfnamefont {T.~J.}\ \bibnamefont {Martinez}},\ }\bibfield  {title} {\enquote {\bibinfo {title} {1b1 splitting in the x-ray emission spectrum of liquid water is dominated by ultrafast dissociation},}\ }\href {https://doi.org/10.26434/chemrxiv-2023-5jvl5} {\bibfield  {journal} {\bibinfo  {journal} {ChemRxiv}\ } (\bibinfo {year} {2023}),\ 10.26434/chemrxiv-2023-5jvl5}\BibitemShut {NoStop}%
\bibitem [{\citenamefont {Barreau}\ \emph {et~al.}(2023)\citenamefont {Barreau}, \citenamefont {Ross}, \citenamefont {Kimberg}, \citenamefont {Krasnov}, \citenamefont {Blinov}, \citenamefont {Neumark},\ and\ \citenamefont {Leone}}]{barreau2023core}%
  \BibitemOpen
  \bibfield  {author} {\bibinfo {author} {\bibfnamefont {L.}~\bibnamefont {Barreau}}, \bibinfo {author} {\bibfnamefont {A.~D.}\ \bibnamefont {Ross}}, \bibinfo {author} {\bibfnamefont {V.}~\bibnamefont {Kimberg}}, \bibinfo {author} {\bibfnamefont {P.}~\bibnamefont {Krasnov}}, \bibinfo {author} {\bibfnamefont {S.}~\bibnamefont {Blinov}}, \bibinfo {author} {\bibfnamefont {D.~M.}\ \bibnamefont {Neumark}},\ and\ \bibinfo {author} {\bibfnamefont {S.~R.}\ \bibnamefont {Leone}},\ }\bibfield  {title} {\enquote {\bibinfo {title} {Core-excited states of sf 6 probed with soft-x-ray femtosecond transient absorption of vibrational wave packets},}\ }\href@noop {} {\bibfield  {journal} {\bibinfo  {journal} {Physical Review A}\ }\textbf {\bibinfo {volume} {108}},\ \bibinfo {pages} {012805} (\bibinfo {year} {2023})}\BibitemShut {NoStop}%
\bibitem [{\citenamefont {Yan}, \citenamefont {Gamble~Jr},\ and\ \citenamefont {Nelson}(1985)}]{yan1985impulsive}%
  \BibitemOpen
  \bibfield  {author} {\bibinfo {author} {\bibfnamefont {Y.-X.}\ \bibnamefont {Yan}}, \bibinfo {author} {\bibfnamefont {E.~B.}\ \bibnamefont {Gamble~Jr}},\ and\ \bibinfo {author} {\bibfnamefont {K.~A.}\ \bibnamefont {Nelson}},\ }\bibfield  {title} {\enquote {\bibinfo {title} {Impulsive stimulated scattering: General importance in femtosecond laser pulse interactions with matter, and spectroscopic applications},}\ }\href@noop {} {\bibfield  {journal} {\bibinfo  {journal} {The Journal of chemical physics}\ }\textbf {\bibinfo {volume} {83}},\ \bibinfo {pages} {5391--5399} (\bibinfo {year} {1985})}\BibitemShut {NoStop}%
\bibitem [{\citenamefont {Yan}\ and\ \citenamefont {Nelson}(1987)}]{yan1987impulsive}%
  \BibitemOpen
  \bibfield  {author} {\bibinfo {author} {\bibfnamefont {Y.-X.}\ \bibnamefont {Yan}}\ and\ \bibinfo {author} {\bibfnamefont {K.~A.}\ \bibnamefont {Nelson}},\ }\bibfield  {title} {\enquote {\bibinfo {title} {Impulsive stimulated light scattering. i. general theory},}\ }\href@noop {} {\bibfield  {journal} {\bibinfo  {journal} {The Journal of chemical physics}\ }\textbf {\bibinfo {volume} {87}},\ \bibinfo {pages} {6240--6256} (\bibinfo {year} {1987})}\BibitemShut {NoStop}%
\bibitem [{\citenamefont {Miron}\ \emph {et~al.}(2012)\citenamefont {Miron}, \citenamefont {Nicolas}, \citenamefont {Travnikova}, \citenamefont {Morin}, \citenamefont {Sun}, \citenamefont {Gel’mukhanov}, \citenamefont {Kosugi},\ and\ \citenamefont {Kimberg}}]{miron2012imaging}%
  \BibitemOpen
  \bibfield  {author} {\bibinfo {author} {\bibfnamefont {C.}~\bibnamefont {Miron}}, \bibinfo {author} {\bibfnamefont {C.}~\bibnamefont {Nicolas}}, \bibinfo {author} {\bibfnamefont {O.}~\bibnamefont {Travnikova}}, \bibinfo {author} {\bibfnamefont {P.}~\bibnamefont {Morin}}, \bibinfo {author} {\bibfnamefont {Y.}~\bibnamefont {Sun}}, \bibinfo {author} {\bibfnamefont {F.}~\bibnamefont {Gel’mukhanov}}, \bibinfo {author} {\bibfnamefont {N.}~\bibnamefont {Kosugi}},\ and\ \bibinfo {author} {\bibfnamefont {V.}~\bibnamefont {Kimberg}},\ }\bibfield  {title} {\enquote {\bibinfo {title} {Imaging molecular potentials using ultrahigh-resolution resonant photoemission},}\ }\href@noop {} {\bibfield  {journal} {\bibinfo  {journal} {Nature Physics}\ }\textbf {\bibinfo {volume} {8}},\ \bibinfo {pages} {135--138} (\bibinfo {year} {2012})}\BibitemShut {NoStop}%
\bibitem [{\citenamefont {Kimberg}\ and\ \citenamefont {Miron}(2014)}]{kimberg2014molecular}%
  \BibitemOpen
  \bibfield  {author} {\bibinfo {author} {\bibfnamefont {V.}~\bibnamefont {Kimberg}}\ and\ \bibinfo {author} {\bibfnamefont {C.}~\bibnamefont {Miron}},\ }\bibfield  {title} {\enquote {\bibinfo {title} {Molecular potentials and wave function mapping by high-resolution electron spectroscopy and ab initio calculations},}\ }\href@noop {} {\bibfield  {journal} {\bibinfo  {journal} {Journal of Electron Spectroscopy and Related Phenomena}\ }\textbf {\bibinfo {volume} {195}},\ \bibinfo {pages} {301--306} (\bibinfo {year} {2014})}\BibitemShut {NoStop}%
\bibitem [{\citenamefont {Schreck}\ \emph {et~al.}(2016)\citenamefont {Schreck}, \citenamefont {Pietzsch}, \citenamefont {Kennedy}, \citenamefont {S{\aa}the}, \citenamefont {Miedema}, \citenamefont {Techert}, \citenamefont {Strocov}, \citenamefont {Schmitt}, \citenamefont {Hennies}, \citenamefont {Rubensson} \emph {et~al.}}]{schreck2016ground}%
  \BibitemOpen
  \bibfield  {author} {\bibinfo {author} {\bibfnamefont {S.}~\bibnamefont {Schreck}}, \bibinfo {author} {\bibfnamefont {A.}~\bibnamefont {Pietzsch}}, \bibinfo {author} {\bibfnamefont {B.}~\bibnamefont {Kennedy}}, \bibinfo {author} {\bibfnamefont {C.}~\bibnamefont {S{\aa}the}}, \bibinfo {author} {\bibfnamefont {P.~S.}\ \bibnamefont {Miedema}}, \bibinfo {author} {\bibfnamefont {S.}~\bibnamefont {Techert}}, \bibinfo {author} {\bibfnamefont {V.~N.}\ \bibnamefont {Strocov}}, \bibinfo {author} {\bibfnamefont {T.}~\bibnamefont {Schmitt}}, \bibinfo {author} {\bibfnamefont {F.}~\bibnamefont {Hennies}}, \bibinfo {author} {\bibfnamefont {J.-E.}\ \bibnamefont {Rubensson}}, \emph {et~al.},\ }\bibfield  {title} {\enquote {\bibinfo {title} {Ground state potential energy surfaces around selected atoms from resonant inelastic x-ray scattering},}\ }\href@noop {} {\bibfield  {journal} {\bibinfo  {journal} {Scientific reports}\ }\textbf {\bibinfo {volume} {6}},\ \bibinfo {pages} {1--8} (\bibinfo {year} {2016})}\BibitemShut
  {NoStop}%
\bibitem [{\citenamefont {Eckert}\ \emph {et~al.}(2018)\citenamefont {Eckert}, \citenamefont {da~Cruz}, \citenamefont {Gel'mukhanov}, \citenamefont {Ertan}, \citenamefont {Ignatova}, \citenamefont {Polyutov}, \citenamefont {Couto}, \citenamefont {Fondell}, \citenamefont {Dantz}, \citenamefont {Kennedy} \emph {et~al.}}]{eckert2018one}%
  \BibitemOpen
  \bibfield  {author} {\bibinfo {author} {\bibfnamefont {S.}~\bibnamefont {Eckert}}, \bibinfo {author} {\bibfnamefont {V.~V.}\ \bibnamefont {da~Cruz}}, \bibinfo {author} {\bibfnamefont {F.}~\bibnamefont {Gel'mukhanov}}, \bibinfo {author} {\bibfnamefont {E.}~\bibnamefont {Ertan}}, \bibinfo {author} {\bibfnamefont {N.}~\bibnamefont {Ignatova}}, \bibinfo {author} {\bibfnamefont {S.}~\bibnamefont {Polyutov}}, \bibinfo {author} {\bibfnamefont {R.~C.}\ \bibnamefont {Couto}}, \bibinfo {author} {\bibfnamefont {M.}~\bibnamefont {Fondell}}, \bibinfo {author} {\bibfnamefont {M.}~\bibnamefont {Dantz}}, \bibinfo {author} {\bibfnamefont {B.}~\bibnamefont {Kennedy}}, \emph {et~al.},\ }\bibfield  {title} {\enquote {\bibinfo {title} {One-dimensional cuts through multidimensional potential-energy surfaces by tunable x rays},}\ }\href@noop {} {\bibfield  {journal} {\bibinfo  {journal} {Physical Review A}\ }\textbf {\bibinfo {volume} {97}},\ \bibinfo {pages} {053410} (\bibinfo {year} {2018})}\BibitemShut {NoStop}%
\bibitem [{\citenamefont {Vaz~da Cruz}\ \emph {et~al.}(2019{\natexlab{b}})\citenamefont {Vaz~da Cruz}, \citenamefont {Gel’mukhanov}, \citenamefont {Eckert}, \citenamefont {Iannuzzi}, \citenamefont {Ertan}, \citenamefont {Pietzsch}, \citenamefont {Couto}, \citenamefont {Niskanen}, \citenamefont {Fondell}, \citenamefont {Dantz} \emph {et~al.}}]{vaz2019probing}%
  \BibitemOpen
  \bibfield  {author} {\bibinfo {author} {\bibfnamefont {V.}~\bibnamefont {Vaz~da Cruz}}, \bibinfo {author} {\bibfnamefont {F.}~\bibnamefont {Gel’mukhanov}}, \bibinfo {author} {\bibfnamefont {S.}~\bibnamefont {Eckert}}, \bibinfo {author} {\bibfnamefont {M.}~\bibnamefont {Iannuzzi}}, \bibinfo {author} {\bibfnamefont {E.}~\bibnamefont {Ertan}}, \bibinfo {author} {\bibfnamefont {A.}~\bibnamefont {Pietzsch}}, \bibinfo {author} {\bibfnamefont {R.~C.}\ \bibnamefont {Couto}}, \bibinfo {author} {\bibfnamefont {J.}~\bibnamefont {Niskanen}}, \bibinfo {author} {\bibfnamefont {M.}~\bibnamefont {Fondell}}, \bibinfo {author} {\bibfnamefont {M.}~\bibnamefont {Dantz}}, \emph {et~al.},\ }\bibfield  {title} {\enquote {\bibinfo {title} {Probing hydrogen bond strength in liquid water by resonant inelastic x-ray scattering},}\ }\href@noop {} {\bibfield  {journal} {\bibinfo  {journal} {Nature communications}\ }\textbf {\bibinfo {volume} {10}},\ \bibinfo {pages} {1--9} (\bibinfo {year} {2019}{\natexlab{b}})}\BibitemShut {NoStop}%
\bibitem [{\citenamefont {Saito}\ \emph {et~al.}(2019)\citenamefont {Saito}, \citenamefont {Sannohe}, \citenamefont {Ishii}, \citenamefont {Kanai}, \citenamefont {Kosugi}, \citenamefont {Wu}, \citenamefont {Chew}, \citenamefont {Han}, \citenamefont {Chang},\ and\ \citenamefont {Itatani}}]{saito2019real}%
  \BibitemOpen
  \bibfield  {author} {\bibinfo {author} {\bibfnamefont {N.}~\bibnamefont {Saito}}, \bibinfo {author} {\bibfnamefont {H.}~\bibnamefont {Sannohe}}, \bibinfo {author} {\bibfnamefont {N.}~\bibnamefont {Ishii}}, \bibinfo {author} {\bibfnamefont {T.}~\bibnamefont {Kanai}}, \bibinfo {author} {\bibfnamefont {N.}~\bibnamefont {Kosugi}}, \bibinfo {author} {\bibfnamefont {Y.}~\bibnamefont {Wu}}, \bibinfo {author} {\bibfnamefont {A.}~\bibnamefont {Chew}}, \bibinfo {author} {\bibfnamefont {S.}~\bibnamefont {Han}}, \bibinfo {author} {\bibfnamefont {Z.}~\bibnamefont {Chang}},\ and\ \bibinfo {author} {\bibfnamefont {J.}~\bibnamefont {Itatani}},\ }\bibfield  {title} {\enquote {\bibinfo {title} {Real-time observation of electronic, vibrational, and rotational dynamics in nitric oxide with attosecond soft x-ray pulses at 400 ev},}\ }\href@noop {} {\bibfield  {journal} {\bibinfo  {journal} {Optica}\ }\textbf {\bibinfo {volume} {6}},\ \bibinfo {pages} {1542--1546} (\bibinfo {year} {2019})}\BibitemShut {NoStop}%
\bibitem [{\citenamefont {Weisshaupt}\ \emph {et~al.}(2017)\citenamefont {Weisshaupt}, \citenamefont {Rouz{\'e}e}, \citenamefont {Woerner}, \citenamefont {Vrakking}, \citenamefont {Elsaesser}, \citenamefont {Shirley},\ and\ \citenamefont {Borgschulte}}]{weisshaupt2017ultrafast}%
  \BibitemOpen
  \bibfield  {author} {\bibinfo {author} {\bibfnamefont {J.}~\bibnamefont {Weisshaupt}}, \bibinfo {author} {\bibfnamefont {A.}~\bibnamefont {Rouz{\'e}e}}, \bibinfo {author} {\bibfnamefont {M.}~\bibnamefont {Woerner}}, \bibinfo {author} {\bibfnamefont {M.}~\bibnamefont {Vrakking}}, \bibinfo {author} {\bibfnamefont {T.}~\bibnamefont {Elsaesser}}, \bibinfo {author} {\bibfnamefont {E.}~\bibnamefont {Shirley}},\ and\ \bibinfo {author} {\bibfnamefont {A.}~\bibnamefont {Borgschulte}},\ }\bibfield  {title} {\enquote {\bibinfo {title} {Ultrafast modulation of electronic structure by coherent phonon excitations},}\ }\href@noop {} {\bibfield  {journal} {\bibinfo  {journal} {Physical Review B}\ }\textbf {\bibinfo {volume} {95}},\ \bibinfo {pages} {081101} (\bibinfo {year} {2017})}\BibitemShut {NoStop}%
\bibitem [{\citenamefont {Cammarata}\ \emph {et~al.}(2014)\citenamefont {Cammarata}, \citenamefont {Bertoni}, \citenamefont {Lorenc}, \citenamefont {Cailleau}, \citenamefont {Di~Matteo}, \citenamefont {Mauriac}, \citenamefont {Matar}, \citenamefont {Lemke}, \citenamefont {Chollet}, \citenamefont {Ravy} \emph {et~al.}}]{cammarata2014sequential}%
  \BibitemOpen
  \bibfield  {author} {\bibinfo {author} {\bibfnamefont {M.}~\bibnamefont {Cammarata}}, \bibinfo {author} {\bibfnamefont {R.}~\bibnamefont {Bertoni}}, \bibinfo {author} {\bibfnamefont {M.}~\bibnamefont {Lorenc}}, \bibinfo {author} {\bibfnamefont {H.}~\bibnamefont {Cailleau}}, \bibinfo {author} {\bibfnamefont {S.}~\bibnamefont {Di~Matteo}}, \bibinfo {author} {\bibfnamefont {C.}~\bibnamefont {Mauriac}}, \bibinfo {author} {\bibfnamefont {S.~F.}\ \bibnamefont {Matar}}, \bibinfo {author} {\bibfnamefont {H.}~\bibnamefont {Lemke}}, \bibinfo {author} {\bibfnamefont {M.}~\bibnamefont {Chollet}}, \bibinfo {author} {\bibfnamefont {S.}~\bibnamefont {Ravy}}, \emph {et~al.},\ }\bibfield  {title} {\enquote {\bibinfo {title} {Sequential activation of molecular breathing and bending during spin-crossover photoswitching revealed by femtosecond optical and x-ray absorption spectroscopy},}\ }\href@noop {} {\bibfield  {journal} {\bibinfo  {journal} {Physical review letters}\ }\textbf {\bibinfo {volume} {113}},\ \bibinfo {pages}
  {227402} (\bibinfo {year} {2014})}\BibitemShut {NoStop}%
\bibitem [{\citenamefont {Rupprecht}\ \emph {et~al.}(2023)\citenamefont {Rupprecht}, \citenamefont {Aufleger}, \citenamefont {Heinze}, \citenamefont {Magunia}, \citenamefont {Ding}, \citenamefont {Rebholz}, \citenamefont {Amberg}, \citenamefont {Mollov}, \citenamefont {Henrich}, \citenamefont {Haverkort} \emph {et~al.}}]{rupprecht2023resolving}%
  \BibitemOpen
  \bibfield  {author} {\bibinfo {author} {\bibfnamefont {P.}~\bibnamefont {Rupprecht}}, \bibinfo {author} {\bibfnamefont {L.}~\bibnamefont {Aufleger}}, \bibinfo {author} {\bibfnamefont {S.}~\bibnamefont {Heinze}}, \bibinfo {author} {\bibfnamefont {A.}~\bibnamefont {Magunia}}, \bibinfo {author} {\bibfnamefont {T.}~\bibnamefont {Ding}}, \bibinfo {author} {\bibfnamefont {M.}~\bibnamefont {Rebholz}}, \bibinfo {author} {\bibfnamefont {S.}~\bibnamefont {Amberg}}, \bibinfo {author} {\bibfnamefont {N.}~\bibnamefont {Mollov}}, \bibinfo {author} {\bibfnamefont {F.}~\bibnamefont {Henrich}}, \bibinfo {author} {\bibfnamefont {M.~W.}\ \bibnamefont {Haverkort}}, \emph {et~al.},\ }\bibfield  {title} {\enquote {\bibinfo {title} {Resolving vibrations in a polyatomic molecule with femtometer precision via x-ray spectroscopy},}\ }\href@noop {} {\bibfield  {journal} {\bibinfo  {journal} {Physical Review A}\ }\textbf {\bibinfo {volume} {108}},\ \bibinfo {pages} {032816} (\bibinfo {year} {2023})}\BibitemShut {NoStop}%
\bibitem [{\citenamefont {Lu}\ \emph {et~al.}(2009)\citenamefont {Lu}, \citenamefont {Chen}, \citenamefont {Lee}, \citenamefont {Haw}, \citenamefont {Chou}, \citenamefont {Chen},\ and\ \citenamefont {Chen}}]{lu2009core}%
  \BibitemOpen
  \bibfield  {author} {\bibinfo {author} {\bibfnamefont {K.}~\bibnamefont {Lu}}, \bibinfo {author} {\bibfnamefont {J.}~\bibnamefont {Chen}}, \bibinfo {author} {\bibfnamefont {J.}~\bibnamefont {Lee}}, \bibinfo {author} {\bibfnamefont {S.}~\bibnamefont {Haw}}, \bibinfo {author} {\bibfnamefont {T.}~\bibnamefont {Chou}}, \bibinfo {author} {\bibfnamefont {S.}~\bibnamefont {Chen}},\ and\ \bibinfo {author} {\bibfnamefont {T.}~\bibnamefont {Chen}},\ }\bibfield  {title} {\enquote {\bibinfo {title} {Core-level anionic photofragmentation of gaseous ccl 4 and solid-state analogs},}\ }\href@noop {} {\bibfield  {journal} {\bibinfo  {journal} {Physical Review A}\ }\textbf {\bibinfo {volume} {80}},\ \bibinfo {pages} {033406} (\bibinfo {year} {2009})}\BibitemShut {NoStop}%
\bibitem [{\citenamefont {Fock}\ and\ \citenamefont {Koch}(1985)}]{fock1985shape}%
  \BibitemOpen
  \bibfield  {author} {\bibinfo {author} {\bibfnamefont {J.-H.}\ \bibnamefont {Fock}}\ and\ \bibinfo {author} {\bibfnamefont {E.-E.}\ \bibnamefont {Koch}},\ }\bibfield  {title} {\enquote {\bibinfo {title} {Shape resonances and partial photoemission cross sections of solid sf6 and ccl4},}\ }\href@noop {} {\bibfield  {journal} {\bibinfo  {journal} {Chemical physics}\ }\textbf {\bibinfo {volume} {96}},\ \bibinfo {pages} {125--143} (\bibinfo {year} {1985})}\BibitemShut {NoStop}%
\bibitem [{\citenamefont {Frank}\ \emph {et~al.}(1998)\citenamefont {Frank}, \citenamefont {Hutter}, \citenamefont {Marx},\ and\ \citenamefont {Parrinello}}]{frank1998molecular}%
  \BibitemOpen
  \bibfield  {author} {\bibinfo {author} {\bibfnamefont {I.}~\bibnamefont {Frank}}, \bibinfo {author} {\bibfnamefont {J.}~\bibnamefont {Hutter}}, \bibinfo {author} {\bibfnamefont {D.}~\bibnamefont {Marx}},\ and\ \bibinfo {author} {\bibfnamefont {M.}~\bibnamefont {Parrinello}},\ }\bibfield  {title} {\enquote {\bibinfo {title} {Molecular dynamics in low-spin excited states},}\ }\href@noop {} {\bibfield  {journal} {\bibinfo  {journal} {J. Chem. Phys.}\ }\textbf {\bibinfo {volume} {108}},\ \bibinfo {pages} {4060--4069} (\bibinfo {year} {1998})}\BibitemShut {NoStop}%
\bibitem [{\citenamefont {Barreau}\ \emph {et~al.}(2020)\citenamefont {Barreau}, \citenamefont {Ross}, \citenamefont {Garg}, \citenamefont {Kraus}, \citenamefont {Neumark},\ and\ \citenamefont {Leone}}]{Barreau2020}%
  \BibitemOpen
  \bibfield  {author} {\bibinfo {author} {\bibfnamefont {L.}~\bibnamefont {Barreau}}, \bibinfo {author} {\bibfnamefont {A.~D.}\ \bibnamefont {Ross}}, \bibinfo {author} {\bibfnamefont {S.}~\bibnamefont {Garg}}, \bibinfo {author} {\bibfnamefont {P.~M.}\ \bibnamefont {Kraus}}, \bibinfo {author} {\bibfnamefont {D.~M.}\ \bibnamefont {Neumark}},\ and\ \bibinfo {author} {\bibfnamefont {S.~R.}\ \bibnamefont {Leone}},\ }\bibfield  {title} {\enquote {\bibinfo {title} {Efficient table-top dual-wavelength beamline for ultrafast transient absorption spectroscopy in the soft x-ray region},}\ }\href {https://doi.org/10.1038/s41598-020-62461-6} {\bibfield  {journal} {\bibinfo  {journal} {Scientific Reports}\ }\textbf {\bibinfo {volume} {10}},\ \bibinfo {pages} {5773} (\bibinfo {year} {2020})}\BibitemShut {NoStop}%
\bibitem [{\citenamefont {Fidler}\ \emph {et~al.}(2019)\citenamefont {Fidler}, \citenamefont {Camp}, \citenamefont {Warrick}, \citenamefont {Bloch}, \citenamefont {Marroux}, \citenamefont {Neumark}, \citenamefont {Schafer}, \citenamefont {Gaarde},\ and\ \citenamefont {Leone}}]{Fidler2019FWM}%
  \BibitemOpen
  \bibfield  {author} {\bibinfo {author} {\bibfnamefont {A.~P.}\ \bibnamefont {Fidler}}, \bibinfo {author} {\bibfnamefont {S.~J.}\ \bibnamefont {Camp}}, \bibinfo {author} {\bibfnamefont {E.~R.}\ \bibnamefont {Warrick}}, \bibinfo {author} {\bibfnamefont {E.}~\bibnamefont {Bloch}}, \bibinfo {author} {\bibfnamefont {H.~J.~B.}\ \bibnamefont {Marroux}}, \bibinfo {author} {\bibfnamefont {D.~M.}\ \bibnamefont {Neumark}}, \bibinfo {author} {\bibfnamefont {K.~J.}\ \bibnamefont {Schafer}}, \bibinfo {author} {\bibfnamefont {M.~B.}\ \bibnamefont {Gaarde}},\ and\ \bibinfo {author} {\bibfnamefont {S.~R.}\ \bibnamefont {Leone}},\ }\bibfield  {title} {\enquote {\bibinfo {title} {Nonlinear xuv signal generation probed by transient grating spectroscopy with attosecond pulses},}\ }\href {https://doi.org/10.1038/s41467-019-09317-4} {\bibfield  {journal} {\bibinfo  {journal} {Nature Communications}\ }\textbf {\bibinfo {volume} {10}},\ \bibinfo {pages} {1384} (\bibinfo {year} {2019})}\BibitemShut {NoStop}%
\bibitem [{\citenamefont {G\'{e}neaux}\ \emph {et~al.}(2021)\citenamefont {G\'{e}neaux}, \citenamefont {Chang}, \citenamefont {Schwartzberg},\ and\ \citenamefont {Marroux}}]{Geneaux2021XUVNoise}%
  \BibitemOpen
  \bibfield  {author} {\bibinfo {author} {\bibfnamefont {R.}~\bibnamefont {G\'{e}neaux}}, \bibinfo {author} {\bibfnamefont {H.-T.}\ \bibnamefont {Chang}}, \bibinfo {author} {\bibfnamefont {A.~M.}\ \bibnamefont {Schwartzberg}},\ and\ \bibinfo {author} {\bibfnamefont {H.~J.~B.}\ \bibnamefont {Marroux}},\ }\bibfield  {title} {\enquote {\bibinfo {title} {Source noise suppression in attosecond transient absorption spectroscopy by edge-pixel referencing},}\ }\href {https://doi.org/10.1364/OE.412117} {\bibfield  {journal} {\bibinfo  {journal} {Opt. Express}\ }\textbf {\bibinfo {volume} {29}},\ \bibinfo {pages} {951--960} (\bibinfo {year} {2021})}\BibitemShut {NoStop}%
\bibitem [{\citenamefont {Ott}\ \emph {et~al.}(2014)\citenamefont {Ott}, \citenamefont {Kaldun}, \citenamefont {Argenti}, \citenamefont {Raith}, \citenamefont {Meyer}, \citenamefont {Laux}, \citenamefont {Zhang}, \citenamefont {Bl{\"a}ttermann}, \citenamefont {Hagstotz}, \citenamefont {Ding} \emph {et~al.}}]{ottNature2014}%
  \BibitemOpen
  \bibfield  {author} {\bibinfo {author} {\bibfnamefont {C.}~\bibnamefont {Ott}}, \bibinfo {author} {\bibfnamefont {A.}~\bibnamefont {Kaldun}}, \bibinfo {author} {\bibfnamefont {L.}~\bibnamefont {Argenti}}, \bibinfo {author} {\bibfnamefont {P.}~\bibnamefont {Raith}}, \bibinfo {author} {\bibfnamefont {K.}~\bibnamefont {Meyer}}, \bibinfo {author} {\bibfnamefont {M.}~\bibnamefont {Laux}}, \bibinfo {author} {\bibfnamefont {Y.}~\bibnamefont {Zhang}}, \bibinfo {author} {\bibfnamefont {A.}~\bibnamefont {Bl{\"a}ttermann}}, \bibinfo {author} {\bibfnamefont {S.}~\bibnamefont {Hagstotz}}, \bibinfo {author} {\bibfnamefont {T.}~\bibnamefont {Ding}}, \emph {et~al.},\ }\bibfield  {title} {\enquote {\bibinfo {title} {Reconstruction and control of a time-dependent two-electron wave packet},}\ }\href {https://doi.org/https://doi.org/10.1038/nature14026} {\bibfield  {journal} {\bibinfo  {journal} {Nature}\ }\textbf {\bibinfo {volume} {516}},\ \bibinfo {pages} {374} (\bibinfo {year} {2014})}\BibitemShut {NoStop}%
\bibitem [{\citenamefont {Chakraborty}\ and\ \citenamefont {Verma}(2002)}]{CHAKRABORTY2002RamanCCl4}%
  \BibitemOpen
  \bibfield  {author} {\bibinfo {author} {\bibfnamefont {T.}~\bibnamefont {Chakraborty}}\ and\ \bibinfo {author} {\bibfnamefont {A.}~\bibnamefont {Verma}},\ }\bibfield  {title} {\enquote {\bibinfo {title} {Vibrational spectra of ccl4: isotopic components and hot bands. part i},}\ }\href {https://doi.org/https://doi.org/10.1016/S1386-1425(01)00571-6} {\bibfield  {journal} {\bibinfo  {journal} {Spectrochimica Acta Part A: Molecular and Biomolecular Spectroscopy}\ }\textbf {\bibinfo {volume} {58}},\ \bibinfo {pages} {1013--1023} (\bibinfo {year} {2002})}\BibitemShut {NoStop}%
\bibitem [{\citenamefont {Epifanovsky}\ \emph {et~al.}(2021)\citenamefont {Epifanovsky} \emph {et~al.}}]{epifanovsky2021software}%
  \BibitemOpen
  \bibfield  {author} {\bibinfo {author} {\bibfnamefont {E.}~\bibnamefont {Epifanovsky}} \emph {et~al.},\ }\bibfield  {title} {\enquote {\bibinfo {title} {Software for the frontiers of quantum chemistry: An overview of developments in the q-chem 5 package},}\ }\href@noop {} {\bibfield  {journal} {\bibinfo  {journal} {J. Chem. Phys.}\ }\textbf {\bibinfo {volume} {155}},\ \bibinfo {pages} {084801} (\bibinfo {year} {2021})}\BibitemShut {NoStop}%
\bibitem [{\citenamefont {Hui}\ and\ \citenamefont {Chai}(2016)}]{scan0}%
  \BibitemOpen
  \bibfield  {author} {\bibinfo {author} {\bibfnamefont {K.}~\bibnamefont {Hui}}\ and\ \bibinfo {author} {\bibfnamefont {J.-D.}\ \bibnamefont {Chai}},\ }\bibfield  {title} {\enquote {\bibinfo {title} {{SCAN-based hybrid and double-hybrid density functionals from models without fitted parameters}},}\ }\href@noop {} {\bibfield  {journal} {\bibinfo  {journal} {J. Chem. Phys.}\ }\textbf {\bibinfo {volume} {144}},\ \bibinfo {pages} {044114} (\bibinfo {year} {2016})}\BibitemShut {NoStop}%
\bibitem [{\citenamefont {Hait}\ and\ \citenamefont {Head-Gordon}(2021)}]{hait2021orbital}%
  \BibitemOpen
  \bibfield  {author} {\bibinfo {author} {\bibfnamefont {D.}~\bibnamefont {Hait}}\ and\ \bibinfo {author} {\bibfnamefont {M.}~\bibnamefont {Head-Gordon}},\ }\bibfield  {title} {\enquote {\bibinfo {title} {Orbital optimized density functional theory for electronic excited states},}\ }\href@noop {} {\bibfield  {journal} {\bibinfo  {journal} {J. Phys. Chem. Lett.}\ }\textbf {\bibinfo {volume} {12}},\ \bibinfo {pages} {4517--4529} (\bibinfo {year} {2021})}\BibitemShut {NoStop}%
\bibitem [{\citenamefont {Hait}\ and\ \citenamefont {Head-Gordon}(2018)}]{hait2018accurate}%
  \BibitemOpen
  \bibfield  {author} {\bibinfo {author} {\bibfnamefont {D.}~\bibnamefont {Hait}}\ and\ \bibinfo {author} {\bibfnamefont {M.}~\bibnamefont {Head-Gordon}},\ }\bibfield  {title} {\enquote {\bibinfo {title} {How accurate are static polarizability predictions from density functional theory? an assessment over 132 species at equilibrium geometry},}\ }\href@noop {} {\bibfield  {journal} {\bibinfo  {journal} {Physical Chemistry Chemical Physics}\ }\textbf {\bibinfo {volume} {20}},\ \bibinfo {pages} {19800--19810} (\bibinfo {year} {2018})}\BibitemShut {NoStop}%
\bibitem [{\citenamefont {Jensen}(2014)}]{jensen2014unifying}%
  \BibitemOpen
  \bibfield  {author} {\bibinfo {author} {\bibfnamefont {F.}~\bibnamefont {Jensen}},\ }\bibfield  {title} {\enquote {\bibinfo {title} {Unifying general and segmented contracted basis sets. segmented polarization consistent basis sets},}\ }\href@noop {} {\bibfield  {journal} {\bibinfo  {journal} {J. Chem. Theory Comput.}\ }\textbf {\bibinfo {volume} {10}},\ \bibinfo {pages} {1074--1085} (\bibinfo {year} {2014})}\BibitemShut {NoStop}%
\bibitem [{\citenamefont {Ambroise}\ and\ \citenamefont {Jensen}(2018)}]{ambroise2018probing}%
  \BibitemOpen
  \bibfield  {author} {\bibinfo {author} {\bibfnamefont {M.~A.}\ \bibnamefont {Ambroise}}\ and\ \bibinfo {author} {\bibfnamefont {F.}~\bibnamefont {Jensen}},\ }\bibfield  {title} {\enquote {\bibinfo {title} {Probing basis set requirements for calculating core ionization and core excitation spectroscopy by the $\delta$ self-consistent-field approach},}\ }\href@noop {} {\bibfield  {journal} {\bibinfo  {journal} {J. Chem. Theory Comput.}\ }\textbf {\bibinfo {volume} {15}},\ \bibinfo {pages} {325--337} (\bibinfo {year} {2018})}\BibitemShut {NoStop}%
\bibitem [{\citenamefont {Cunha}\ \emph {et~al.}(2022)\citenamefont {Cunha}, \citenamefont {Hait}, \citenamefont {Kang}, \citenamefont {Mao},\ and\ \citenamefont {Head-Gordon}}]{cunha2022relativistic}%
  \BibitemOpen
  \bibfield  {author} {\bibinfo {author} {\bibfnamefont {L.~A.}\ \bibnamefont {Cunha}}, \bibinfo {author} {\bibfnamefont {D.}~\bibnamefont {Hait}}, \bibinfo {author} {\bibfnamefont {R.}~\bibnamefont {Kang}}, \bibinfo {author} {\bibfnamefont {Y.}~\bibnamefont {Mao}},\ and\ \bibinfo {author} {\bibfnamefont {M.}~\bibnamefont {Head-Gordon}},\ }\bibfield  {title} {\enquote {\bibinfo {title} {Relativistic orbital-optimized density functional theory for accurate core-level spectroscopy},}\ }\href@noop {} {\bibfield  {journal} {\bibinfo  {journal} {J. Phys. Chem. Lett.}\ }\textbf {\bibinfo {volume} {13}},\ \bibinfo {pages} {3438--3449} (\bibinfo {year} {2022})}\BibitemShut {NoStop}%
\bibitem [{\citenamefont {Saue}(2011)}]{saue2011relativistic}%
  \BibitemOpen
  \bibfield  {author} {\bibinfo {author} {\bibfnamefont {T.}~\bibnamefont {Saue}},\ }\bibfield  {title} {\enquote {\bibinfo {title} {Relativistic hamiltonians for chemistry: A primer},}\ }\href@noop {} {\bibfield  {journal} {\bibinfo  {journal} {ChemPhysChem}\ }\textbf {\bibinfo {volume} {12}},\ \bibinfo {pages} {3077--3094} (\bibinfo {year} {2011})}\BibitemShut {NoStop}%
\bibitem [{\citenamefont {Hait}\ and\ \citenamefont {Head-Gordon}(2020{\natexlab{b}})}]{hait2020excited}%
  \BibitemOpen
  \bibfield  {author} {\bibinfo {author} {\bibfnamefont {D.}~\bibnamefont {Hait}}\ and\ \bibinfo {author} {\bibfnamefont {M.}~\bibnamefont {Head-Gordon}},\ }\bibfield  {title} {\enquote {\bibinfo {title} {Excited state orbital optimization via minimizing the square of the gradient: General approach and application to singly and doubly excited states via density functional theory},}\ }\href@noop {} {\bibfield  {journal} {\bibinfo  {journal} {J. Chem. Theory Comput.}\ }\textbf {\bibinfo {volume} {16}},\ \bibinfo {pages} {1699--1710} (\bibinfo {year} {2020}{\natexlab{b}})}\BibitemShut {NoStop}%
\bibitem [{\citenamefont {Teodorescu}\ \emph {et~al.}(1994)\citenamefont {Teodorescu}, \citenamefont {Esteva}, \citenamefont {Karnatak},\ and\ \citenamefont {El~Afif}}]{teodorescu1994approximation}%
  \BibitemOpen
  \bibfield  {author} {\bibinfo {author} {\bibfnamefont {C.}~\bibnamefont {Teodorescu}}, \bibinfo {author} {\bibfnamefont {J.}~\bibnamefont {Esteva}}, \bibinfo {author} {\bibfnamefont {R.}~\bibnamefont {Karnatak}},\ and\ \bibinfo {author} {\bibfnamefont {A.}~\bibnamefont {El~Afif}},\ }\bibfield  {title} {\enquote {\bibinfo {title} {An approximation of the voigt i profile for the fitting of experimental x-ray absorption data},}\ }\href@noop {} {\bibfield  {journal} {\bibinfo  {journal} {Nuclear Instruments and Methods in Physics Research Section A: Accelerators, Spectrometers, Detectors and Associated Equipment}\ }\textbf {\bibinfo {volume} {345}},\ \bibinfo {pages} {141--147} (\bibinfo {year} {1994})}\BibitemShut {NoStop}%
\bibitem [{\citenamefont {Iwamitsu}\ \emph {et~al.}(2020)\citenamefont {Iwamitsu}, \citenamefont {Yokota}, \citenamefont {Murata}, \citenamefont {Kamezaki}, \citenamefont {Mizumaki}, \citenamefont {Uruga},\ and\ \citenamefont {Akai}}]{iwamitsu2020spectral}%
  \BibitemOpen
  \bibfield  {author} {\bibinfo {author} {\bibfnamefont {K.}~\bibnamefont {Iwamitsu}}, \bibinfo {author} {\bibfnamefont {T.}~\bibnamefont {Yokota}}, \bibinfo {author} {\bibfnamefont {K.}~\bibnamefont {Murata}}, \bibinfo {author} {\bibfnamefont {M.}~\bibnamefont {Kamezaki}}, \bibinfo {author} {\bibfnamefont {M.}~\bibnamefont {Mizumaki}}, \bibinfo {author} {\bibfnamefont {T.}~\bibnamefont {Uruga}},\ and\ \bibinfo {author} {\bibfnamefont {I.}~\bibnamefont {Akai}},\ }\bibfield  {title} {\enquote {\bibinfo {title} {Spectral analysis of x-ray absorption near edge structure in $\alpha$-fe2o3 based on bayesian spectroscopy},}\ }\href@noop {} {\bibfield  {journal} {\bibinfo  {journal} {physica status solidi (b)}\ }\textbf {\bibinfo {volume} {257}},\ \bibinfo {pages} {2000107} (\bibinfo {year} {2020})}\BibitemShut {NoStop}%
\bibitem [{\citenamefont {Briggs}\ and\ \citenamefont {Grant}(2012)}]{briggs2012surface}%
  \BibitemOpen
  \bibfield  {author} {\bibinfo {author} {\bibfnamefont {D.}~\bibnamefont {Briggs}}\ and\ \bibinfo {author} {\bibfnamefont {J.~T.}\ \bibnamefont {Grant}},\ }\href@noop {} {\emph {\bibinfo {title} {Surface analysis by Auger and X-ray photoelectron spectroscopy}}}\ (\bibinfo  {publisher} {SurfaceSpectra},\ \bibinfo {year} {2012})\BibitemShut {NoStop}%
\bibitem [{\citenamefont {Hudson}\ \emph {et~al.}(1993)\citenamefont {Hudson}, \citenamefont {Shirley}, \citenamefont {Domke}, \citenamefont {Remmers}, \citenamefont {Puschmann}, \citenamefont {Mandel}, \citenamefont {Xue},\ and\ \citenamefont {Kaindl}}]{Hudson1993SF6Absorption}%
  \BibitemOpen
  \bibfield  {author} {\bibinfo {author} {\bibfnamefont {E.}~\bibnamefont {Hudson}}, \bibinfo {author} {\bibfnamefont {D.~A.}\ \bibnamefont {Shirley}}, \bibinfo {author} {\bibfnamefont {M.}~\bibnamefont {Domke}}, \bibinfo {author} {\bibfnamefont {G.}~\bibnamefont {Remmers}}, \bibinfo {author} {\bibfnamefont {A.}~\bibnamefont {Puschmann}}, \bibinfo {author} {\bibfnamefont {T.}~\bibnamefont {Mandel}}, \bibinfo {author} {\bibfnamefont {C.}~\bibnamefont {Xue}},\ and\ \bibinfo {author} {\bibfnamefont {G.}~\bibnamefont {Kaindl}},\ }\bibfield  {title} {\enquote {\bibinfo {title} {High-resolution measurements of near-edge resonances in the core-level photoionization spectra of ${\mathrm{sf}}_{6}$},}\ }\href {https://doi.org/10.1103/PhysRevA.47.361} {\bibfield  {journal} {\bibinfo  {journal} {Phys. Rev. A}\ }\textbf {\bibinfo {volume} {47}},\ \bibinfo {pages} {361--373} (\bibinfo {year} {1993})}\BibitemShut {NoStop}%
\bibitem [{\citenamefont {Hitchcock}\ and\ \citenamefont {Brion}(1978)}]{hitchcock1978inner}%
  \BibitemOpen
  \bibfield  {author} {\bibinfo {author} {\bibfnamefont {A.}~\bibnamefont {Hitchcock}}\ and\ \bibinfo {author} {\bibfnamefont {C.}~\bibnamefont {Brion}},\ }\bibfield  {title} {\enquote {\bibinfo {title} {Inner-shell excitation and exafs-type phenomena in the chloromethanes},}\ }\href@noop {} {\bibfield  {journal} {\bibinfo  {journal} {Journal of Electron Spectroscopy and Related Phenomena}\ }\textbf {\bibinfo {volume} {14}},\ \bibinfo {pages} {417--441} (\bibinfo {year} {1978})}\BibitemShut {NoStop}%
\bibitem [{\citenamefont {Lu}\ \emph {et~al.}(2008)\citenamefont {Lu}, \citenamefont {Chen}, \citenamefont {Lee}, \citenamefont {Chen}, \citenamefont {Chou},\ and\ \citenamefont {Chen}}]{lu2008state}%
  \BibitemOpen
  \bibfield  {author} {\bibinfo {author} {\bibfnamefont {K.}~\bibnamefont {Lu}}, \bibinfo {author} {\bibfnamefont {J.}~\bibnamefont {Chen}}, \bibinfo {author} {\bibfnamefont {J.}~\bibnamefont {Lee}}, \bibinfo {author} {\bibfnamefont {C.}~\bibnamefont {Chen}}, \bibinfo {author} {\bibfnamefont {T.}~\bibnamefont {Chou}},\ and\ \bibinfo {author} {\bibfnamefont {H.}~\bibnamefont {Chen}},\ }\bibfield  {title} {\enquote {\bibinfo {title} {State-specific dissociation enhancement of ionic and excited neutral photofragments of gaseous ccl4 and solid-state analogs following cl 2p core-level excitation},}\ }\href@noop {} {\bibfield  {journal} {\bibinfo  {journal} {New Journal of Physics}\ }\textbf {\bibinfo {volume} {10}},\ \bibinfo {pages} {053009} (\bibinfo {year} {2008})}\BibitemShut {NoStop}%
\bibitem [{\citenamefont {Chakraborty}\ and\ \citenamefont {Rai}(2006)}]{chakraborty2006comparative}%
  \BibitemOpen
  \bibfield  {author} {\bibinfo {author} {\bibfnamefont {T.}~\bibnamefont {Chakraborty}}\ and\ \bibinfo {author} {\bibfnamefont {S.~N.}\ \bibnamefont {Rai}},\ }\bibfield  {title} {\enquote {\bibinfo {title} {Comparative study of infrared and raman spectra of ccl4 in vapour and condensed phases: Effect of lo--to splitting resulting from hetero-isotopic td--td interactions},}\ }\href@noop {} {\bibfield  {journal} {\bibinfo  {journal} {Spectrochimica Acta Part A: Molecular and Biomolecular Spectroscopy}\ }\textbf {\bibinfo {volume} {65}},\ \bibinfo {pages} {406--413} (\bibinfo {year} {2006})}\BibitemShut {NoStop}%
\bibitem [{\citenamefont {Gaynor}\ \emph {et~al.}(2015)\citenamefont {Gaynor}, \citenamefont {Wetterer}, \citenamefont {Cochran}, \citenamefont {Valente},\ and\ \citenamefont {Mayer}}]{gaynor2015vibrational}%
  \BibitemOpen
  \bibfield  {author} {\bibinfo {author} {\bibfnamefont {J.~D.}\ \bibnamefont {Gaynor}}, \bibinfo {author} {\bibfnamefont {A.~M.}\ \bibnamefont {Wetterer}}, \bibinfo {author} {\bibfnamefont {R.~M.}\ \bibnamefont {Cochran}}, \bibinfo {author} {\bibfnamefont {E.~J.}\ \bibnamefont {Valente}},\ and\ \bibinfo {author} {\bibfnamefont {S.~G.}\ \bibnamefont {Mayer}},\ }\bibfield  {title} {\enquote {\bibinfo {title} {Vibrational spectroscopy of the ccl4 v 1 mode: Theoretical prediction of isotopic effects},}\ }\href@noop {} {\bibfield  {journal} {\bibinfo  {journal} {Journal of Chemical Education}\ }\textbf {\bibinfo {volume} {92}},\ \bibinfo {pages} {1081--1085} (\bibinfo {year} {2015})}\BibitemShut {NoStop}%
\bibitem [{\citenamefont {Schmidt}\ and\ \citenamefont {Hartmann}(2018)}]{schmidt2018wavepacket}%
  \BibitemOpen
  \bibfield  {author} {\bibinfo {author} {\bibfnamefont {B.}~\bibnamefont {Schmidt}}\ and\ \bibinfo {author} {\bibfnamefont {C.}~\bibnamefont {Hartmann}},\ }\bibfield  {title} {\enquote {\bibinfo {title} {Wavepacket: A matlab package for numerical quantum dynamics. ii: Open quantum systems, optimal control, and model reduction},}\ }\href@noop {} {\bibfield  {journal} {\bibinfo  {journal} {Computer Physics Communications}\ }\textbf {\bibinfo {volume} {228}},\ \bibinfo {pages} {229--244} (\bibinfo {year} {2018})}\BibitemShut {NoStop}%
\bibitem [{\citenamefont {Hong}\ and\ \citenamefont {Nam}(2004)}]{hong2004adaptive}%
  \BibitemOpen
  \bibfield  {author} {\bibinfo {author} {\bibfnamefont {K.-H.}\ \bibnamefont {Hong}}\ and\ \bibinfo {author} {\bibfnamefont {C.~H.}\ \bibnamefont {Nam}},\ }\bibfield  {title} {\enquote {\bibinfo {title} {Adaptive pulse compression of femtosecond laser pulses using a low-loss pulse shaper},}\ }\href@noop {} {\bibfield  {journal} {\bibinfo  {journal} {Japanese journal of applied physics}\ }\textbf {\bibinfo {volume} {43}},\ \bibinfo {pages} {5289} (\bibinfo {year} {2004})}\BibitemShut {NoStop}%
\bibitem [{\citenamefont {Meng}\ \emph {et~al.}(2010)\citenamefont {Meng}, \citenamefont {Zhang}, \citenamefont {Jin}, \citenamefont {Li},\ and\ \citenamefont {Wang}}]{meng2010enhanced}%
  \BibitemOpen
  \bibfield  {author} {\bibinfo {author} {\bibfnamefont {Y.}~\bibnamefont {Meng}}, \bibinfo {author} {\bibfnamefont {S.}~\bibnamefont {Zhang}}, \bibinfo {author} {\bibfnamefont {C.}~\bibnamefont {Jin}}, \bibinfo {author} {\bibfnamefont {H.}~\bibnamefont {Li}},\ and\ \bibinfo {author} {\bibfnamefont {X.}~\bibnamefont {Wang}},\ }\bibfield  {title} {\enquote {\bibinfo {title} {Enhanced compression of femtosecond pulse in hollow-core photonic bandgap fibers},}\ }\href@noop {} {\bibfield  {journal} {\bibinfo  {journal} {Optics communications}\ }\textbf {\bibinfo {volume} {283}},\ \bibinfo {pages} {2411--2415} (\bibinfo {year} {2010})}\BibitemShut {NoStop}%
\bibitem [{\citenamefont {Miranda}\ \emph {et~al.}(2012)\citenamefont {Miranda}, \citenamefont {Arnold}, \citenamefont {Fordell}, \citenamefont {Silva}, \citenamefont {Alonso}, \citenamefont {Weigand}, \citenamefont {L’Huillier},\ and\ \citenamefont {Crespo}}]{miranda2012characterization}%
  \BibitemOpen
  \bibfield  {author} {\bibinfo {author} {\bibfnamefont {M.}~\bibnamefont {Miranda}}, \bibinfo {author} {\bibfnamefont {C.~L.}\ \bibnamefont {Arnold}}, \bibinfo {author} {\bibfnamefont {T.}~\bibnamefont {Fordell}}, \bibinfo {author} {\bibfnamefont {F.}~\bibnamefont {Silva}}, \bibinfo {author} {\bibfnamefont {B.}~\bibnamefont {Alonso}}, \bibinfo {author} {\bibfnamefont {R.}~\bibnamefont {Weigand}}, \bibinfo {author} {\bibfnamefont {A.}~\bibnamefont {L’Huillier}},\ and\ \bibinfo {author} {\bibfnamefont {H.}~\bibnamefont {Crespo}},\ }\bibfield  {title} {\enquote {\bibinfo {title} {Characterization of broadband few-cycle laser pulses with the d-scan technique},}\ }\href@noop {} {\bibfield  {journal} {\bibinfo  {journal} {Optics express}\ }\textbf {\bibinfo {volume} {20}},\ \bibinfo {pages} {18732--18743} (\bibinfo {year} {2012})}\BibitemShut {NoStop}%
\bibitem [{\citenamefont {Ehrenfest}(1927)}]{ehrenfest1927bemerkung}%
  \BibitemOpen
  \bibfield  {author} {\bibinfo {author} {\bibfnamefont {P.}~\bibnamefont {Ehrenfest}},\ }\bibfield  {title} {\enquote {\bibinfo {title} {Bemerkung {\"u}ber die angen{\"a}herte g{\"u}ltigkeit der klassischen mechanik innerhalb der quantenmechanik},}\ }\href@noop {} {\bibfield  {journal} {\bibinfo  {journal} {Zeitschrift f{\"u}r physik}\ }\textbf {\bibinfo {volume} {45}},\ \bibinfo {pages} {455--457} (\bibinfo {year} {1927})}\BibitemShut {NoStop}%
\bibitem [{\citenamefont {Shankar}(2012)}]{shankar2012principles}%
  \BibitemOpen
  \bibfield  {author} {\bibinfo {author} {\bibfnamefont {R.}~\bibnamefont {Shankar}},\ }\href@noop {} {\emph {\bibinfo {title} {Principles of quantum mechanics}}}\ (\bibinfo  {publisher} {Springer Science \& Business Media},\ \bibinfo {year} {2012})\BibitemShut {NoStop}%
\bibitem [{\citenamefont {L{\"o}wdin}(1950)}]{lowdin1950non}%
  \BibitemOpen
  \bibfield  {author} {\bibinfo {author} {\bibfnamefont {P.-O.}\ \bibnamefont {L{\"o}wdin}},\ }\bibfield  {title} {\enquote {\bibinfo {title} {On the non-orthogonality problem connected with the use of atomic wave functions in the theory of molecules and crystals},}\ }\href@noop {} {\bibfield  {journal} {\bibinfo  {journal} {The Journal of Chemical Physics}\ }\textbf {\bibinfo {volume} {18}},\ \bibinfo {pages} {365--375} (\bibinfo {year} {1950})}\BibitemShut {NoStop}%
\bibitem [{\citenamefont {Yamamoto}, \citenamefont {Takami},\ and\ \citenamefont {Kuchitsu}(1984)}]{yamamoto1984diode}%
  \BibitemOpen
  \bibfield  {author} {\bibinfo {author} {\bibfnamefont {S.}~\bibnamefont {Yamamoto}}, \bibinfo {author} {\bibfnamefont {M.}~\bibnamefont {Takami}},\ and\ \bibinfo {author} {\bibfnamefont {K.}~\bibnamefont {Kuchitsu}},\ }\bibfield  {title} {\enquote {\bibinfo {title} {Diode laser spectroscopy of the $\nu$3 band of carbon tetrachloride (c35cl4): Stark modulation and cold jet infrared absorption spectrum},}\ }\href@noop {} {\bibfield  {journal} {\bibinfo  {journal} {The Journal of chemical physics}\ }\textbf {\bibinfo {volume} {81}},\ \bibinfo {pages} {3800--3804} (\bibinfo {year} {1984})}\BibitemShut {NoStop}%
\bibitem [{\citenamefont {Hait}\ and\ \citenamefont {Head-Gordon}(2023)}]{hait2023bond}%
  \BibitemOpen
  \bibfield  {author} {\bibinfo {author} {\bibfnamefont {D.}~\bibnamefont {Hait}}\ and\ \bibinfo {author} {\bibfnamefont {M.}~\bibnamefont {Head-Gordon}},\ }\bibfield  {title} {\enquote {\bibinfo {title} {When is a bond broken? the polarizability perspective.}}\ }\href@noop {} {\bibfield  {journal} {\bibinfo  {journal} {Angewandte Chemie International Edition}\ }\textbf {\bibinfo {volume} {62}},\ \bibinfo {pages} {e202312078} (\bibinfo {year} {2023})}\BibitemShut {NoStop}%
\bibitem [{\citenamefont {Jahn}\ and\ \citenamefont {Teller}(1937)}]{jahn1937stability}%
  \BibitemOpen
  \bibfield  {author} {\bibinfo {author} {\bibfnamefont {H.~A.}\ \bibnamefont {Jahn}}\ and\ \bibinfo {author} {\bibfnamefont {E.}~\bibnamefont {Teller}},\ }\bibfield  {title} {\enquote {\bibinfo {title} {Stability of polyatomic molecules in degenerate electronic states-i—orbital degeneracy},}\ }\href@noop {} {\bibfield  {journal} {\bibinfo  {journal} {Proceedings of the Royal Society of London. Series A-Mathematical and Physical Sciences}\ }\textbf {\bibinfo {volume} {161}},\ \bibinfo {pages} {220--235} (\bibinfo {year} {1937})}\BibitemShut {NoStop}%
\bibitem [{\citenamefont {Kowalczyk}\ \emph {et~al.}(2013)\citenamefont {Kowalczyk}, \citenamefont {Tsuchimochi}, \citenamefont {Chen}, \citenamefont {Top},\ and\ \citenamefont {Van~Voorhis}}]{kowalczyk2013excitation}%
  \BibitemOpen
  \bibfield  {author} {\bibinfo {author} {\bibfnamefont {T.}~\bibnamefont {Kowalczyk}}, \bibinfo {author} {\bibfnamefont {T.}~\bibnamefont {Tsuchimochi}}, \bibinfo {author} {\bibfnamefont {P.-T.}\ \bibnamefont {Chen}}, \bibinfo {author} {\bibfnamefont {L.}~\bibnamefont {Top}},\ and\ \bibinfo {author} {\bibfnamefont {T.}~\bibnamefont {Van~Voorhis}},\ }\bibfield  {title} {\enquote {\bibinfo {title} {Excitation energies and stokes shifts from a restricted open-shell kohn-sham approach},}\ }\href@noop {} {\bibfield  {journal} {\bibinfo  {journal} {The Journal of chemical physics}\ }\textbf {\bibinfo {volume} {138}} (\bibinfo {year} {2013})}\BibitemShut {NoStop}%
\bibitem [{\citenamefont {Ross}\ \emph {et~al.}(2024)\citenamefont {Ross}, \citenamefont {Hait}, \citenamefont {Scutelnic}, \citenamefont {Neumark}, \citenamefont {Head-Gordon},\ and\ \citenamefont {Leone}}]{ross_2024_11153002}%
  \BibitemOpen
  \bibfield  {author} {\bibinfo {author} {\bibfnamefont {A.}~\bibnamefont {Ross}}, \bibinfo {author} {\bibfnamefont {D.}~\bibnamefont {Hait}}, \bibinfo {author} {\bibfnamefont {V.}~\bibnamefont {Scutelnic}}, \bibinfo {author} {\bibfnamefont {D.}~\bibnamefont {Neumark}}, \bibinfo {author} {\bibfnamefont {M.}~\bibnamefont {Head-Gordon}},\ and\ \bibinfo {author} {\bibfnamefont {S.}~\bibnamefont {Leone}},\ }\bibfield  {title} {\enquote {\bibinfo {title} {{Computational data for `Measurement of Coherent Vibrational Dynamics with X-ray Transient Absorption Spectroscopy Simultaneously at the Carbon K- and Chlorine L$_{2,3}$- Edges'}},}\ }\href {https://doi.org/10.5281/zenodo.11153002} {10.5281/zenodo.11153002} (\bibinfo {year} {2024})\BibitemShut {NoStop}%
\end{thebibliography}%


\begin{thebibliography}{19}%
\makeatletter
\providecommand \@ifxundefined [1]{%
 \@ifx{#1\undefined}
}%
\providecommand \@ifnum [1]{%
 \ifnum #1\expandafter \@firstoftwo
 \else \expandafter \@secondoftwo
 \fi
}%
\providecommand \@ifx [1]{%
 \ifx #1\expandafter \@firstoftwo
 \else \expandafter \@secondoftwo
 \fi
}%
\providecommand \natexlab [1]{#1}%
\providecommand \enquote  [1]{``#1''}%
\providecommand \bibnamefont  [1]{#1}%
\providecommand \bibfnamefont [1]{#1}%
\providecommand \citenamefont [1]{#1}%
\providecommand \href@noop [0]{\@secondoftwo}%
\providecommand \href [0]{\begingroup \@sanitize@url \@href}%
\providecommand \@href[1]{\@@startlink{#1}\@@href}%
\providecommand \@@href[1]{\endgroup#1\@@endlink}%
\providecommand \@sanitize@url [0]{\catcode `\\12\catcode `\$12\catcode `\&12\catcode `\#12\catcode `\^12\catcode `\_12\catcode `\%12\relax}%
\providecommand \@@startlink[1]{}%
\providecommand \@@endlink[0]{}%
\providecommand \url  [0]{\begingroup\@sanitize@url \@url }%
\providecommand \@url [1]{\endgroup\@href {#1}{\urlprefix }}%
\providecommand \urlprefix  [0]{URL }%
\providecommand \Eprint [0]{\href }%
\providecommand \doibase [0]{https://doi.org/}%
\providecommand \selectlanguage [0]{\@gobble}%
\providecommand \bibinfo  [0]{\@secondoftwo}%
\providecommand \bibfield  [0]{\@secondoftwo}%
\providecommand \translation [1]{[#1]}%
\providecommand \BibitemOpen [0]{}%
\providecommand \bibitemStop [0]{}%
\providecommand \bibitemNoStop [0]{.\EOS\space}%
\providecommand \EOS [0]{\spacefactor3000\relax}%
\providecommand \BibitemShut  [1]{\csname bibitem#1\endcsname}%
\let\auto@bib@innerbib\@empty
\bibitem [{\citenamefont {G\'{e}neaux}\ \emph {et~al.}(2021)\citenamefont {G\'{e}neaux}, \citenamefont {Chang}, \citenamefont {Schwartzberg},\ and\ \citenamefont {Marroux}}]{Geneaux2021XUVNoise}%
  \BibitemOpen
  \bibfield  {author} {\bibinfo {author} {\bibfnamefont {R.}~\bibnamefont {G\'{e}neaux}}, \bibinfo {author} {\bibfnamefont {H.-T.}\ \bibnamefont {Chang}}, \bibinfo {author} {\bibfnamefont {A.~M.}\ \bibnamefont {Schwartzberg}},\ and\ \bibinfo {author} {\bibfnamefont {H.~J.~B.}\ \bibnamefont {Marroux}},\ }\bibfield  {title} {\enquote {\bibinfo {title} {Source noise suppression in attosecond transient absorption spectroscopy by edge-pixel referencing},}\ }\href {https://doi.org/10.1364/OE.412117} {\bibfield  {journal} {\bibinfo  {journal} {Opt. Express}\ }\textbf {\bibinfo {volume} {29}},\ \bibinfo {pages} {951--960} (\bibinfo {year} {2021})}\BibitemShut {NoStop}%
\bibitem [{\citenamefont {Geneaux}\ \emph {et~al.}(2019)\citenamefont {Geneaux}, \citenamefont {Marroux}, \citenamefont {Guggenmos}, \citenamefont {Neumark},\ and\ \citenamefont {Leone}}]{Geneaux2019ATASReview}%
  \BibitemOpen
  \bibfield  {author} {\bibinfo {author} {\bibfnamefont {R.}~\bibnamefont {Geneaux}}, \bibinfo {author} {\bibfnamefont {H.~J.~B.}\ \bibnamefont {Marroux}}, \bibinfo {author} {\bibfnamefont {A.}~\bibnamefont {Guggenmos}}, \bibinfo {author} {\bibfnamefont {D.~M.}\ \bibnamefont {Neumark}},\ and\ \bibinfo {author} {\bibfnamefont {S.~R.}\ \bibnamefont {Leone}},\ }\bibfield  {title} {\enquote {\bibinfo {title} {Transient absorption spectroscopy using high harmonic generation: a review of ultrafast x-ray dynamics in molecules and solids},}\ }\href {https://doi.org/10.1098/rsta.2017.0463} {\bibfield  {journal} {\bibinfo  {journal} {Philosophical Transactions of the Royal Society A: Mathematical, Physical and Engineering Sciences}\ }\textbf {\bibinfo {volume} {377}},\ \bibinfo {pages} {20170463} (\bibinfo {year} {2019})}\BibitemShut {NoStop}%
\bibitem [{\citenamefont {Ott}\ \emph {et~al.}(2014)\citenamefont {Ott}, \citenamefont {Kaldun}, \citenamefont {Argenti}, \citenamefont {Raith}, \citenamefont {Meyer}, \citenamefont {Laux}, \citenamefont {Zhang}, \citenamefont {Bl{\"a}ttermann}, \citenamefont {Hagstotz}, \citenamefont {Ding} \emph {et~al.}}]{ottNature2014}%
  \BibitemOpen
  \bibfield  {author} {\bibinfo {author} {\bibfnamefont {C.}~\bibnamefont {Ott}}, \bibinfo {author} {\bibfnamefont {A.}~\bibnamefont {Kaldun}}, \bibinfo {author} {\bibfnamefont {L.}~\bibnamefont {Argenti}}, \bibinfo {author} {\bibfnamefont {P.}~\bibnamefont {Raith}}, \bibinfo {author} {\bibfnamefont {K.}~\bibnamefont {Meyer}}, \bibinfo {author} {\bibfnamefont {M.}~\bibnamefont {Laux}}, \bibinfo {author} {\bibfnamefont {Y.}~\bibnamefont {Zhang}}, \bibinfo {author} {\bibfnamefont {A.}~\bibnamefont {Bl{\"a}ttermann}}, \bibinfo {author} {\bibfnamefont {S.}~\bibnamefont {Hagstotz}}, \bibinfo {author} {\bibfnamefont {T.}~\bibnamefont {Ding}}, \emph {et~al.},\ }\bibfield  {title} {\enquote {\bibinfo {title} {Reconstruction and control of a time-dependent two-electron wave packet},}\ }\href {https://doi.org/https://doi.org/10.1038/nature14026} {\bibfield  {journal} {\bibinfo  {journal} {Nature}\ }\textbf {\bibinfo {volume} {516}},\ \bibinfo {pages} {374} (\bibinfo {year} {2014})}\BibitemShut {NoStop}%
\bibitem [{\citenamefont {Bhattacherjee}\ and\ \citenamefont {Leone}(2018)}]{bhattacherjee2018ultrafast}%
  \BibitemOpen
  \bibfield  {author} {\bibinfo {author} {\bibfnamefont {A.}~\bibnamefont {Bhattacherjee}}\ and\ \bibinfo {author} {\bibfnamefont {S.~R.}\ \bibnamefont {Leone}},\ }\bibfield  {title} {\enquote {\bibinfo {title} {Ultrafast x-ray transient absorption spectroscopy of gas-phase photochemical reactions: A new universal probe of photoinduced molecular dynamics},}\ }\href@noop {} {\bibfield  {journal} {\bibinfo  {journal} {Accounts of chemical research}\ }\textbf {\bibinfo {volume} {51}},\ \bibinfo {pages} {3203--3211} (\bibinfo {year} {2018})}\BibitemShut {NoStop}%
\bibitem [{\citenamefont {Sidiropoulos}\ \emph {et~al.}(2021)\citenamefont {Sidiropoulos}, \citenamefont {Di~Palo}, \citenamefont {Rivas}, \citenamefont {Severino}, \citenamefont {Reduzzi}, \citenamefont {Nandy}, \citenamefont {Bauerhenne}, \citenamefont {Krylow}, \citenamefont {Vasileiadis}, \citenamefont {Danz} \emph {et~al.}}]{sidiropoulos2021probing}%
  \BibitemOpen
  \bibfield  {author} {\bibinfo {author} {\bibfnamefont {T.}~\bibnamefont {Sidiropoulos}}, \bibinfo {author} {\bibfnamefont {N.}~\bibnamefont {Di~Palo}}, \bibinfo {author} {\bibfnamefont {D.}~\bibnamefont {Rivas}}, \bibinfo {author} {\bibfnamefont {S.}~\bibnamefont {Severino}}, \bibinfo {author} {\bibfnamefont {M.}~\bibnamefont {Reduzzi}}, \bibinfo {author} {\bibfnamefont {B.}~\bibnamefont {Nandy}}, \bibinfo {author} {\bibfnamefont {B.}~\bibnamefont {Bauerhenne}}, \bibinfo {author} {\bibfnamefont {S.}~\bibnamefont {Krylow}}, \bibinfo {author} {\bibfnamefont {T.}~\bibnamefont {Vasileiadis}}, \bibinfo {author} {\bibfnamefont {T.}~\bibnamefont {Danz}}, \emph {et~al.},\ }\bibfield  {title} {\enquote {\bibinfo {title} {Probing the energy conversion pathways between light, carriers, and lattice in real time with attosecond core-level spectroscopy},}\ }\href@noop {} {\bibfield  {journal} {\bibinfo  {journal} {Physical Review X}\ }\textbf {\bibinfo {volume} {11}},\ \bibinfo {pages} {041060} (\bibinfo {year}
  {2021})}\BibitemShut {NoStop}%
\bibitem [{\citenamefont {Liao}\ \emph {et~al.}(2017)\citenamefont {Liao}, \citenamefont {Li}, \citenamefont {Haxton}, \citenamefont {Rescigno}, \citenamefont {Lucchese}, \citenamefont {McCurdy},\ and\ \citenamefont {Sandhu}}]{liao2017probing}%
  \BibitemOpen
  \bibfield  {author} {\bibinfo {author} {\bibfnamefont {C.-T.}\ \bibnamefont {Liao}}, \bibinfo {author} {\bibfnamefont {X.}~\bibnamefont {Li}}, \bibinfo {author} {\bibfnamefont {D.~J.}\ \bibnamefont {Haxton}}, \bibinfo {author} {\bibfnamefont {T.~N.}\ \bibnamefont {Rescigno}}, \bibinfo {author} {\bibfnamefont {R.~R.}\ \bibnamefont {Lucchese}}, \bibinfo {author} {\bibfnamefont {C.~W.}\ \bibnamefont {McCurdy}},\ and\ \bibinfo {author} {\bibfnamefont {A.}~\bibnamefont {Sandhu}},\ }\bibfield  {title} {\enquote {\bibinfo {title} {Probing autoionizing states of molecular oxygen with xuv transient absorption: Electronic-symmetry-dependent line shapes and laser-induced modifications},}\ }\href@noop {} {\bibfield  {journal} {\bibinfo  {journal} {Physical Review A}\ }\textbf {\bibinfo {volume} {95}},\ \bibinfo {pages} {043427} (\bibinfo {year} {2017})}\BibitemShut {NoStop}%
\bibitem [{\citenamefont {Ross}\ \emph {et~al.}(2022)\citenamefont {Ross}, \citenamefont {Hait}, \citenamefont {Scutelnic}, \citenamefont {Haugen}, \citenamefont {Ridente}, \citenamefont {Balkew}, \citenamefont {Neumark}, \citenamefont {Head-Gordon},\ and\ \citenamefont {Leone}}]{ross2022jahn}%
  \BibitemOpen
  \bibfield  {author} {\bibinfo {author} {\bibfnamefont {A.~D.}\ \bibnamefont {Ross}}, \bibinfo {author} {\bibfnamefont {D.}~\bibnamefont {Hait}}, \bibinfo {author} {\bibfnamefont {V.}~\bibnamefont {Scutelnic}}, \bibinfo {author} {\bibfnamefont {E.~A.}\ \bibnamefont {Haugen}}, \bibinfo {author} {\bibfnamefont {E.}~\bibnamefont {Ridente}}, \bibinfo {author} {\bibfnamefont {M.~B.}\ \bibnamefont {Balkew}}, \bibinfo {author} {\bibfnamefont {D.~M.}\ \bibnamefont {Neumark}}, \bibinfo {author} {\bibfnamefont {M.}~\bibnamefont {Head-Gordon}},\ and\ \bibinfo {author} {\bibfnamefont {S.~R.}\ \bibnamefont {Leone}},\ }\bibfield  {title} {\enquote {\bibinfo {title} {Jahn-teller distortion and dissociation of ccl 4+ by transient x-ray spectroscopy simultaneously at the carbon k-and chlorine l-edge},}\ }\href@noop {} {\bibfield  {journal} {\bibinfo  {journal} {Chemical science}\ }\textbf {\bibinfo {volume} {13}},\ \bibinfo {pages} {9310--9320} (\bibinfo {year} {2022})}\BibitemShut {NoStop}%
\bibitem [{\citenamefont {Korn}, \citenamefont {D{\"u}hr},\ and\ \citenamefont {Nazarkin}(1998)}]{korn1998observation}%
  \BibitemOpen
  \bibfield  {author} {\bibinfo {author} {\bibfnamefont {G.}~\bibnamefont {Korn}}, \bibinfo {author} {\bibfnamefont {O.}~\bibnamefont {D{\"u}hr}},\ and\ \bibinfo {author} {\bibfnamefont {A.}~\bibnamefont {Nazarkin}},\ }\bibfield  {title} {\enquote {\bibinfo {title} {Observation of raman self-conversion of fs-pulse frequency due to impulsive excitation of molecular vibrations},}\ }\href@noop {} {\bibfield  {journal} {\bibinfo  {journal} {Physical review letters}\ }\textbf {\bibinfo {volume} {81}},\ \bibinfo {pages} {1215} (\bibinfo {year} {1998})}\BibitemShut {NoStop}%
\bibitem [{\citenamefont {Chakraborty}\ and\ \citenamefont {Verma}(2002)}]{CHAKRABORTY2002RamanCCl4}%
  \BibitemOpen
  \bibfield  {author} {\bibinfo {author} {\bibfnamefont {T.}~\bibnamefont {Chakraborty}}\ and\ \bibinfo {author} {\bibfnamefont {A.}~\bibnamefont {Verma}},\ }\bibfield  {title} {\enquote {\bibinfo {title} {Vibrational spectra of ccl4: isotopic components and hot bands. part i},}\ }\href {https://doi.org/https://doi.org/10.1016/S1386-1425(01)00571-6} {\bibfield  {journal} {\bibinfo  {journal} {Spectrochimica Acta Part A: Molecular and Biomolecular Spectroscopy}\ }\textbf {\bibinfo {volume} {58}},\ \bibinfo {pages} {1013--1023} (\bibinfo {year} {2002})}\BibitemShut {NoStop}%
\bibitem [{\citenamefont {Gaynor}\ \emph {et~al.}(2015)\citenamefont {Gaynor}, \citenamefont {Wetterer}, \citenamefont {Cochran}, \citenamefont {Valente},\ and\ \citenamefont {Mayer}}]{gaynor2015vibrational}%
  \BibitemOpen
  \bibfield  {author} {\bibinfo {author} {\bibfnamefont {J.~D.}\ \bibnamefont {Gaynor}}, \bibinfo {author} {\bibfnamefont {A.~M.}\ \bibnamefont {Wetterer}}, \bibinfo {author} {\bibfnamefont {R.~M.}\ \bibnamefont {Cochran}}, \bibinfo {author} {\bibfnamefont {E.~J.}\ \bibnamefont {Valente}},\ and\ \bibinfo {author} {\bibfnamefont {S.~G.}\ \bibnamefont {Mayer}},\ }\bibfield  {title} {\enquote {\bibinfo {title} {Vibrational spectroscopy of the ccl4 v 1 mode: Theoretical prediction of isotopic effects},}\ }\href@noop {} {\bibfield  {journal} {\bibinfo  {journal} {Journal of Chemical Education}\ }\textbf {\bibinfo {volume} {92}},\ \bibinfo {pages} {1081--1085} (\bibinfo {year} {2015})}\BibitemShut {NoStop}%
\bibitem [{\citenamefont {Ehrenfest}(1927)}]{ehrenfest1927bemerkung}%
  \BibitemOpen
  \bibfield  {author} {\bibinfo {author} {\bibfnamefont {P.}~\bibnamefont {Ehrenfest}},\ }\bibfield  {title} {\enquote {\bibinfo {title} {Bemerkung {\"u}ber die angen{\"a}herte g{\"u}ltigkeit der klassischen mechanik innerhalb der quantenmechanik},}\ }\href@noop {} {\bibfield  {journal} {\bibinfo  {journal} {Zeitschrift f{\"u}r physik}\ }\textbf {\bibinfo {volume} {45}},\ \bibinfo {pages} {455--457} (\bibinfo {year} {1927})}\BibitemShut {NoStop}%
\bibitem [{\citenamefont {Shankar}(2012)}]{shankar2012principles}%
  \BibitemOpen
  \bibfield  {author} {\bibinfo {author} {\bibfnamefont {R.}~\bibnamefont {Shankar}},\ }\href@noop {} {\emph {\bibinfo {title} {Principles of quantum mechanics}}}\ (\bibinfo  {publisher} {Springer Science \& Business Media},\ \bibinfo {year} {2012})\BibitemShut {NoStop}%
\bibitem [{\citenamefont {Swope}\ \emph {et~al.}(1982)\citenamefont {Swope}, \citenamefont {Andersen}, \citenamefont {Berens},\ and\ \citenamefont {Wilson}}]{swope1982computer}%
  \BibitemOpen
  \bibfield  {author} {\bibinfo {author} {\bibfnamefont {W.~C.}\ \bibnamefont {Swope}}, \bibinfo {author} {\bibfnamefont {H.~C.}\ \bibnamefont {Andersen}}, \bibinfo {author} {\bibfnamefont {P.~H.}\ \bibnamefont {Berens}},\ and\ \bibinfo {author} {\bibfnamefont {K.~R.}\ \bibnamefont {Wilson}},\ }\bibfield  {title} {\enquote {\bibinfo {title} {A computer simulation method for the calculation of equilibrium constants for the formation of physical clusters of molecules: Application to small water clusters},}\ }\href@noop {} {\bibfield  {journal} {\bibinfo  {journal} {The Journal of chemical physics}\ }\textbf {\bibinfo {volume} {76}},\ \bibinfo {pages} {637--649} (\bibinfo {year} {1982})}\BibitemShut {NoStop}%
\bibitem [{\citenamefont {Yan}, \citenamefont {Gamble~Jr},\ and\ \citenamefont {Nelson}(1985)}]{yan1985impulsive}%
  \BibitemOpen
  \bibfield  {author} {\bibinfo {author} {\bibfnamefont {Y.-X.}\ \bibnamefont {Yan}}, \bibinfo {author} {\bibfnamefont {E.~B.}\ \bibnamefont {Gamble~Jr}},\ and\ \bibinfo {author} {\bibfnamefont {K.~A.}\ \bibnamefont {Nelson}},\ }\bibfield  {title} {\enquote {\bibinfo {title} {Impulsive stimulated scattering: General importance in femtosecond laser pulse interactions with matter, and spectroscopic applications},}\ }\href@noop {} {\bibfield  {journal} {\bibinfo  {journal} {The Journal of chemical physics}\ }\textbf {\bibinfo {volume} {83}},\ \bibinfo {pages} {5391--5399} (\bibinfo {year} {1985})}\BibitemShut {NoStop}%
\bibitem [{\citenamefont {Lebedev}\ and\ \citenamefont {Skorokhodov}(1992)}]{lebedev1992quadrature}%
  \BibitemOpen
  \bibfield  {author} {\bibinfo {author} {\bibfnamefont {V.~I.}\ \bibnamefont {Lebedev}}\ and\ \bibinfo {author} {\bibfnamefont {A.}~\bibnamefont {Skorokhodov}},\ }\bibfield  {title} {\enquote {\bibinfo {title} {Quadrature formulas of orders 41, 47, and 53 for the sphere},}\ }\href@noop {} {\bibfield  {journal} {\bibinfo  {journal} {Russian Acad. Sci. Dokl. Math}\ }\textbf {\bibinfo {volume} {45}},\ \bibinfo {pages} {587--592} (\bibinfo {year} {1992})}\BibitemShut {NoStop}%
\bibitem [{\citenamefont {Ross}\ \emph {et~al.}(2024)\citenamefont {Ross}, \citenamefont {Hait}, \citenamefont {Scutelnic}, \citenamefont {Neumark}, \citenamefont {Head-Gordon},\ and\ \citenamefont {Leone}}]{ross_2024_11153002}%
  \BibitemOpen
  \bibfield  {author} {\bibinfo {author} {\bibfnamefont {A.}~\bibnamefont {Ross}}, \bibinfo {author} {\bibfnamefont {D.}~\bibnamefont {Hait}}, \bibinfo {author} {\bibfnamefont {V.}~\bibnamefont {Scutelnic}}, \bibinfo {author} {\bibfnamefont {D.}~\bibnamefont {Neumark}}, \bibinfo {author} {\bibfnamefont {M.}~\bibnamefont {Head-Gordon}},\ and\ \bibinfo {author} {\bibfnamefont {S.}~\bibnamefont {Leone}},\ }\bibfield  {title} {\enquote {\bibinfo {title} {{Computational data for `Measurement of Coherent Vibrational Dynamics with X-ray Transient Absorption Spectroscopy Simultaneously at the Carbon K- and Chlorine L$_{2,3}$- Edges'}},}\ }\href {https://doi.org/10.5281/zenodo.11153002} {10.5281/zenodo.11153002} (\bibinfo {year} {2024})\BibitemShut {NoStop}%
\bibitem [{\citenamefont {Kowalczyk}\ \emph {et~al.}(2013)\citenamefont {Kowalczyk}, \citenamefont {Tsuchimochi}, \citenamefont {Chen}, \citenamefont {Top},\ and\ \citenamefont {Van~Voorhis}}]{kowalczyk2013excitation}%
  \BibitemOpen
  \bibfield  {author} {\bibinfo {author} {\bibfnamefont {T.}~\bibnamefont {Kowalczyk}}, \bibinfo {author} {\bibfnamefont {T.}~\bibnamefont {Tsuchimochi}}, \bibinfo {author} {\bibfnamefont {P.-T.}\ \bibnamefont {Chen}}, \bibinfo {author} {\bibfnamefont {L.}~\bibnamefont {Top}},\ and\ \bibinfo {author} {\bibfnamefont {T.}~\bibnamefont {Van~Voorhis}},\ }\bibfield  {title} {\enquote {\bibinfo {title} {Excitation energies and stokes shifts from a restricted open-shell kohn-sham approach},}\ }\href@noop {} {\bibfield  {journal} {\bibinfo  {journal} {The Journal of chemical physics}\ }\textbf {\bibinfo {volume} {138}} (\bibinfo {year} {2013})}\BibitemShut {NoStop}%
\bibitem [{\citenamefont {Hait}\ and\ \citenamefont {Head-Gordon}(2020)}]{hait2020highly}%
  \BibitemOpen
  \bibfield  {author} {\bibinfo {author} {\bibfnamefont {D.}~\bibnamefont {Hait}}\ and\ \bibinfo {author} {\bibfnamefont {M.}~\bibnamefont {Head-Gordon}},\ }\bibfield  {title} {\enquote {\bibinfo {title} {Highly accurate prediction of core spectra of molecules at density functional theory cost: Attaining sub-electronvolt error from a restricted open-shell kohn--sham approach},}\ }\href@noop {} {\bibfield  {journal} {\bibinfo  {journal} {J. Phys. Chem. Lett.}\ }\textbf {\bibinfo {volume} {11}},\ \bibinfo {pages} {775--786} (\bibinfo {year} {2020})}\BibitemShut {NoStop}%
\bibitem [{\citenamefont {Furness}\ \emph {et~al.}(2020)\citenamefont {Furness}, \citenamefont {Kaplan}, \citenamefont {Ning}, \citenamefont {Perdew},\ and\ \citenamefont {Sun}}]{furness2020accurate}%
  \BibitemOpen
  \bibfield  {author} {\bibinfo {author} {\bibfnamefont {J.~W.}\ \bibnamefont {Furness}}, \bibinfo {author} {\bibfnamefont {A.~D.}\ \bibnamefont {Kaplan}}, \bibinfo {author} {\bibfnamefont {J.}~\bibnamefont {Ning}}, \bibinfo {author} {\bibfnamefont {J.~P.}\ \bibnamefont {Perdew}},\ and\ \bibinfo {author} {\bibfnamefont {J.}~\bibnamefont {Sun}},\ }\bibfield  {title} {\enquote {\bibinfo {title} {Accurate and numerically efficient r2scan meta-generalized gradient approximation},}\ }\href@noop {} {\bibfield  {journal} {\bibinfo  {journal} {The journal of physical chemistry letters}\ }\textbf {\bibinfo {volume} {11}},\ \bibinfo {pages} {8208--8215} (\bibinfo {year} {2020})}\BibitemShut {NoStop}%
\end{thebibliography}%

\end{document}


\title{Supplementary Information For: Measurement of Coherent Vibrational Dynamics with X-Ray Transient Absorption Spectroscopy Simultaneously at the Carbon K- and Chlorine L$_{2,3}$- Edges}
\author{Andrew D. Ross}
\altaffiliation{These authors contributed equally to this work.}
\affiliation{%
 Department of Chemistry, University of California, Berkeley, CA, 94720, USA
}
\affiliation{%
 Chemical Sciences Division, Lawrence Berkeley National Laboratory, Berkeley, CA, 94720, USA
}
\affiliation{%
 Current Address: Toptica Photonics, Inc., Pittsford, NY 14534, USA
}

\author{Diptarka Hait}%
\altaffiliation{These authors contributed equally to this work.}
\affiliation{%
 Department of Chemistry, University of California, Berkeley, CA, 94720, USA
}
\affiliation{%
 Chemical Sciences Division, Lawrence Berkeley National Laboratory, Berkeley, CA, 94720, USA
}
\affiliation{%
 Current Address: Department of Chemistry and PULSE Institute, Stanford University, Stanford, CA 94305, USA
}
\author{Valeriu Scutelnic}%
\affiliation{%
 Department of Chemistry, University of California, Berkeley, CA, 94720, USA
}
\affiliation{%
 Chemical Sciences Division, Lawrence Berkeley National Laboratory, Berkeley, CA, 94720, USA
}

\author{Daniel M. Neumark}%
\affiliation{%
 Department of Chemistry, University of California, Berkeley, CA, 94720, USA
}
\affiliation{%
 Chemical Sciences Division, Lawrence Berkeley National Laboratory, Berkeley, CA, 94720, USA
}
\author{Martin Head-Gordon}%
\affiliation{%
 Department of Chemistry, University of California, Berkeley, CA, 94720, USA
}
\affiliation{%
 Chemical Sciences Division, Lawrence Berkeley National Laboratory, Berkeley, CA, 94720, USA
}

\author{Stephen R. Leone}%
\email{srl@berkeley.edu}
\affiliation{%
 Department of Chemistry, 
 University of California, Berkeley, CA, 94720, USA
}
\affiliation{%
 Chemical Sciences Division, 
 Lawrence Berkeley National Laboratory, Berkeley, CA, 94720, USA
}
\affiliation{%
 Department of Physics, 
 University of California, Berkeley, CA, 94720, USA
}%
\date{\today}

\maketitle

\tableofcontents
\newpage 

\section{Code Development for Analysis of Experimental Data}

\begin{figure}[h] \centering
\includegraphics[max height=\textheight/2,max width=\textwidth,keepaspectratio]{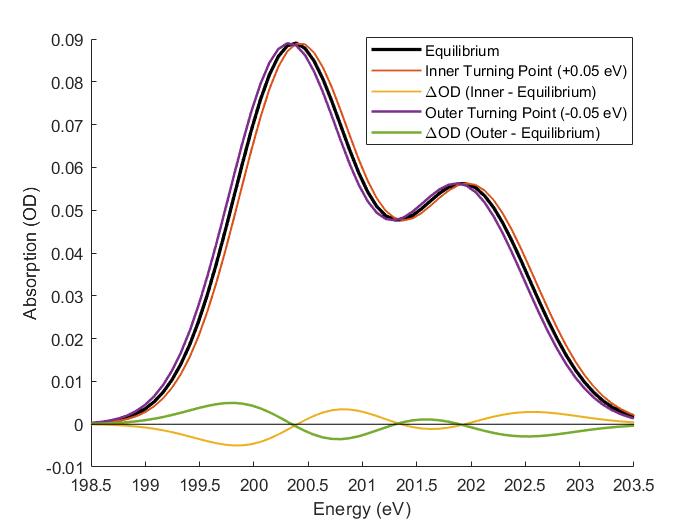}
\caption{\label{fig:Example vib dOD}
An example of the types of data that result from vibration is shown for the Cl 2p$\to 7a_1^*$ of \ce{CCl4}. $\Delta$OD is calculated by taking the difference of the OD at the inner (outer) turning point and the OD of the equilibrium.
The shift of $\pm$0.05 eV corresponds to roughly 0.005 {\AA} of movement along the symmetric stretch. 
Notice that the $\Delta$OD signals are strongest where the slope of the static absorption is the largest, and the $\Delta$OD is smallest where the peaks are centered.
}
\end{figure}

Due to the nature of measuring vibrations, the transient absorption data has correlations between energy that provide overlapping information, which can lead to much stronger signals if properly accounted for. For example, the vibration leads to a net shift of the energy of a state absorption, which leads to a positive change in optical density ($\Delta$OD) on one side of the energies and a negative $\Delta$OD on the other, an example of which is shown in Fig. \ref{fig:Example vib dOD}. Simply adding these two signals together or taking a line out along one of the energies ignores large pieces of this information. However, there is no existing analysis code that can take advantage of the additional information, so code must be developed, according to a model, expressly for this purpose. Applying a model of a shifted static spectrum can allow extraction of the exact energy changes in a more consistent manner. Another consideration that must be taken into account is that at the intensities used in these experiments to excite the vibrations, the molecular samples can also be ionized, and any model used will also need to take these new absorptions from the ionized molecule and its derivatives into account.

\subsection{Noise Reduction by Edge Referencing and Fourier Filtering}

A type of noise that is omnipresent in X-ray transient absorption spectroscopy (XTAS) is high harmonic generation (HHG) energy fluctuations, where the phase-matching conditions are changed and the peaks of HHG flux at the odd-harmonics are shifted\cite{Geneaux2021XUVNoise,Geneaux2019ATASReview}. This is particularly a problem for short fundamental pulses where the phase-matching is not strictly enforced by many X-ray bursts and changes in the carrier envelope pulse (CEP) produce significant changes to the HHG. These fluctuations introduce noise at repeated 2$\omega$ of the fundamental spacing in the $\Delta$OD data.

\begin{figure}[h] \centering
\includegraphics[max height=\textheight/2,max width=\textwidth,keepaspectratio]{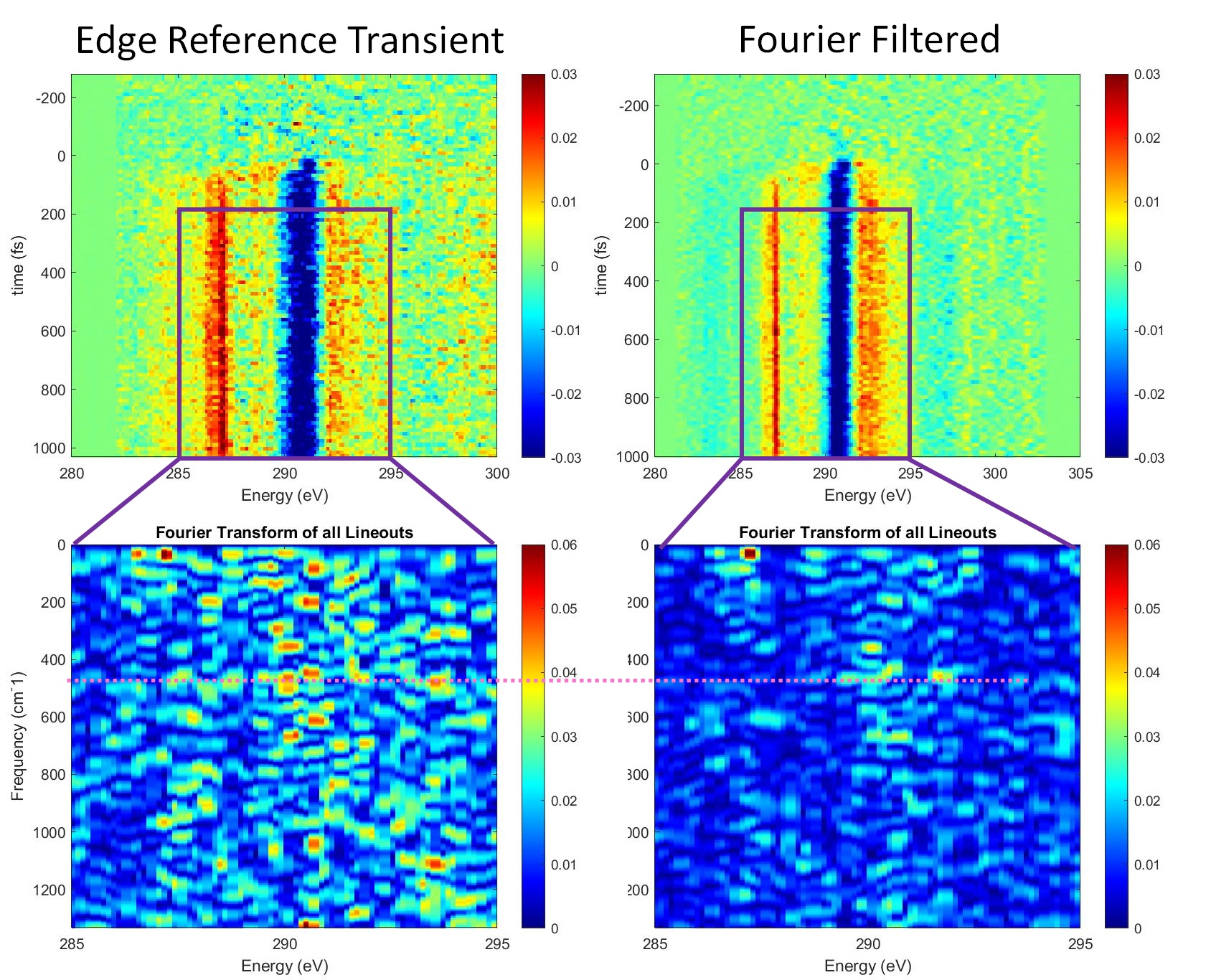}
\caption{\label{fig:FFTFilt CCl4 CK}
The results of noise filtering by edge referencing and Fourier filtering are compared for the C K-edge region. The symmetric vibration frequency is clear only for the Fourier filtered spectrum. Note again that the Fourier filtering applies only on the spectral axis; improvements in the Fourier transform of the temporal axis are only due to reduction of noise.}
\end{figure}

\begin{figure}[h] \centering
\includegraphics[max height=0.5\textheight,max width=\textwidth,keepaspectratio]{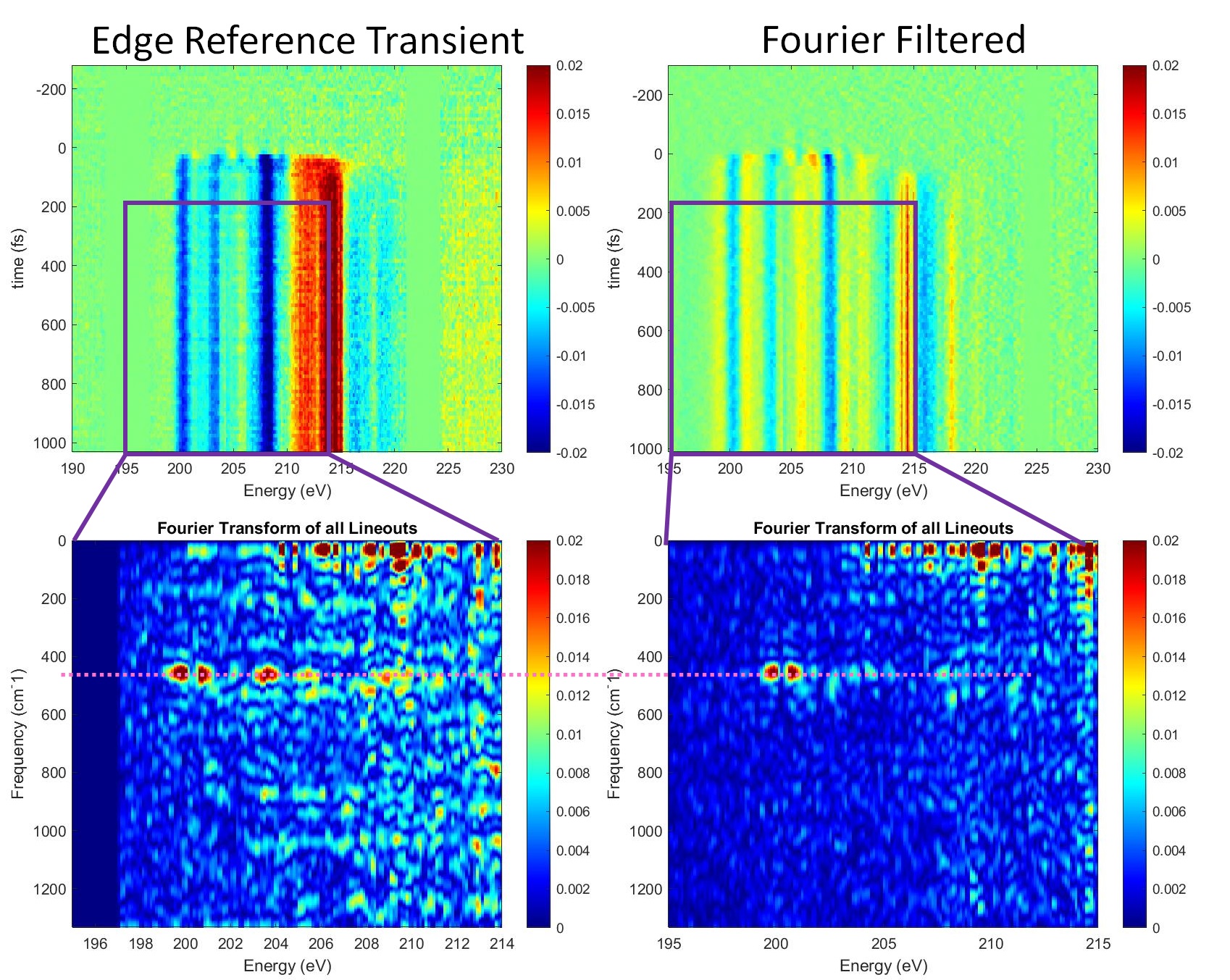}
\caption{\label{fig:Fourier transform FFT Filt CCl4 ClL}
The results of noise filtering by edge referencing and Fourier filtering are compared for the Cl L$_{2,3}$-edge region. The spectral features have the low frequency component removed by the Fourier filtering, which causes it to have a drastically different shape, although the frequency information in the time domain is generally preserved. Note that the frequency peak at 203 eV is smaller, due to the greater low frequency components in that region, compared to 200 eV.}
\end{figure}

This noise source can be corrected out, though. By choosing a region of energies where a $\Delta$OD of 0 is expected, the correlation matrix of that region can be calculated and applied to the rest of the energies, through a previously developed ``edge referencing" algorithm\cite{Geneaux2021XUVNoise}. This severely reduces the noise associated with HHG intensity and energy fluctuations, and it is essentially necessary for working in the soft X-ray regime. Those noise sources are not the only noise in the experiment; there is always random shot noise from the camera, for example, which cannot be removed and will be more significant with the lower photon counts. The data is filtered and large amounts of correlated noise are removed using the edge referencing algorithm discussed previously\cite{Geneaux2021XUVNoise}. 

However for vibration data, where each time step must have minimal noise in order not to introduce new frequencies, elimination of noise from changes in X-ray harmonic intensity is paramount. In the normal algorithm for $\Delta$OD, a spectrum with a pump on is compared to a pump off, \(\Delta{}OD = -\log_{10}(I_{on}/I_{off})\), so camera integrations of X-ray pulses where there is a net change in the X-ray flux between on and off can result in net changes of positive or negative $\Delta$OD, which adds noise between each time point, affecting Fourier space. Edge referencing removes a lot of this noise, but at higher energies these intensity fluctuations are much stronger and require additional filtering.

A way to achieve more complete removal of X-ray harmonic intensity fluctuation is to compare 2 different X-ray shots, such that only the pump on spectra are used for calculating the $\Delta$OD. This can be achieved by using  a previously developed Fourier filtering technique\cite{ottNature2014}. At each delay point $\tau$, the absorbance $OD$ is calculated as $OD(\tau) = -\log_{10} \frac{I(\tau)}{I_0} - -\log_{10} \frac{I_{low-frequency}(\tau)}{I_0}$, where the low frequency spectrum, $I_{low-frequency}$ is reconstructed by using a Fourier low-pass filter along the energy axis directly on the measured pump on signal $I(\tau)$, and the pump off X-ray data are not used. The reference spectrum, $I_0$ cancels out, so it is largely irrelevant, but it must be chosen so that it does not contain negative absorptions, as this will affect the filtering in the Fourier domain. Typically, $I_0$ is just an experimental spectrum with no gas absorption, but a string of large numbers works just as well. This algorithm acts separately on each time point and does not directly influence the separate time points. This procedure allows for removal of the intensity fluctuations of the X-ray spectra\cite{ottNature2014}. It can be modified easily to recreate the standard $\Delta$OD by simply subtracting the signal obtained at a time delay with known zero signal, $\Delta{}OD(\tau) = OD(\tau) - OD(-200 fs)$. This can also be combined with edge referencing to remove noise associated with harmonic modulation, the 2$\omega$ peaks of the fundamental laser for HHG that shift in energy with changing phase matching conditions.

Edge referencing is well-suited in the case of narrow absorption lines. In regions where the static X-ray absorption has few low-frequency components (those consisting of stand alone Voigt-like peaks) the Fourier filtered $\Delta$OD data closely resemble that of the standard algorithm, \(\Delta{}OD = -\log{}_{10}(I_{on}/I_{off})\). A comparison of the 2 noise reduction algorithms is shown for the C K-edge \ce{CCl4} in Fig. \ref{fig:FFTFilt CCl4 CK}. In regions where the static spectrum consists of low-frequency components, the Fourier filtered data appears very different from the standard algorithm.  For all cases, the Fourier transforms of the data along the time axis show reduced noise signal, although the actual signal is also reduced in the regions with low-frequency static components, shown for the Cl L$_{2,3}$-edges of \ce{CCl4} in Fig. \ref{fig:Fourier transform FFT Filt CCl4 ClL}.

\subsection{Physical parameter extraction by large multivariate fitting}

\begin{figure}[h] \centering
\includegraphics[max size={\linewidth}{\textheight/2}]{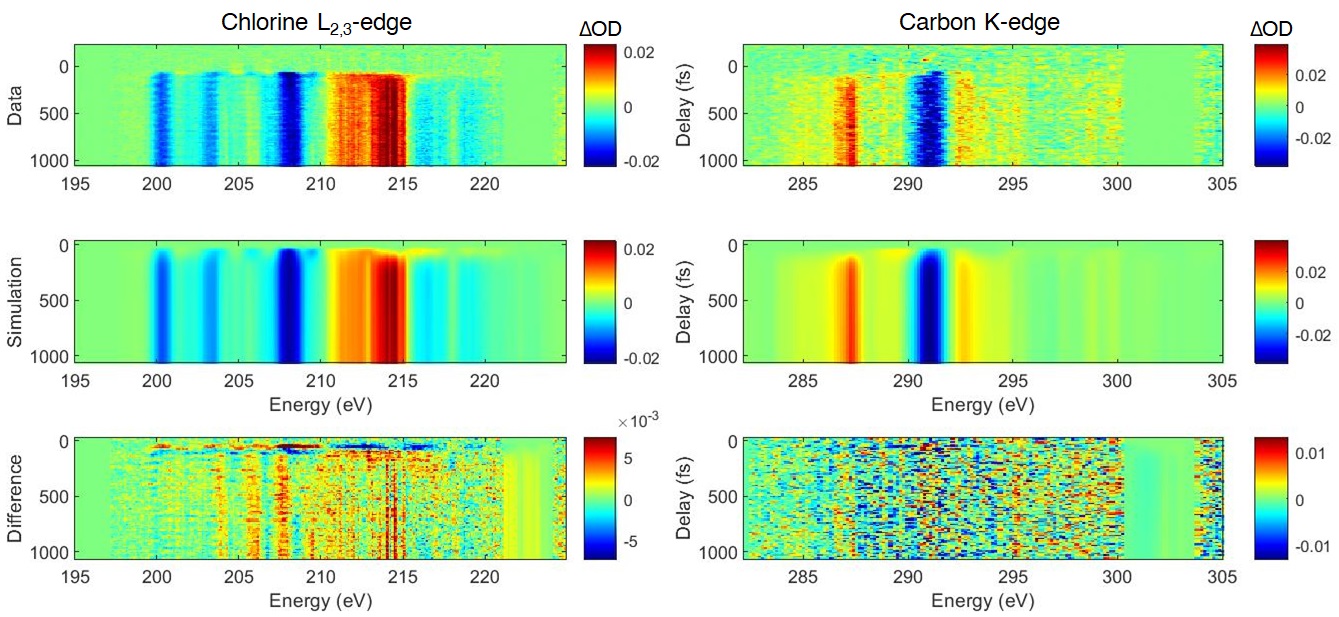}
\caption{\label{fig:exampleMultivariateFitting} 
An example of one dataset fitted by the multivariate fitting algorithm. Both the chlorine and carbon edges are shown in the two separate columns. The top row shows the data. The middle row shows the result of the fitting. The bottom row shows the difference between the data and the fitting (the top and middle rows). Note that the color scale on the bottom row is zoomed by a factor of 2 to highlight errors. This shows only one dataset. For this fitting, 9 datasets are fitted simultaneously. The others are not shown for space considerations.
}
\end{figure}

The usual method for extracting temporal information from XTAS data is to take a lineout at a particular energy along the time axis\cite{bhattacherjee2018ultrafast,Geneaux2019ATASReview,sidiropoulos2021probing,liao2017probing}. This is useful in that it does not require a correct model of the temporal evolution. However, in order to make use of all of the information provided, the data must be able to be fully recreated according to a model that represents the physical parameters of the experiment. This is attempted by making a multivariate fit, and this particular implementation does this fitting for multiple datasets at once. An example of a finished fit for one dataset is shown in Fig. \ref{fig:exampleMultivariateFitting}.

After the data for each dataset are prepared, it is saved in files containing the $\Delta$OD, the static absorption of that measurement, delay axis, and pixel-to-energy calibration axis. A dataset is the set of data taken continuously with no changes to the experiment, usually datasets represent data from different days, but they may also be taken on single days with multiple conditions. Slight differences in energy calibration and sample pressure are determined by comparing the static absorption of each dataset to a master static spectrum. Changes are made to internal parameters according to energy calibration and pressure to make each dataset comparable to each other. These are all loaded into a single program to fit all of them to a model.

The model is concerned with 2 different parts: 1: non-oscillatory changes to different electronic states, including ionization, excited states, or extreme nuclear changes, and 2: oscillatory changes, in this case due to vibration. A complete model of both are necessary for extracting information about the core-excited potential energy surface (CEPES).

The first part of the model for the non-oscillatory changes concerns signals associated with ionization and dissociation of \ce{CCl4+}, and is fully explained in the SI of Ref \citenum{ross2022jahn}. Briefly, it assumes each of the states of dissociation follow an exponential population transfer with unique spectra. These lead to large signals in the $\Delta$OD. 

The second part of the model is concerned with the oscillatory features due to the vibrations. The main assumption of this part of the model is that described in the introduction that the energy of absorption is equal to the Potential Energy Surface of a specific core-excited state minus that of the ground state:
\[E_{photon}(q) = PES_{core-excited}(q) - PES_{ground}(q)\]
where q is the movement along the vibrational mode. In the model, E$_{photon}$(q) is represented as a linear polynomial. Another assumption was that the vibration occurs in a part of the potential well that is approximately harmonic, i.e. that the vibration q position can be denoted by a sine function \(q(t) = A\times{}\sin(\omega{}(t - c) - \phi{})\), where A is the amplitude of vibration, proportional to the input power, $\omega$ is the frequency of vibration, c allows for slightly shifted time zeros in each of the individual datasets, and $\phi$ is the phase of the vibration, relative to time zero. The difference between the c and $\phi$ parameters is that c affects the rise of excited and ion signals and is allowed to change for each dataset, while $\phi$ is held constant across all datasets. The phase parameter is allowed to change as a fit, and it was found to be close to $\pi$/3, although this is may be because the program defines time zero as the start of the ground state bleach (when the leading edge of the pump pulse arrives). Time zero is better defined as when the pump pulse is at maximum intensity. The evolution of the absorption with this assumption is simply that of the Voigt function of static absorption changes in energy without changing its shape:
\[E_{photon}(t) = E_{polynomial}(q)(A\times{}\sin(\omega{}(t - c) - \phi{})\]
To note, while this assumption is a simplification, a more chemically accurate model was created and used that defined the vibration q position as a superposition of Morse vibrational levels, which could lead to different absorption shapes, based on the vibrational wavepacket. However, the result of this fitting was that the superposition and resultant wavepacket approximated the simplified assumption result, but required much more computational power. The simplified model provided exceptionally accurate fits, so the chemically accurate model was abandoned. One last assumption was made, that all the molecules in the electronic ground state experience the same electric field and vibrational dynamics. This does not affect the relative slopes of the core-excited states within a measurement and such an assumption is generally made for experiments like this\cite{korn1998observation}.

The $\Delta$OD as a function of time is then calculated based on this model and is broadened by convolution with a Gaussian to account for the temporal resolution of our experiment. 

The model is subtracted from the data and fed into a minimization of least squares fitting algorithm, lsqnonlin in Matlab. The parameters that are needed and common to all datasets are:
\begin{enumerate}
  \item OD of each new state
  \item Frequency and phase of the neutral vibration
  \item Core-Excited state slopes for the vibration of each transition
  \item The lifetime constants, $\tau_{in}$, $\tau_{out}$, and \(d\), for each of the new states
\end{enumerate}
The parameters that are unique to each individual dataset are:
\begin{enumerate}
  \item Timing delay offset, \(c\)
  \item Population parameters, \(a\), for each state in each dataset
  \item Vibrational amplitude of the neutral
  \item Temporal broadening amounts
\end{enumerate}

\begin{figure}[htb!] \centering
\includegraphics[max height=\textheight/2,max width=\textwidth]{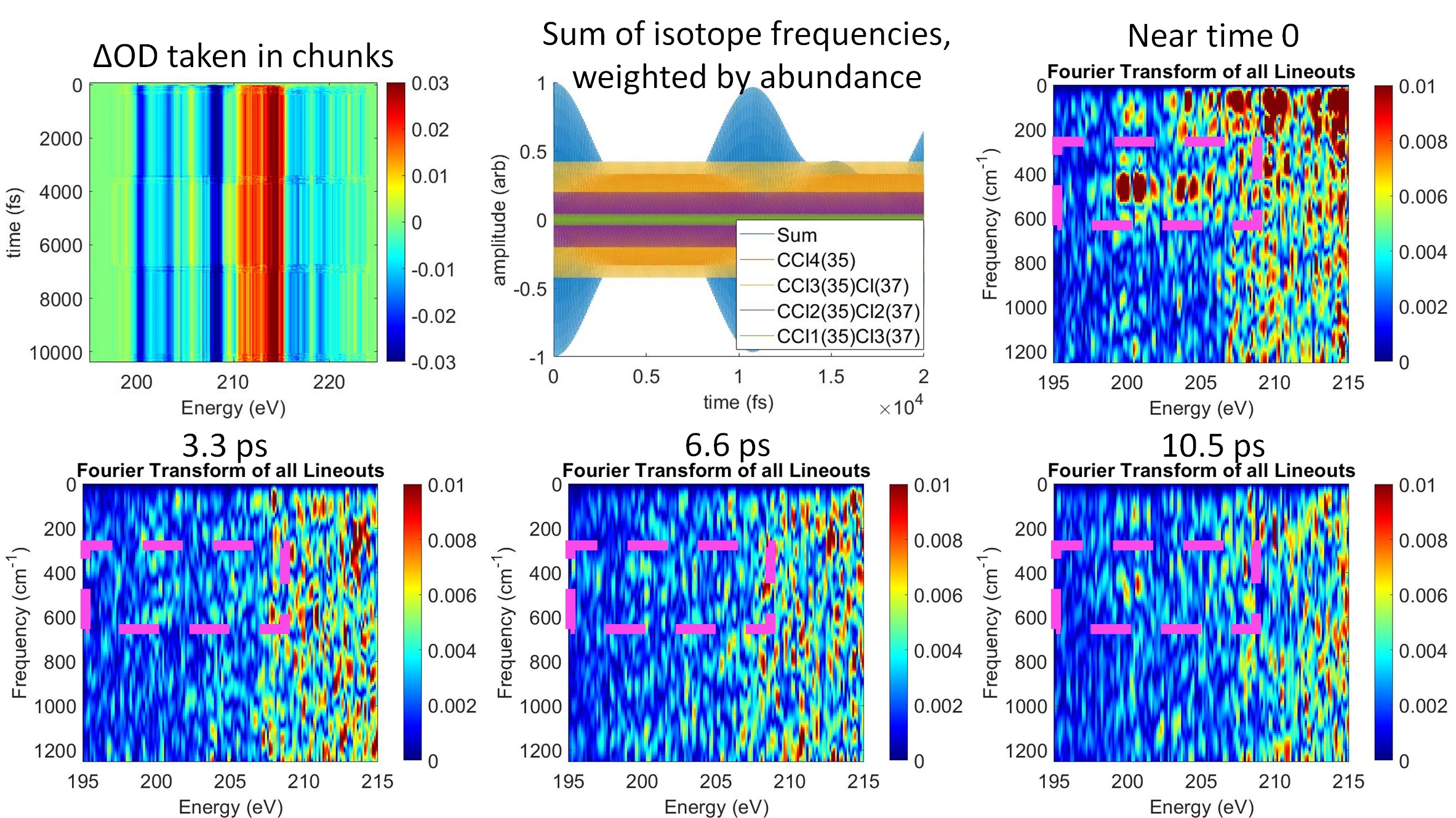}
\caption{\label{fig:CCl4 isotope 10 ps measurement}
The data of a scan that covers 10 ps is shown in the upper left. The scan is broken into 4 equal chunks with no time points in between. The middle shows cosine waves of the frequencies of the $\nu_1$ mode of the isotope combinations with amplitude proportional to their natural abundances; the case of 4 Cl$^{37}$ is not shown due to low abundance. The sum of all other isotope combinations is the blue curve, which is the constructive interference between them and the expected coherence of the vibration signal. The rest are Fourier transforms of the 4 chunks, which show a strong signal at time zero, when all isotopic combinations are in phase. This is followed by no signal at 3.3 and 6.6 ps, where the isotopic combinations are out of phase. Finally, at 10.5 ps, the signal reappears with Fourier signal 40$\%$ of the time zero signal.}
\end{figure}

\subsection{Isotope Effect}

Because \ce{CCl4} contains chlorine, there are multiple isotopes that affect the frequency of vibration. The isotope effect has been measured previously\cite{CHAKRABORTY2002RamanCCl4,gaynor2015vibrational}, and shows the $\nu_1$ mode consists of vibrations at 453.7, 456.6, 459.5, and 462.7 cm$^{-1}$ for molecules containing 0, 1, 2, and 3 Cl$^{37}$ atoms, compared to the more abundant Cl$^{35}$. With these values and the known abundances of Cl$^{35}$ and Cl$^{37}$, the amount of constructive and destructive interference is predicted, which is shown in Fig. \ref{fig:CCl4 isotope 10 ps measurement}. Based on this, the X-ray $\Delta$OD signal should become less clear as the time approaches 3 ps. This destructive interference is expected to cause broadening of the static signal, as it corresponds to one population being red shifted and another blue shifted. The populations are expected to come back in phase at 10.5 ps, so an experiment was performed that scanned 37 time points near time zero, 3.3 ps, 6.6 ps, and 10.5 ps, the results of which are shown in Fig. \ref{fig:CCl4 isotope 10 ps measurement}. Indeed, a rephasing of the vibrational signal is apparent at 10 ps, although the expected rephasing of the Fourier signal is 95$\%$ of the original, and the measured is $\sim$40$\%$. This difference may be due to redistribution of vibrational energy, dephasing of the vibration from other sources, or spatial drift between pump and probe over the time difference.

\section{Ehrenfest's Theorem for Pump-Induced Normal Mode Displacements}
Ehrenfest's theorem\cite{ehrenfest1927bemerkung,shankar2012principles} relates the dynamics for the expectation values of the positions $\{\hat{q}_i\}$ and momenta $\{\hat{p}_i\}$ of particles experiencing a time-dependent potential $ V\left(\{\hat{q}_i\},t\right)$ through:
\begin{align}
    \dfrac{d}{dt} \langle \hat{q}_i \rangle &= \dfrac{\langle \hat{p}_i\rangle}{m_i}\\
    \dfrac{d}{dt} \langle \hat{p}_i \rangle &= -\left\langle\dfrac{ \partial V\left(\{\hat{q}_i\},t\right)}{\partial \hat{q}_i}\right\rangle
\end{align}
where $\{m_i\}$ are the masses. Within the harmonic approximation and assuming that the pump pulse field does not couple normal modes together, we have $V(q_i,t)=\dfrac{1}{2}m_i \omega^2 q_i^2- \dfrac{1}{2}\alpha_{iso}(q_i)\left|\mathcal{E}(t)\right|^2$ for a given mode $i$ after averaging over all possible orientations of the field relative to the molecule ($m_i$ being the reduced mass of the mode). If we furthermore approximate $\alpha_{iso}(q_i)$ with a quadratic polynomial $\alpha_0+\alpha_1 q_i + \alpha_2 q_i^2$, we obtain $\dfrac{ \partial V({q}_i,t)}{\partial {q}_i}= m_i\omega_i^2 q_i -\dfrac{1}{2}\left|\mathcal{E}(t)\right|^2\left(\alpha_1+2\alpha_2 q_i\right)$.

The differential equations then simplify to:
\begin{align}
    m_i \dfrac{d^2}{dt^2} \langle \hat{q}_i \rangle &= \dfrac{d}{dt}\langle \hat{p}_i \rangle= -m_i\omega_i^2 \langle \hat{q}_i\rangle +\dfrac{1}{2}\left|\mathcal{E}(t)\right|^2\left(\alpha_1+2\alpha_2 \langle \hat{q}_i\rangle\right)\\
    \implies \dfrac{d^2}{dt^2} \langle \hat{q}_i \rangle + \omega_i^2 \langle \hat{q}_i\rangle &= \dfrac{1}{2m_i}\left|\mathcal{E}(t)\right|^2\left(\alpha_1+2\alpha_2 \langle \hat{q}_i\rangle\right) \label{eq:ehren}
\end{align}
The solution to this second order differential equation in $ \langle \hat{q}_i \rangle$ is completely specified by the parameters and the initial conditions. These initial conditions are trivially obtained by noting that long before the pump pulse (i.e. $t\to-\infty$), we are in the ground state of the corresponding harmonic oscillator and thus have $\langle \hat{q}_i \rangle= \langle \hat{p}_i \rangle=0$. In particular, $\alpha_1=0$ ensures that $\langle \hat{q}_i\rangle (t) = 0$ is the only solution--i.e. the center of the wavepacket is undisplaced by the field. As discussed in the main text and shown in Fig. \ref{fig:polarfit}, $\alpha_1=0$ for the asymmetric modes, leading to no displacement of the center of the vibrational wavefunction along such modes. We note that this discussion assumes a quadratic approximation for the polynomial to be adequate, but cubic or higher odd order terms in the Taylor series expression for $\alpha_{iso}$ could in principle induce some small displacement along asymmetric modes. Fig. \ref{fig:polarfit} however indicates the effective absence of such terms for the asymmetric modes of \ce{CCl4}. 
We note that Eqn. \ref{eq:ehren} can be numerically solved with general molecular dynamics techniques, such as with a velocity Verlet integrator\cite{swope1982computer}. An analytical solution for a Gaussian $\mathcal{E}(t)$ and $\alpha_2=0$ is provided in Ref \citenum{yan1985impulsive}.

\section{Fits of Isotropic Polarizability Against Normal Mode Displacement}
\begin{figure}[htb!]
\begin{minipage}{0.48\textwidth}
    \centering
    \includegraphics[width=\columnwidth]{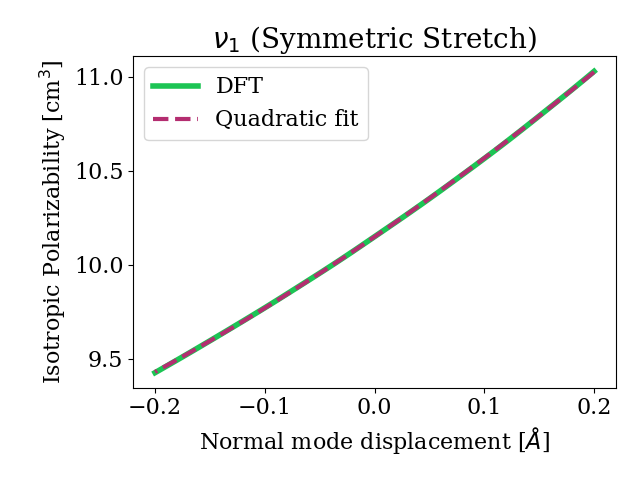}
\end{minipage}
\begin{minipage}{0.48\textwidth}
    \centering
    \includegraphics[width=\columnwidth]{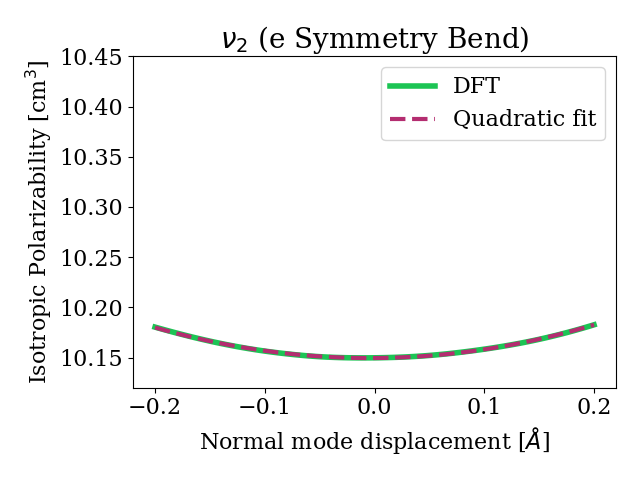}
\end{minipage}
\begin{minipage}{0.48\textwidth}
    \centering
    \includegraphics[width=\columnwidth]{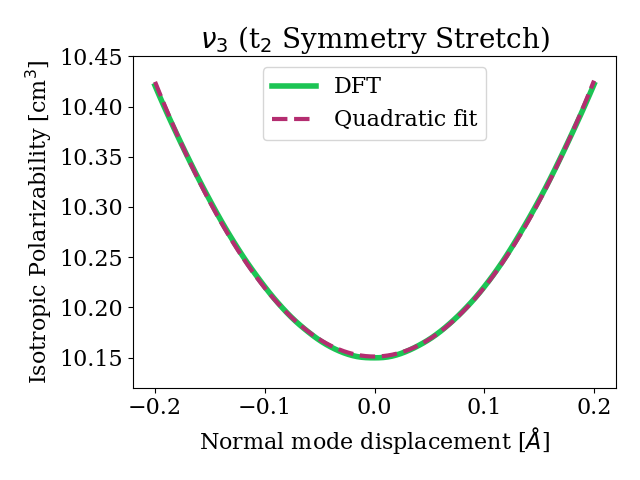}
\end{minipage}
\begin{minipage}{0.48\textwidth}
    \centering
    \includegraphics[width=\columnwidth]{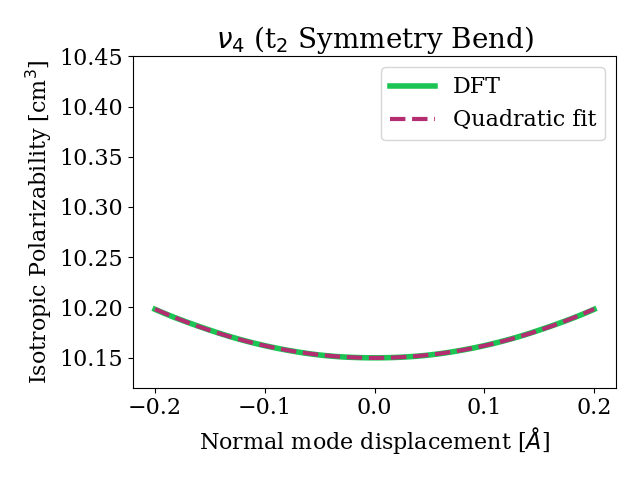}
\end{minipage}
\caption{Computed isotropic polarizabilities ($\alpha_{iso}$) vs normal mode displacement, along with quadratic least-square fits for comparison. The computed normal mode with the lowest frequency in each degenerate set (see discussion in main text) was selected for depiction, the other modes in the set yield essentially equivalent behavior. Note that the plot for the symmetric stretch (top left) has a different y axis range than the other three, as displacement along this mode leads to significantly greater change in $\alpha_{iso}$. We note that the normal mode coordinate for the symmetric stretch is $2\times$bond elongation (i.e., symmetric stretch of all C-Cl bonds by 0.1 {\AA} will change the corresponding normal mode coordinate by $0.2$ {\AA}).}
\label{fig:polarfit}
\end{figure}

The variation in molecular polarizability with normal mode displacements was analytically computed at the SCAN0/aug-pcseg-2 level of theory.
The computed isotropic polarizability profiles are plotted in Fig. \ref{fig:polarfit}, along with quadratic fits for comparison. It is clear that the quadratic fits are quite adequate for fitting over the chosen range of displacements (-0.2 to 0.2 {\AA}) and $\alpha_1=\left(\dfrac{d \alpha_{iso}}{d q_i}\right)_{q_i=0}$ vanishes for the asymmetric modes, as discussed in the main text. In addition, the polarizability profiles for the asymmetric modes appear to resemble even functions over the plotted range of normal mode displacements, and thus there are no substantial higher odd order terms that can produce a force on the wavepacket center within the Ehrenfest picture for field-induced dynamics discussed earlier. Furthermore, Fig. \ref{fig:polarfit} clearly shows that the isotropic polarizability has the greatest variation along the symmetric stretchic mode and does not change as much over displacements along other modes, on the scale of spatial displacements considered (which are roughly an order of magnitude larger than the amplitude for the symmetric stretch oscillations, as shown in Fig. \ref{fig:exptq}).

\section{Wavepacket Simulation of Raman Process}
We carried out a time-dependent Schr{\"o}dinger equation simulation of the pump-molecule interaction to model the vibrational dynamics and estimate oscillation amplitudes from first principles. All normal modes were assumed to be independent of each other, and subject to the time-dependent potentials:
\begin{align}
    V(q_i,t)&= \dfrac{1}{2}m_i \omega_i^2 q_i^2-\vec{\mu}(q_i)\cdot\vec{\mathcal{E}}(t) -\dfrac{1}{2}\vec{\mathcal{E}}(t)\cdot\bm{\alpha}(q_i)\cdot\vec{\mathcal{E}}(t) 
\end{align}
where $\vec{\mu}(q_i)$ and $\bm{\alpha}(q_i)$ are quadratic fits to the dipole moment vector and the full polarizability tensor (computed analytically with SCAN0/aug-pcseg-2) vs normal mode displacements, respectively. Ansiotropic effects are therefore considered in the simulation. The dipole $\vec{\mu}(q_i)$ is zero at the equilibrium geometry, but displacement along asymmetric modes can generate a nonzero dipole. We therefore included this term in our simulations, even though it had no perceptible effect on the simulation results due to the light frequency not being resonant with any of the molecular vibrational degrees of freedom.  The time-dependent electric field is given by: 
\begin{align}
    \vec{\mathcal{E}}(t)&=E_0 e^{-t^2/(4\sigma^2)}\cos\omega_0 t\,  \hat{n}
\end{align}
 where $\sigma$ is the temporal standard deviation of the intensity profile envelope (2.55 fs for a 6 fs FWHM pump pulse), $\omega_0$ is the angular frequency for 800 nm light (16.77 rad/fs) and $E_0=\sqrt{c\varepsilon_0 I}$ is the peak magnitude of the electric field, which depends on the peak pump intensity $I$, speed of light $c$ and permittivity of free space $\varepsilon_0$. For the experimental $I=3\times 10^{14}$ W/cm$^2$, $E_0=3.36$ V/{\AA} (0.065 a.u.), which is the value used in the simulations unless specified otherwise. The unit vector $\hat{n}$ is randomly oriented in space, as the molecule has no preferred direction of orientation. For a given $\hat{n}$, the time-dependent wavefunction $\psi_i(t;\hat{n})$ (starting from the harmonic oscillator ground state at initial $t=-50$ fs) is propagated in time. The time propagation is done in a basis spanning the lowest eight eigenstates of the unperturbed harmonic oscillator corresponding to the mode. The interaction picture\cite{shankar2012principles} is employed to separate the unperturbed harmonic oscillator dynamics (reference) from the field-dependent terms (perturbation). The midpoint method is used to integrate the interaction picture time-dependent Schr{\"o}dinger equation with timesteps of 2 atomic units ($\sim0.048$ fs). The resulting interaction picture wavefunction was subsequently transformed with the exact propagator for the harmonic oscillator reference Hamiltonian to obtain the true $\psi_i(t;\hat{n})$.
 
 The value of this wavefunction is subsequently evaluated on a spatial grid of points $\{x\}$ to obtain a probability density $\left|\psi_i(x,t;\hat{n})\right|^2$. The densities resulting from independent simulations for all the field directions $\hat{n}$ corresponding to a 590 point Lebedev grid\cite{lebedev1992quadrature} are then summed together (with the corresponding quadrature weights) to yield the rotationally averaged behavior. All further properties (such as oscillations in the wavepacket center) are subsequently computed using the final, rotationally averaged time-dependent probability density $\left|\psi_i(x,t)\right|^2$. Further details about the simulations can be obtained from the provided python code in the Zenodo repository\cite{ross_2024_11153002}.

\begin{figure}[htb!]
\begin{minipage}{0.48\textwidth}
    \centering
    \includegraphics[width=\columnwidth]{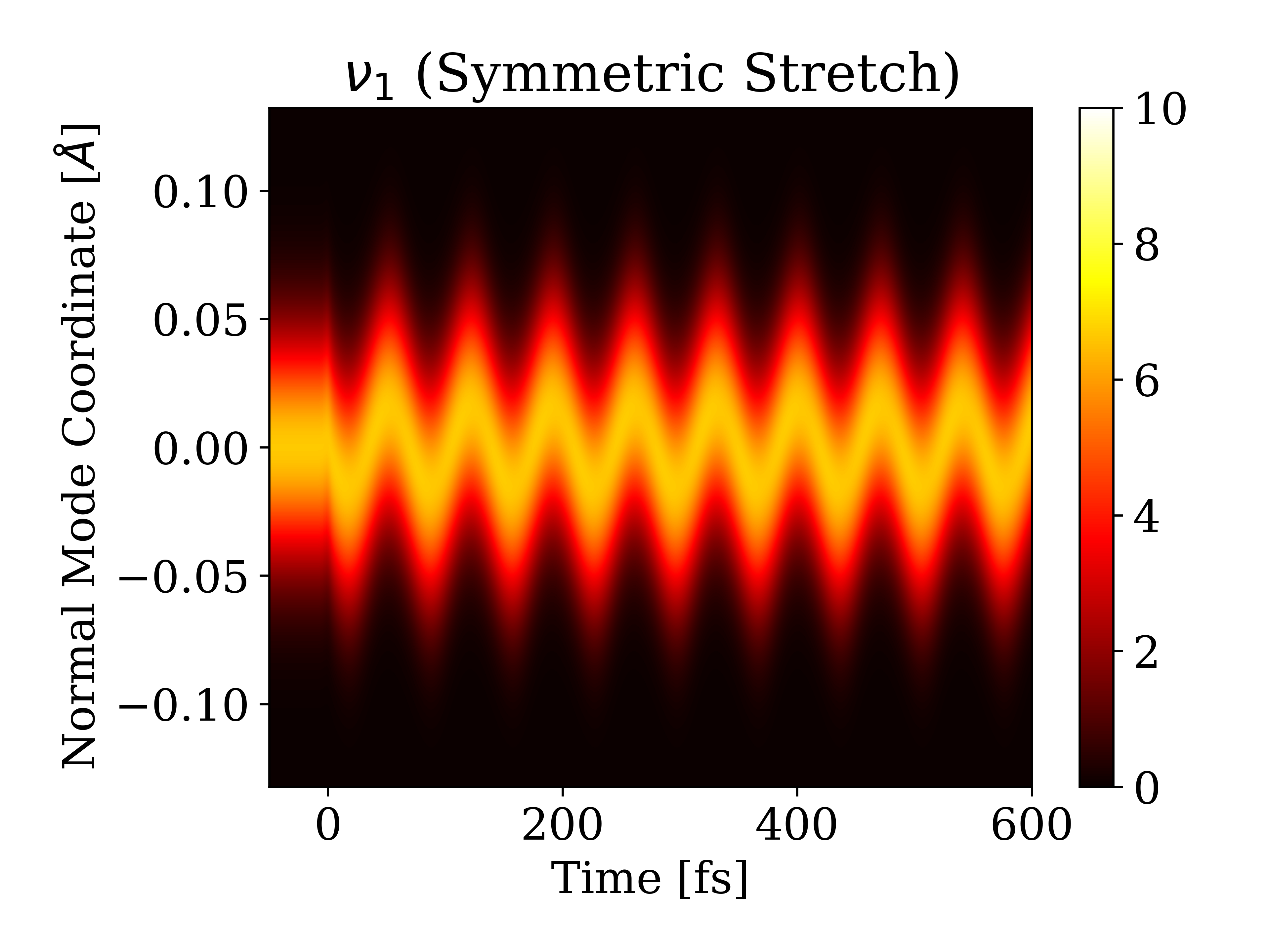}
\end{minipage}
\begin{minipage}{0.48\textwidth}
    \centering
    \includegraphics[width=\columnwidth]{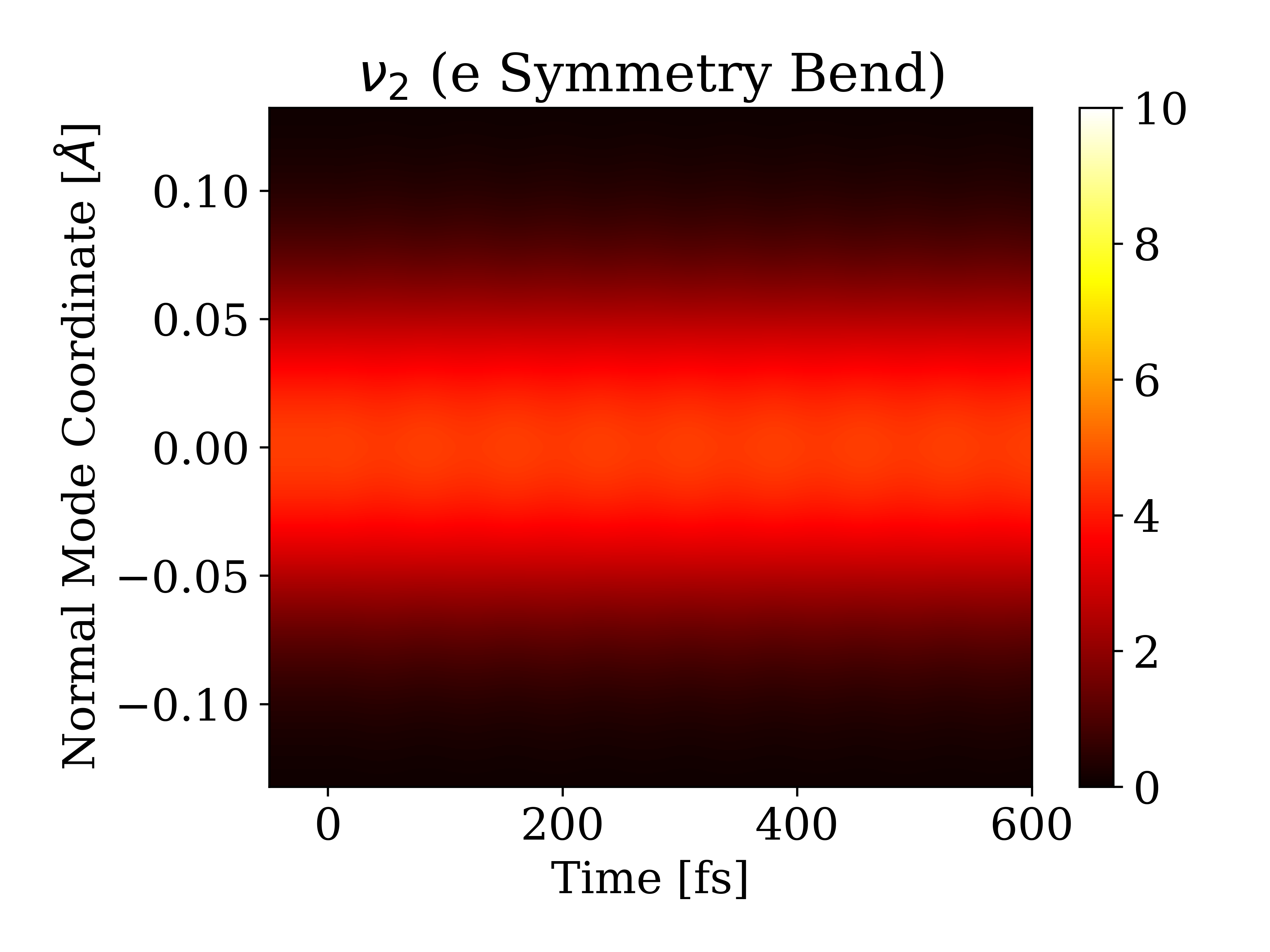}
\end{minipage}
\begin{minipage}{0.48\textwidth}
    \centering
    \includegraphics[width=\columnwidth]{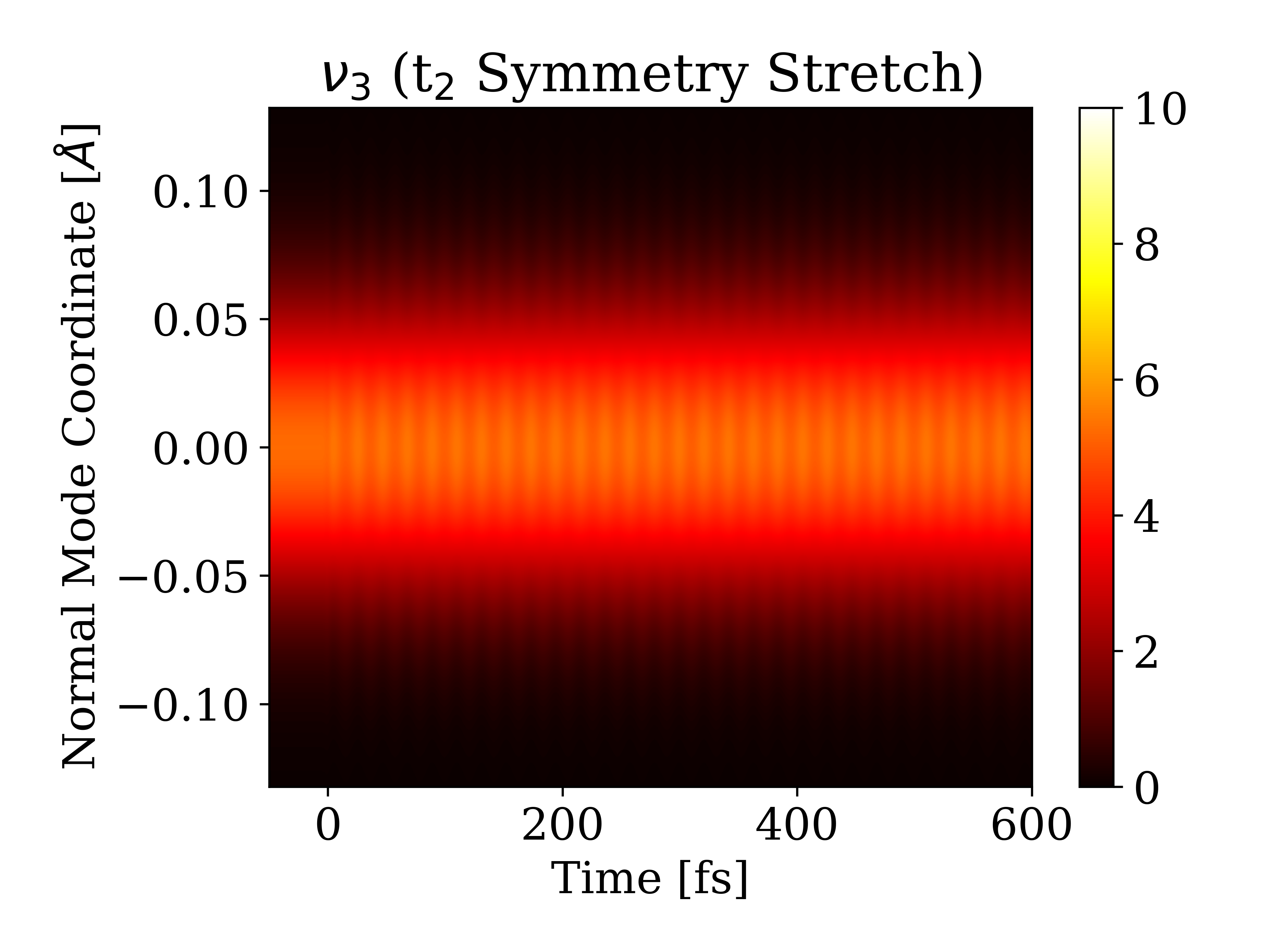}
\end{minipage}
\begin{minipage}{0.48\textwidth}
    \centering
    \includegraphics[width=\columnwidth]{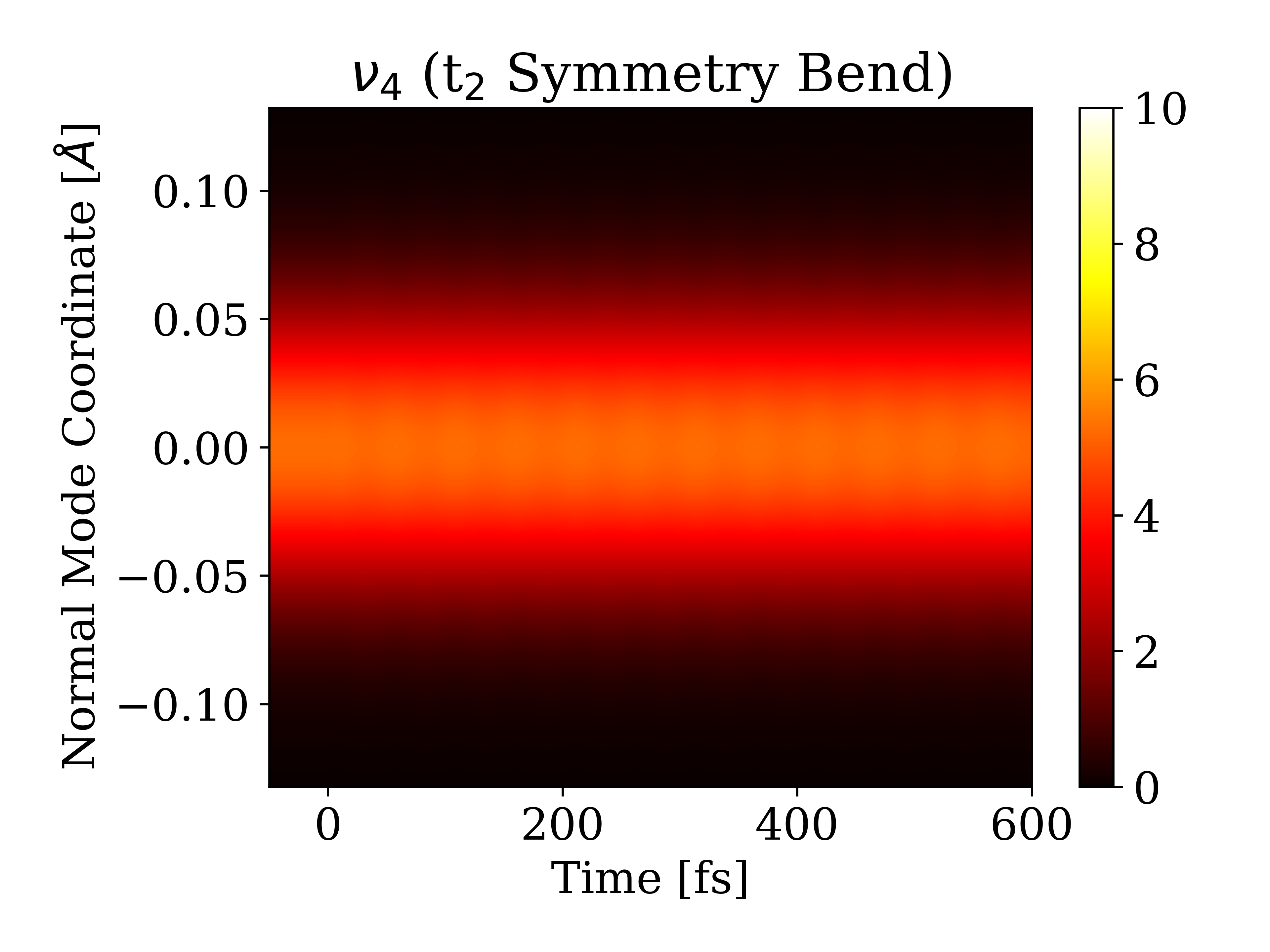}
\end{minipage}
\caption{Simulated $\left|\psi_i(x,t)\right|^2$ along different normal modes from the time-dependent Schr{\"o}dinger equation simulations. The computed normal mode with the lowest frequency in each degenerate set (see discussion in main text) was selected for depiction, the other modes in the set yield essentially identical behavior. Only the symmetric stretch (top left) shows perceptible displacement of the wavepacket center. The asymmetric ($t_2$) stretch shows wavepacket broadening and contraction at twice the normal mode frequency, without \textit{net} displacement to the wavepacket center. This likely arises from the anisotropic terms in the polarizability tensor, which contribute to time-dependent broadening after rotational averaging. The bending modes appear to show little time evolution in these plots, but also undergo change in the wavepacket width like the asymmetric stretch (albeit on a smaller scale), as shown in Fig. \ref{fig:exptwidth}.
}
\label{fig:psisq}
\end{figure}

 \begin{figure}[htb!]
\begin{minipage}{0.48\textwidth}
    \centering
    \includegraphics[width=\columnwidth]{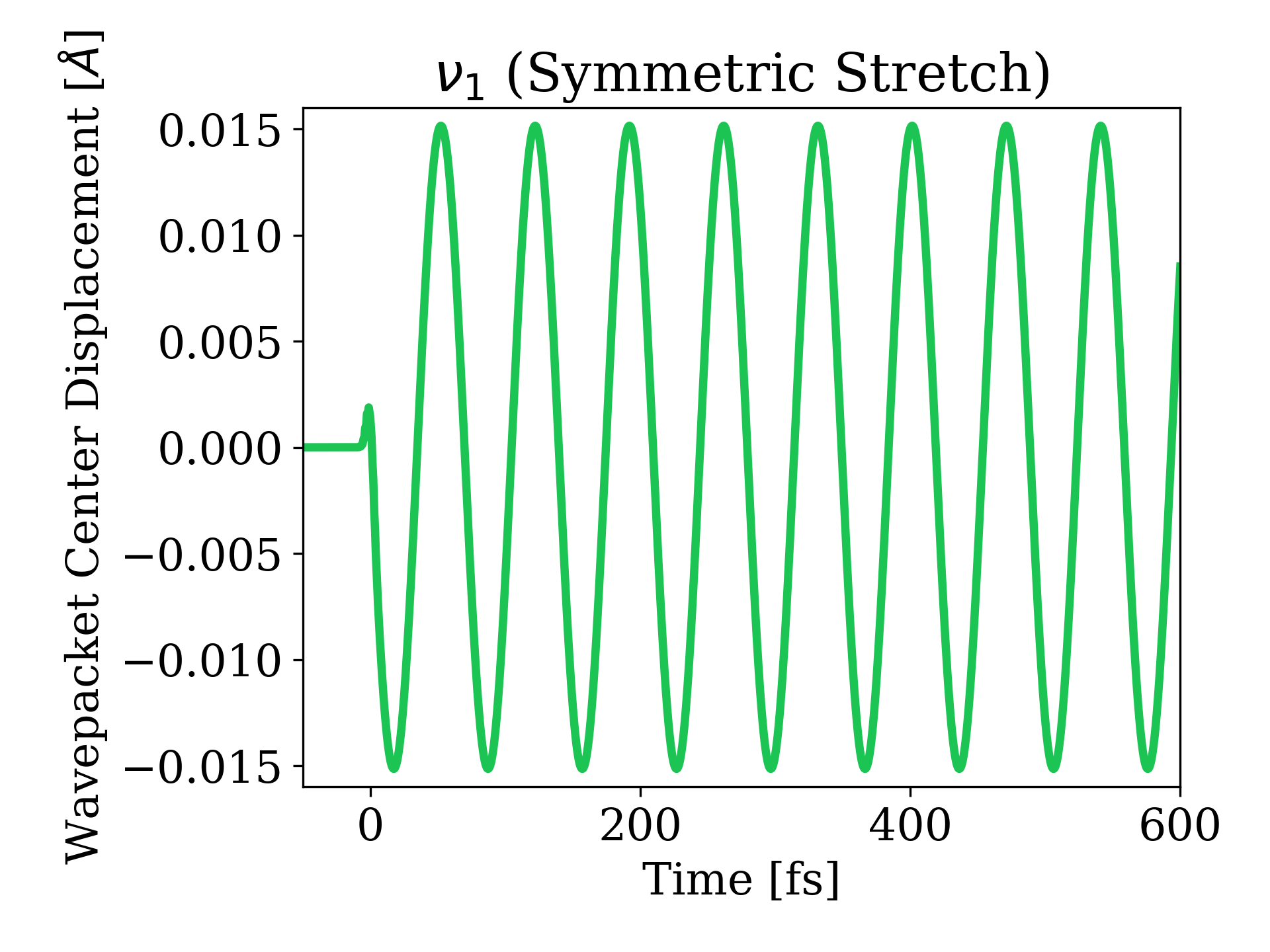}
\end{minipage}
\begin{minipage}{0.48\textwidth}
    \centering
    \includegraphics[width=\columnwidth]{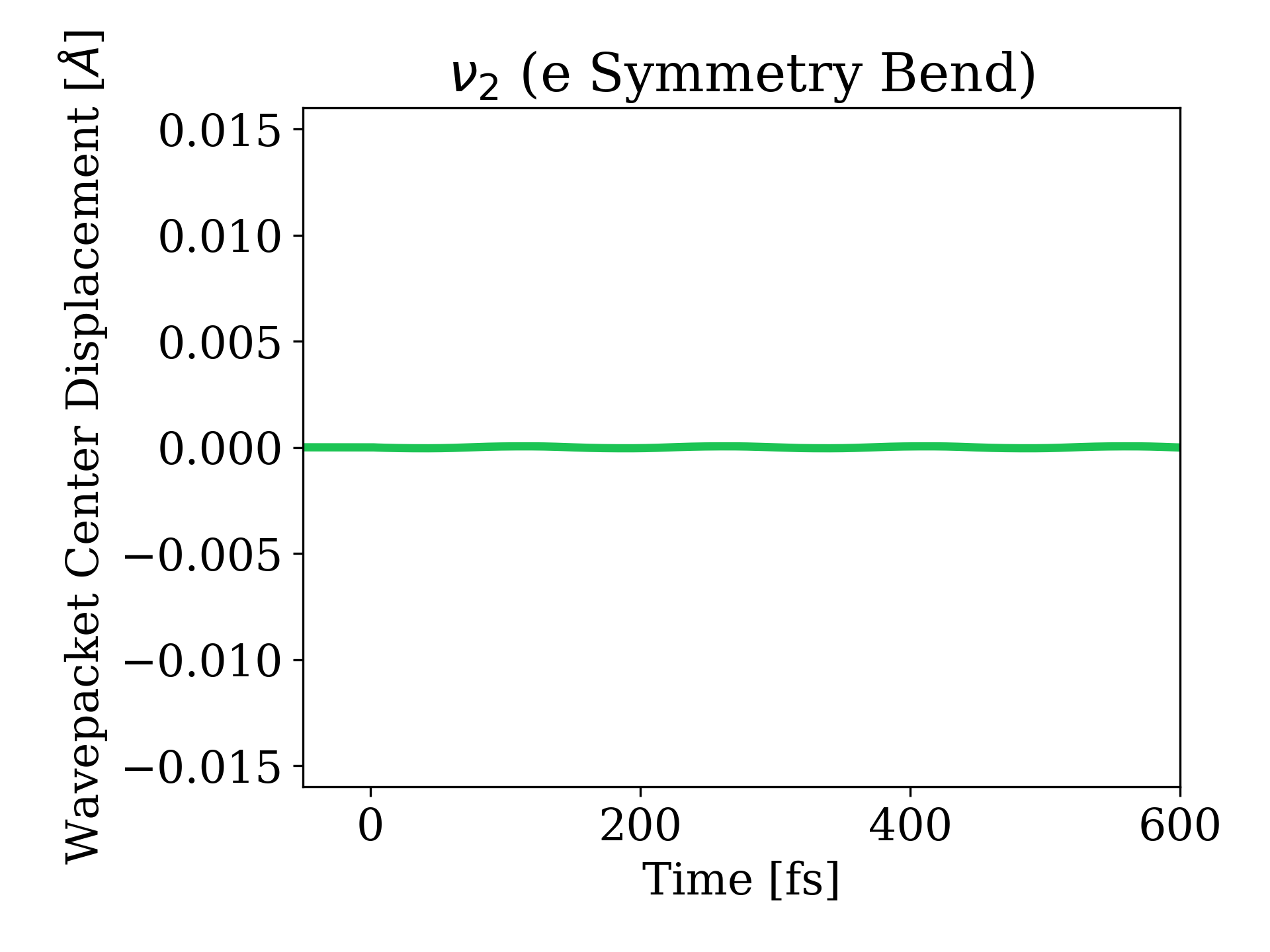}
\end{minipage}
\begin{minipage}{0.48\textwidth}
    \centering
    \includegraphics[width=\columnwidth]{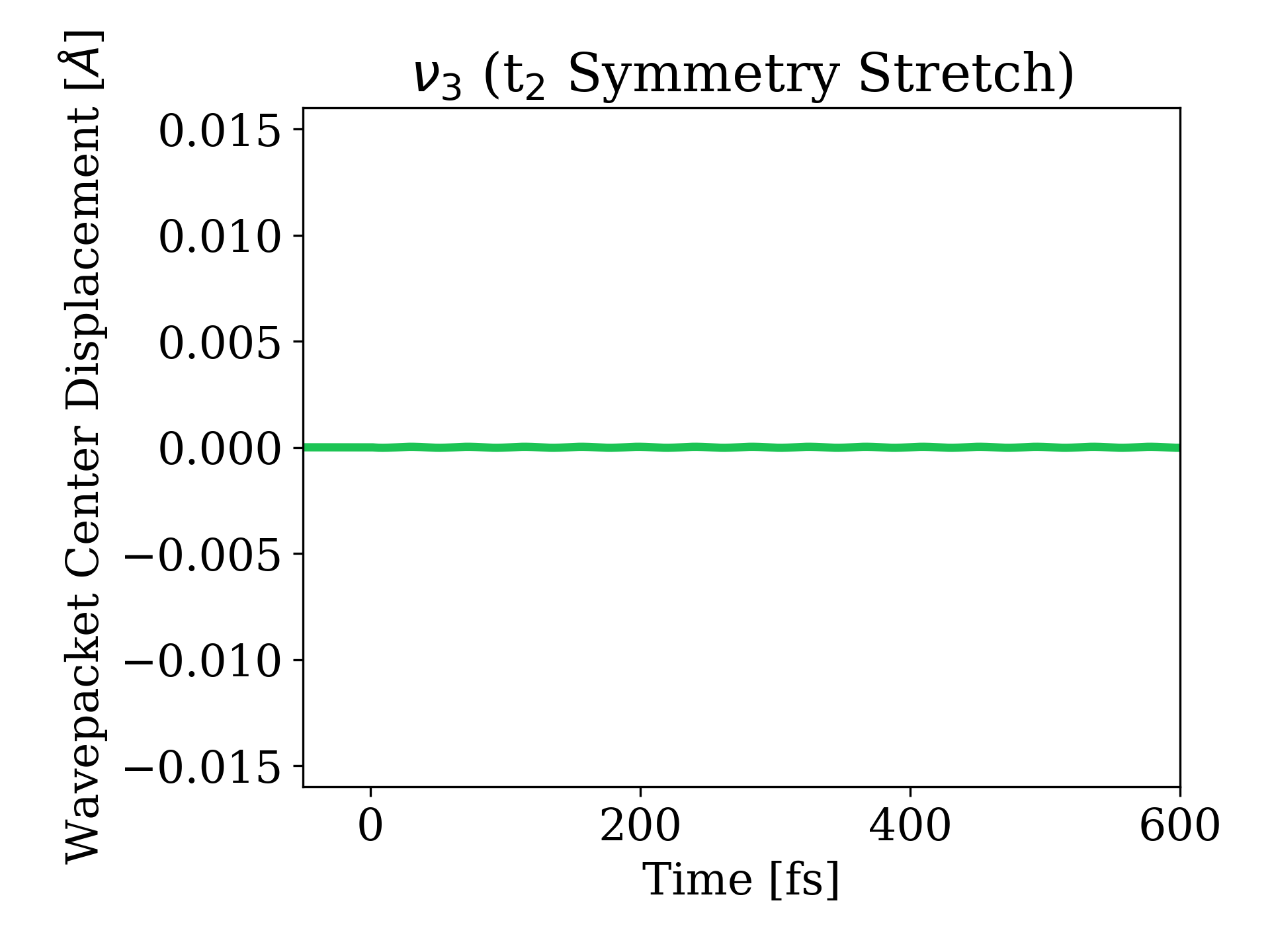}
\end{minipage}
\begin{minipage}{0.48\textwidth}
    \centering
    \includegraphics[width=\columnwidth]{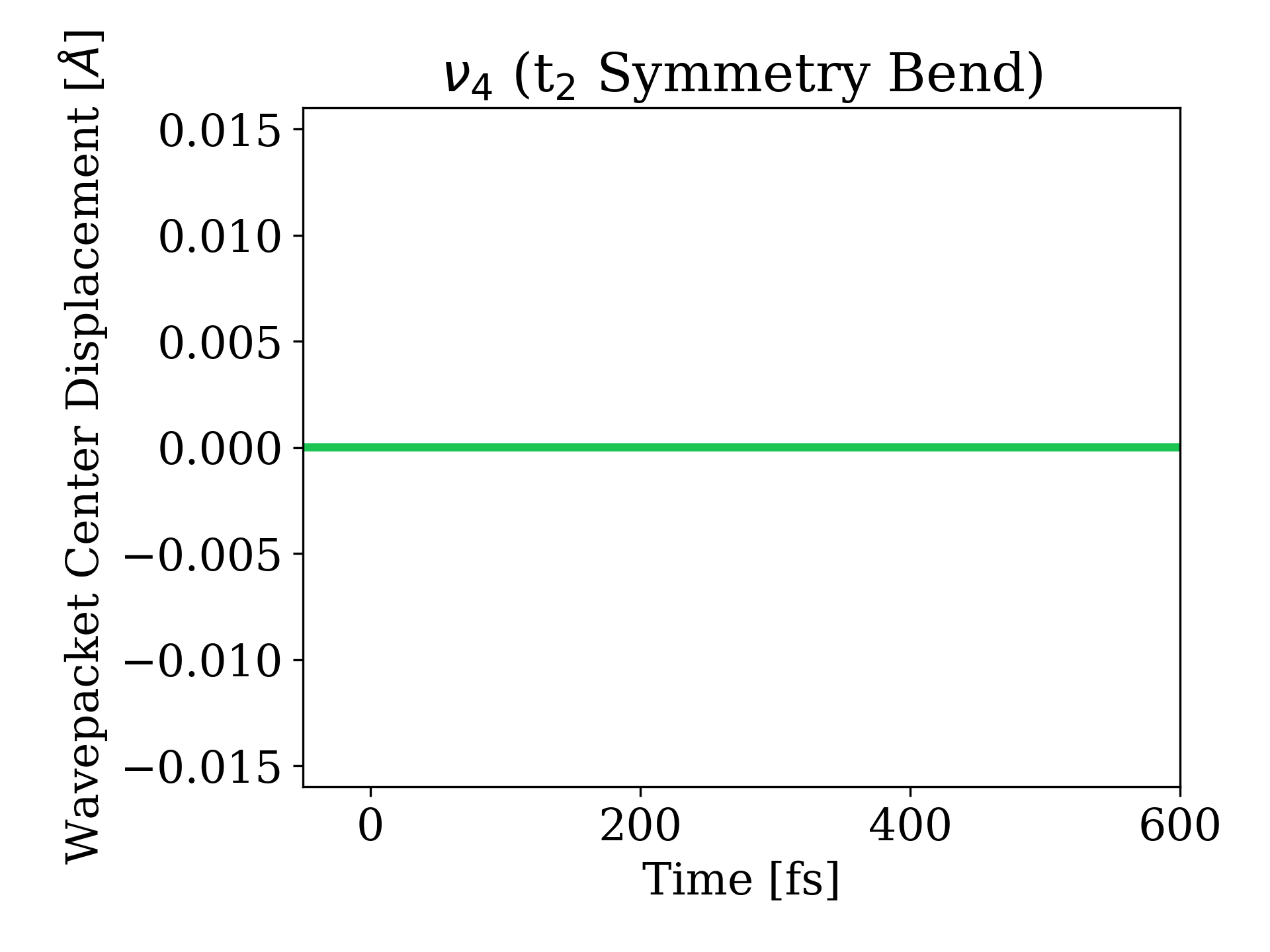}
\end{minipage}
\caption{Wavepacket center displacements $\langle q_i \rangle (t)= \displaystyle\int\limits_{-\infty}^{\infty} x\left|\psi_i(x,t)\right|^2 dx$ along different normal modes from the time-dependent Schr{\"o}dinger equation simulations, vs time. The computed normal mode with the lowest frequency in each degenerate set (see discussion in main text) was selected for depiction, the other modes in the set yield essentially equivalent behavior. Only the symmetric stretch (top left) shows perceptible displacement of the wavepacket center, and the other modes show no displacement of the wavepacket center (beyond numerical noise on the scale of $\sim 1\times 10^{-4}$ {\AA}), as is expected from the lack of a slope for isotropic polarizability along these modes. Note that the normal mode coordinate for the symmetric stretch differs from the stretching of individual C-Cl bonds by a factor of two; the C-Cl bond lengths are predicted to oscillate with an amplitude of $0.0075$ {\AA} from the simulations, as noted in the main text. 
}
\label{fig:exptq}
\end{figure}

\begin{figure}[htb!]
\begin{minipage}{0.48\textwidth}
    \centering
    \includegraphics[width=\columnwidth]{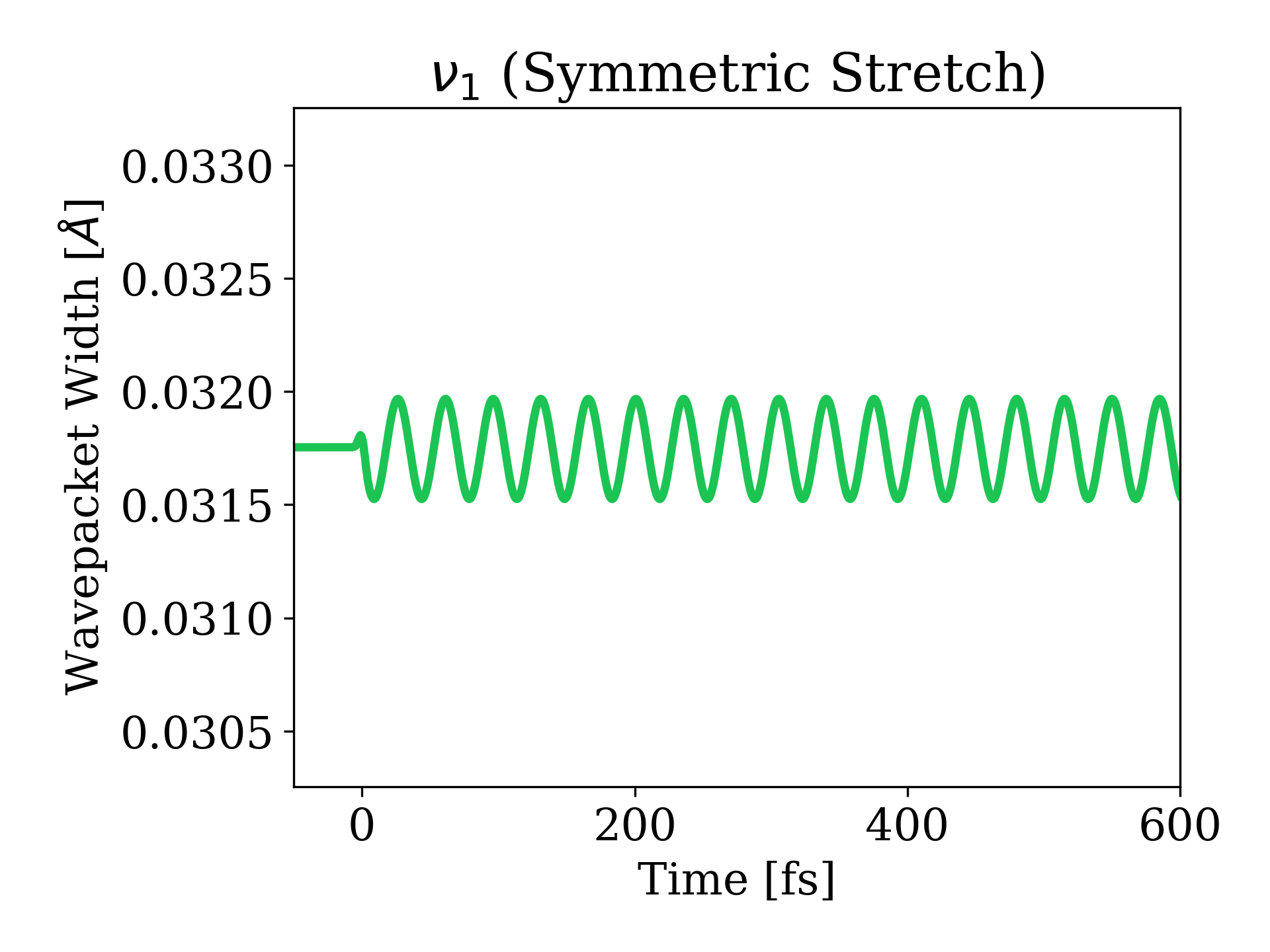}
\end{minipage}
\begin{minipage}{0.48\textwidth}
    \centering
    \includegraphics[width=\columnwidth]{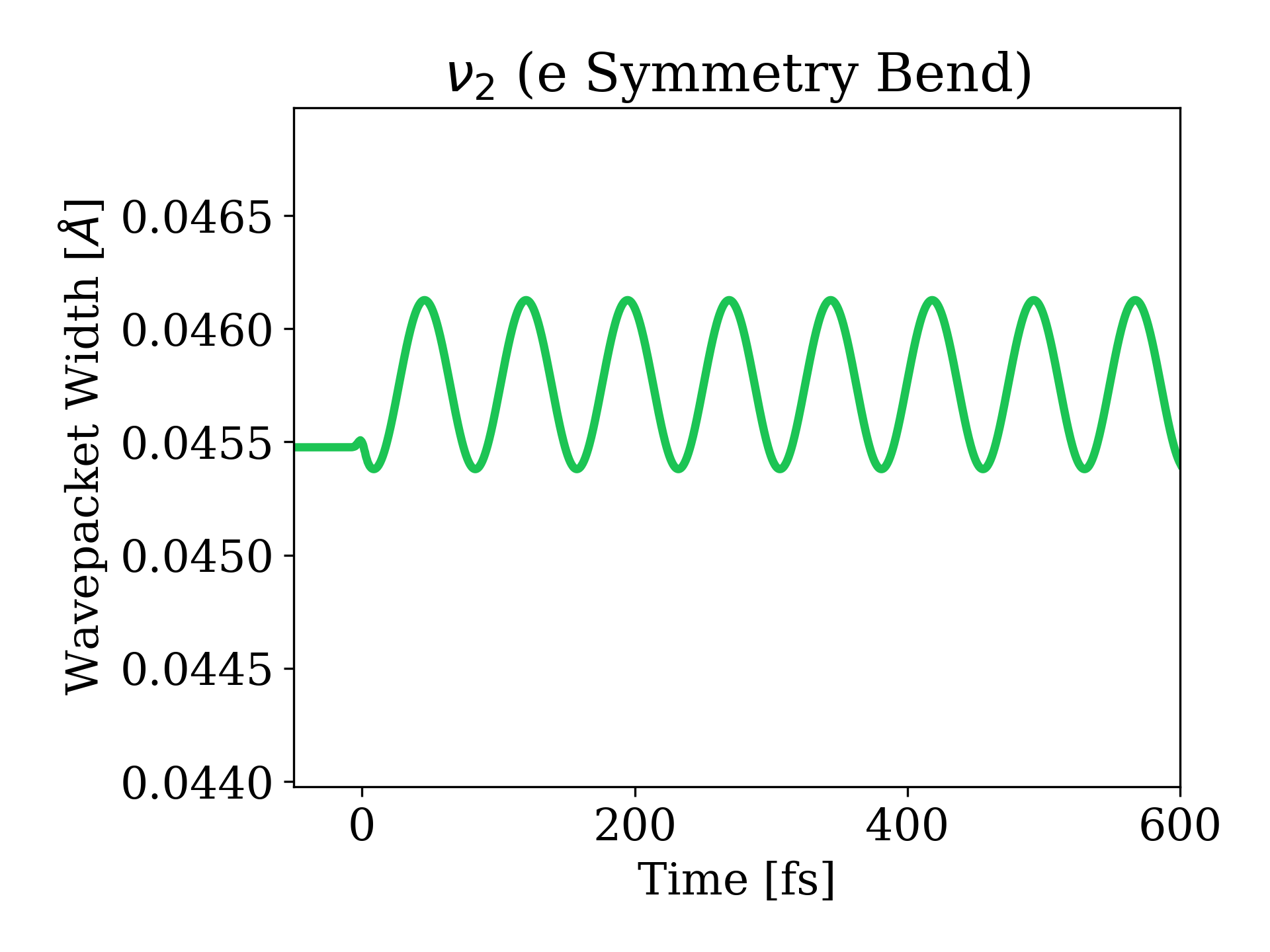}
\end{minipage}
\begin{minipage}{0.48\textwidth}
    \centering
    \includegraphics[width=\columnwidth]{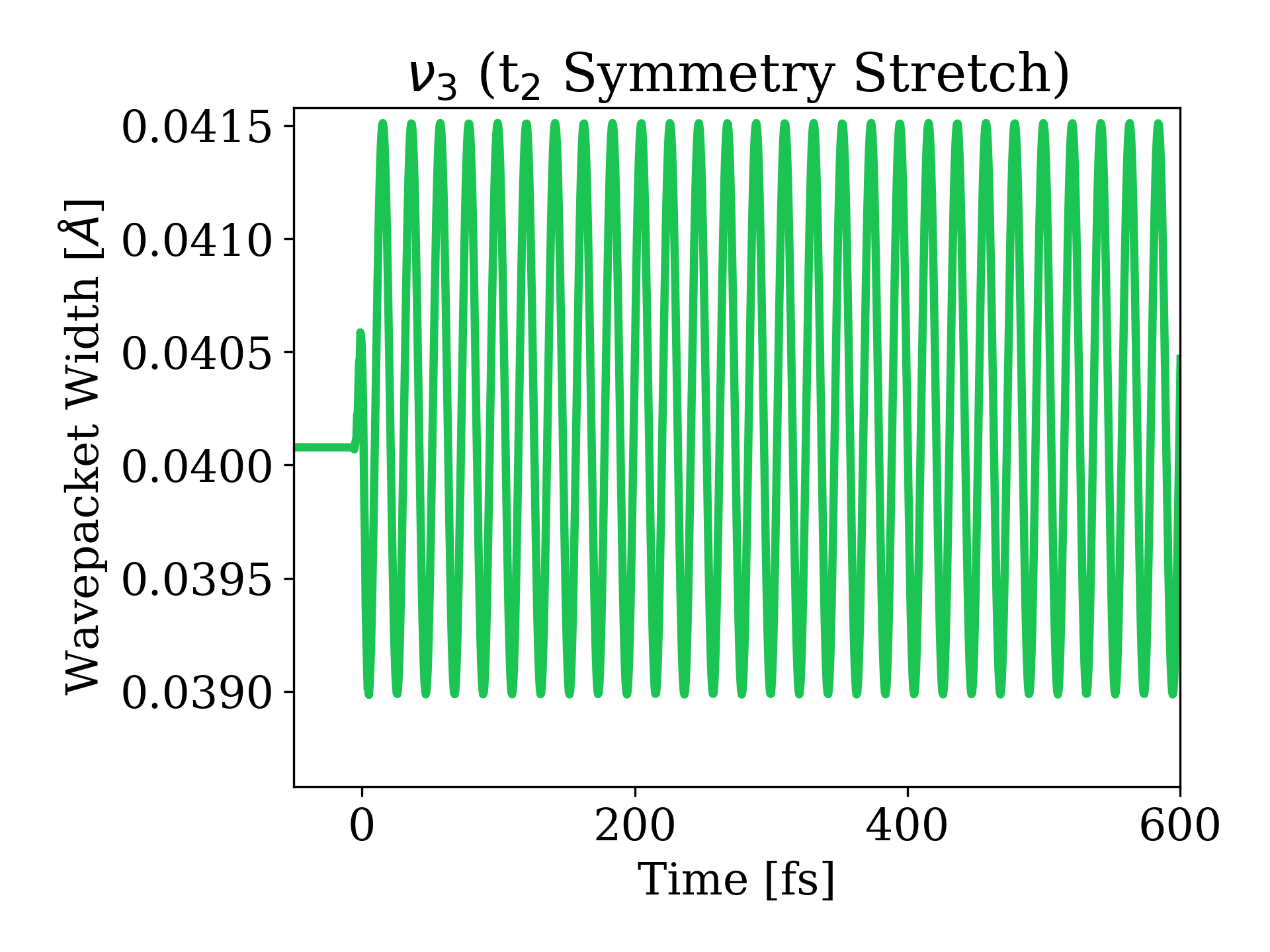}
\end{minipage}
\begin{minipage}{0.48\textwidth}
    \centering
    \includegraphics[width=\columnwidth]{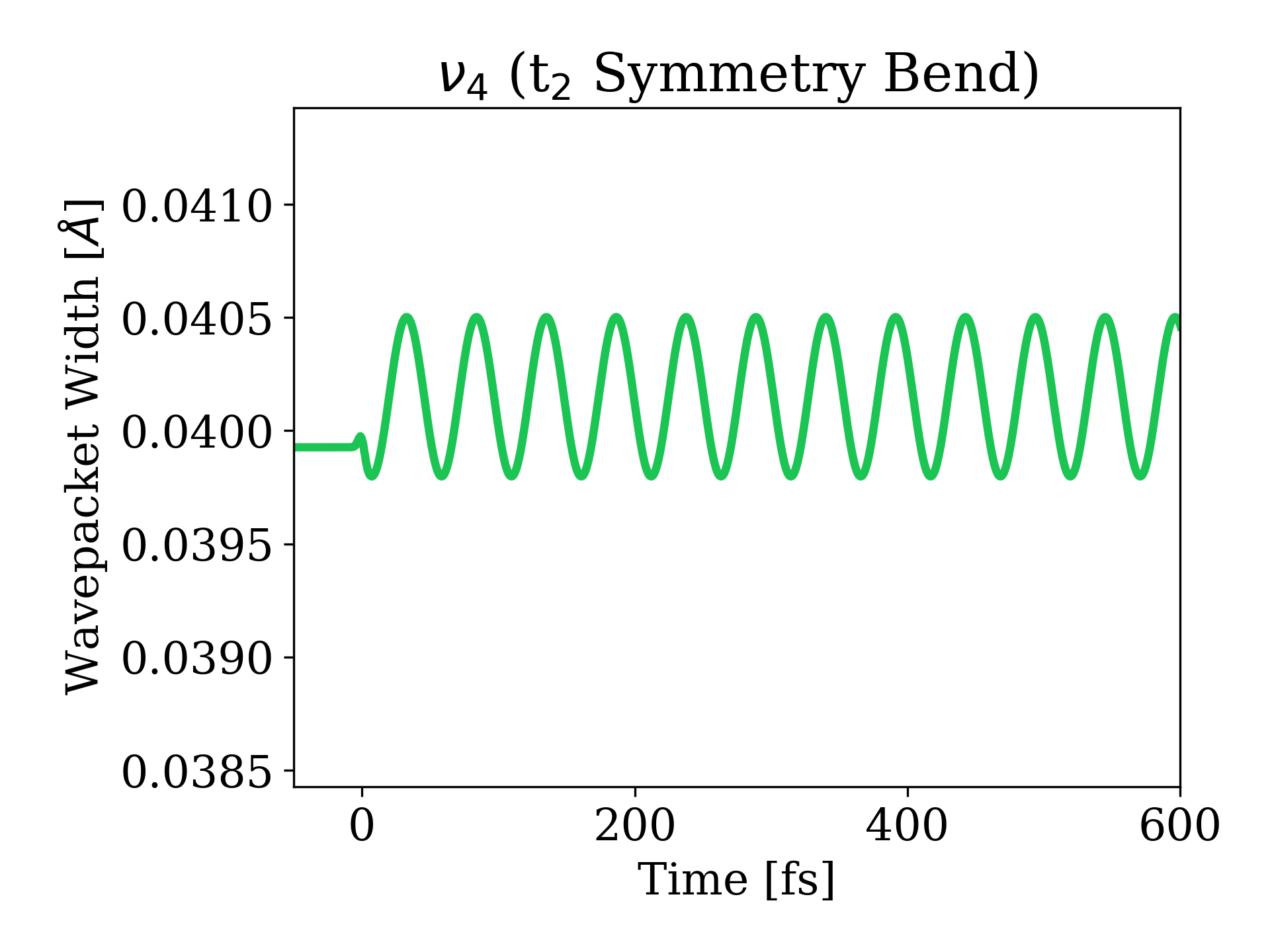}
\end{minipage}
\caption{Wavepacket widths $\langle \sigma_i \rangle (t) =\sqrt{\langle q_i^2 \rangle (t)-\left(\langle q_i \rangle(t)\right)^2}= \sqrt{\displaystyle\int\limits_{-\infty}^{\infty} x^2\left|\psi_i(x,t)\right|^2 dx-\left( \displaystyle\int\limits_{-\infty}^{\infty} x\left|\psi_i(x,t)\right|^2 dx\right)^2}$ along different normal modes from the time-dependent Schr{\"o}dinger equation simulations, vs time. The computed normal mode with the lowest frequency in each degenerate set (see discussion in main text) was selected for depiction, the other modes in the set yield essentially equivalent behavior. The widths show oscillatory behavior with twice the normal mode frequency and amplitudes on the scale of $10^{-4}-10^{-3}$ {\AA}. The antisymmetric (t$_2$) stretch (bottom left) shows the largest amplitude for change in width, as visually suggested by Fig. \ref{fig:psisq}.}
\label{fig:exptwidth}
\end{figure}

 Color plots of the probability densities $\left|\psi_i(x,t)\right|^2$ (depicted in Fig. \ref{fig:psisq}) reveal that only the symmetric stretch coordinate exhibits perceptible displacement of the wavepacket center, as argued in the main text. This is further confirmed by Fig. \ref{fig:exptq}, which depicts the time-dependent displacement of the wavepacket center (given by the expectation value of the normal mode coordinate $\langle q_i \rangle(t)$).  Small oscillations in the spatial width of the wavepacket at twice the fundamental frequency are observed to occur for all normal modes (as shown in Fig. \ref{fig:exptwidth}), but most prominently for the antisymmetric stretch. 
 The spatial extent of the wavepackets also confirm that the range of displacements considered for the polarizability fits ($\pm 0.2$ {\AA}, as shown in Fig. \ref{fig:polarfit}) was adequate.

\section{\label{sec:slope}Projection of CEPES Slopes on Normal Mode Directions}
We computed nonrelativistic restricted open-shell Kohn-Sham (ROKS) CEPES analytical gradients\cite{kowalczyk2013excitation,hait2020highly} at the ground state equilibrium geometry of \ce{CCl4} for the three excited states comprising the C 1s$\to 8t_2^*$ spectral feature and the twelve states (corresponding to holes in all 12 Cl 2p orbitals) that contribute to the Cl 2p$ \to 7a_1^*$ spectral feature. These gradients were projected against unit vectors along the nine  normal modes of the molecule, yielding CEPES slopes along these directions. The slopes for the three C 1s$\to 8t_2^*$ states and twelve Cl 2p$ \to 7a_1^*$ states (spin-orbit effects not included) are respectively reported in Tables \ref{tab:forceprojectionC1s} and \ref{tab:forceprojectionCl2p}, along with the average. It is apparent that the average CEPES slope along the symmetric stretch is much larger than the slopes along the other normal modes (by roughly two orders of magnitude). Indeed, the average slopes along the asymmetric modes are effectively zero for all intents and purposes, considering the errors introduced by the high sensitivity of the local exchange-correlation component of SCAN0 to the DFT integration grid\cite{furness2020accurate}.

The small average asymmtric mode slopes however often arise through cancellation between individual constituent state slopes of different signs, and many individual states have quite large slope \textit{magnitudes} along the asymmetric stretch directions. The observed spectral features may therefore broaden or contract with vibrations in the asymmetric modes (even if the peak center does not significantly shift), as some underlying excitations may redshift while others are blueshifting. Such time-dependent broadening effects would show up at twice the normal mode frequency in the XTAS signal Fourier transform, although we cannot detect any such features above the noise from the experimental data. Fourier transforms of computed XTAS from wavepacket Raman simulations however indicate weak features at $\sim 1600$ cm$^{-1}$ arising from broadening effects of the asymmetric stretch, which may potentially be detectable by experiments with better signal to noise.  

\begin{table}[htb!]
\setlength\tabcolsep{6pt}
\begin{tabular}{lrrrr}
Frequency (cm$^{-1}$) & \multicolumn{4}{c}{CEPES slope (eV/{\AA}) for C 1s$\to 8t_2^*$} \\ \hline
     &   $8t_2^*(x)$                & $8t_2^*(y)$ & $8t_2^*(z)$ & $8t_2^*(avg.)$                \\ \hline 
224.1 & 0.76  & -0.82 & 0.07  & 0.00    \\
225.0 & 0.00  & 0.00  & 0.00  & 0.00    \\
325.6 & 0.00  & 0.00 & 0.00  & 0.00    \\
325.7 & -0.49 & 0.67  & -0.13 & 0.02    \\
326.0 & -0.53 & -0.29 & 0.87  & 0.02    \\
477.9 & -2.63  & -2.69  & -2.56  & -2.63    \\
791.9 & -0.56 & 7.81  & -7.34 & -0.03   \\
794.2 & 0.00  & -0.06 & 0.00  & -0.02   \\
798.7 & 8.16  & -2.21 & -6.05 & -0.03  
\end{tabular}
\caption{Nonrelativistic CEPES gradients (in eV/{\AA}) projected along the nine ground state normal mode vectors for the C 1s$\to 8t_2^*$ excitation, taking into account each of the three possible $8t_2^*$ orbitals ($x,y,z$) and the signal average ($avg.$). Note that the slope for the totally symmetric mode corresponds to projection onto the normal mode vector, and not individual bond lengths, resulting in a $\sim $2 factor difference with the previously reported slope of -5.3 eV/{\AA}. 
}
\label{tab:forceprojectionC1s}
\end{table}

\begin{table}[htb!]
\setlength\tabcolsep{6pt}
\begin{tabular}{lrrrrrrrrrrrrrr}
Frequency (cm$^{-1}$) & \multicolumn{10}{c}{CEPES slope (eV/{\AA}) for Cl 2p$\to 7a_1^*$} \\ \hline
    & \multicolumn{3}{c}{Cl$^1$} & \multicolumn{3}{c}{Cl$^2$} & \multicolumn{3}{c}{Cl$^3$} & \multicolumn{3}{c}{Cl$^4$} & $avg.$   \\ \hline 
      & 2p$_\sigma$  & 2p$_\pi$  & 2p$_\pi$ & 2p$_\sigma$  & 2p$_\pi$  & 2p$_\pi$ & 2p$_\sigma$  & 2p$_\pi$  & 2p$_\pi$ & 2p$_\sigma$  & 2p$_\pi$  & 2p$_\pi$ &       \\ \hline 
224.1 & 0.05   & 0.06   & 0.04  & -0.05  & -0.04  & -0.06 & -0.02  & -0.01  & -0.03 & -0.02  & -0.01  & -0.03 & -0.01 \\
225.0 & 0.00   & 0.00   & 0.00  & 0.00   & 0.00   & 0.00  & -0.02  & -0.02  & -0.03 & 0.02   & 0.02   & 0.03  & 0.00  \\
325.6 & 0.00   & 0.00   & 0.00  & 0.00   & 0.00   & 0.00  & -2.02  & -1.96  & -1.96 & 2.02   & 1.96   & 1.97  & 0.00  \\
325.7 & -0.36  & -0.35  & -0.35 & -2.09  & -2.03  & -2.03 & 1.26   & 1.22   & 1.22  & 1.26   & 1.22   & 1.22  & 0.02  \\
326.0 & 2.39   & 2.32   & 2.33  & -1.17  & -1.14  & -1.14 & -0.58  & -0.57  & -0.56 & -0.58  & -0.57  & -0.56 & 0.01  \\
477.9 & -5.60   & -5.46   & -5.46  & -5.66   & -5.53   & -5.53  & -5.74   & -5.59   & -5.59  & -5.74   & -5.59   & -5.59  & -5.59  \\
791.9 & -9.40  & -9.11  & -9.12 & -4.03  & -3.92  & -3.91 & 6.68   & 6.49   & 6.49  & 6.68   & 6.49   & 6.49  & -0.01 \\
794.2 & 0.00   & 0.00   & 0.00  & 0.00   & 0.00   & 0.00  & -9.82  & -9.53  & -9.53 & 9.82   & 9.52   & 9.53  & 0.00  \\
798.7 & -7.59  & -7.36  & -7.37 & 11.36  & 11.02  & 11.02 & -1.91  & -1.86  & -1.86 & -1.91  & -1.85  & -1.86 & -0.01
\end{tabular}
\caption{Nonrelativistic CEPES gradients (in eV/{\AA}) projected along the nine ground state normal mode vectors for the  Cl 2p$\to 7a_1^*$ excitations, taking into account each of the 4 distinct Cl atoms (Cl$^{\{1-4\}}$) and the three possible $2p$ orbitals for each Cl atom (1 of $\sigma$ symmetry and 2 of $\pi$, as defined by the axis of the corresponding C-Cl bond) and the signal average ($avg.$). Note that the slope for the totally symmetric mode corresponds to projection onto the normal mode vector, and not individual bond lengths, resulting in a $\sim $2 factor difference with the previously reported slope of -11.1 eV/{\AA}.  
}
\label{tab:forceprojectionCl2p}
\end{table}
 \section{Simulated XTAS}

\begin{figure}
\begin{minipage}{0.48\textwidth}
    \centering
    \includegraphics[width=\columnwidth]{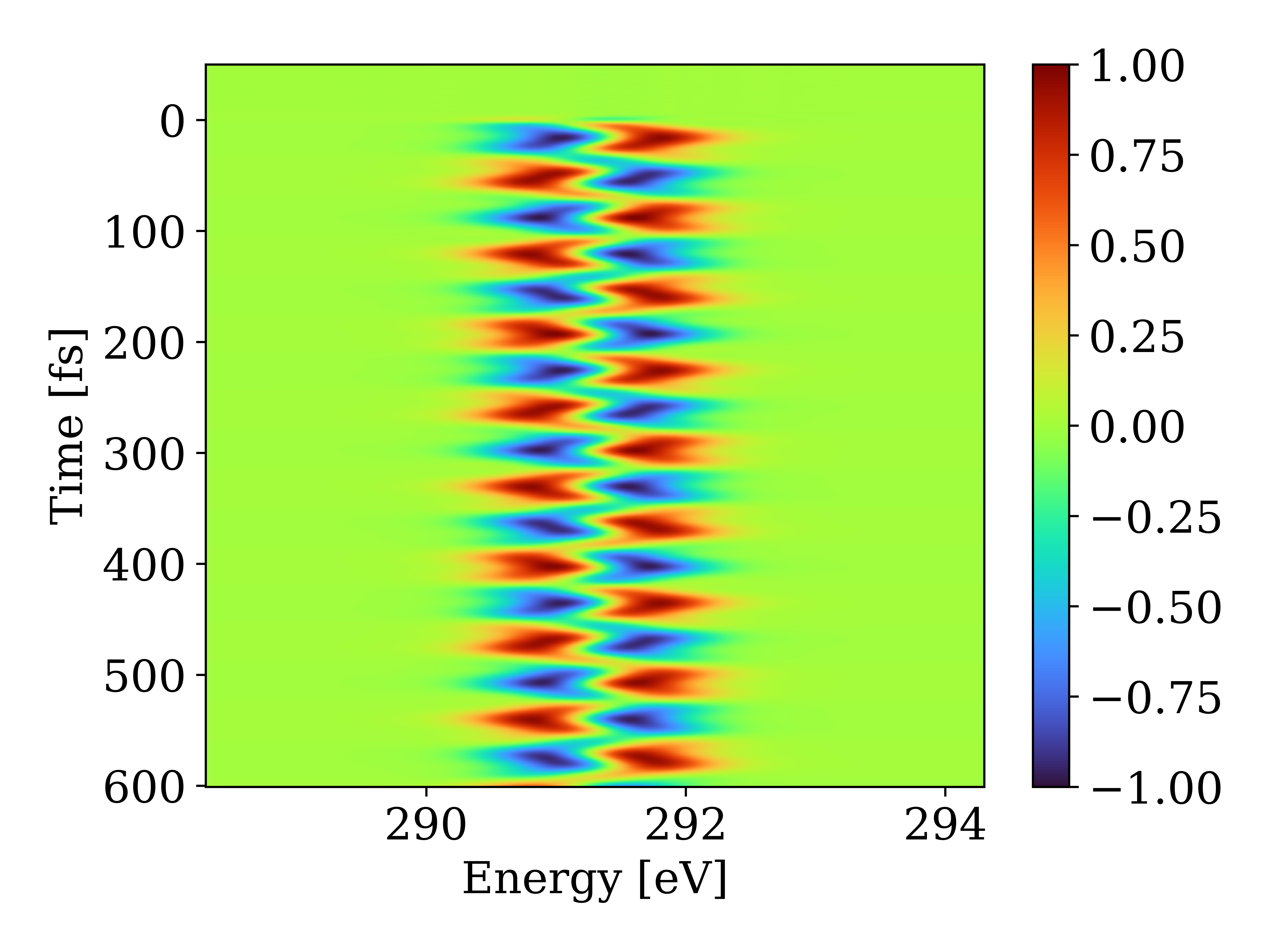}
\end{minipage}
\begin{minipage}{0.48\textwidth}
    \centering
    \includegraphics[width=\columnwidth]{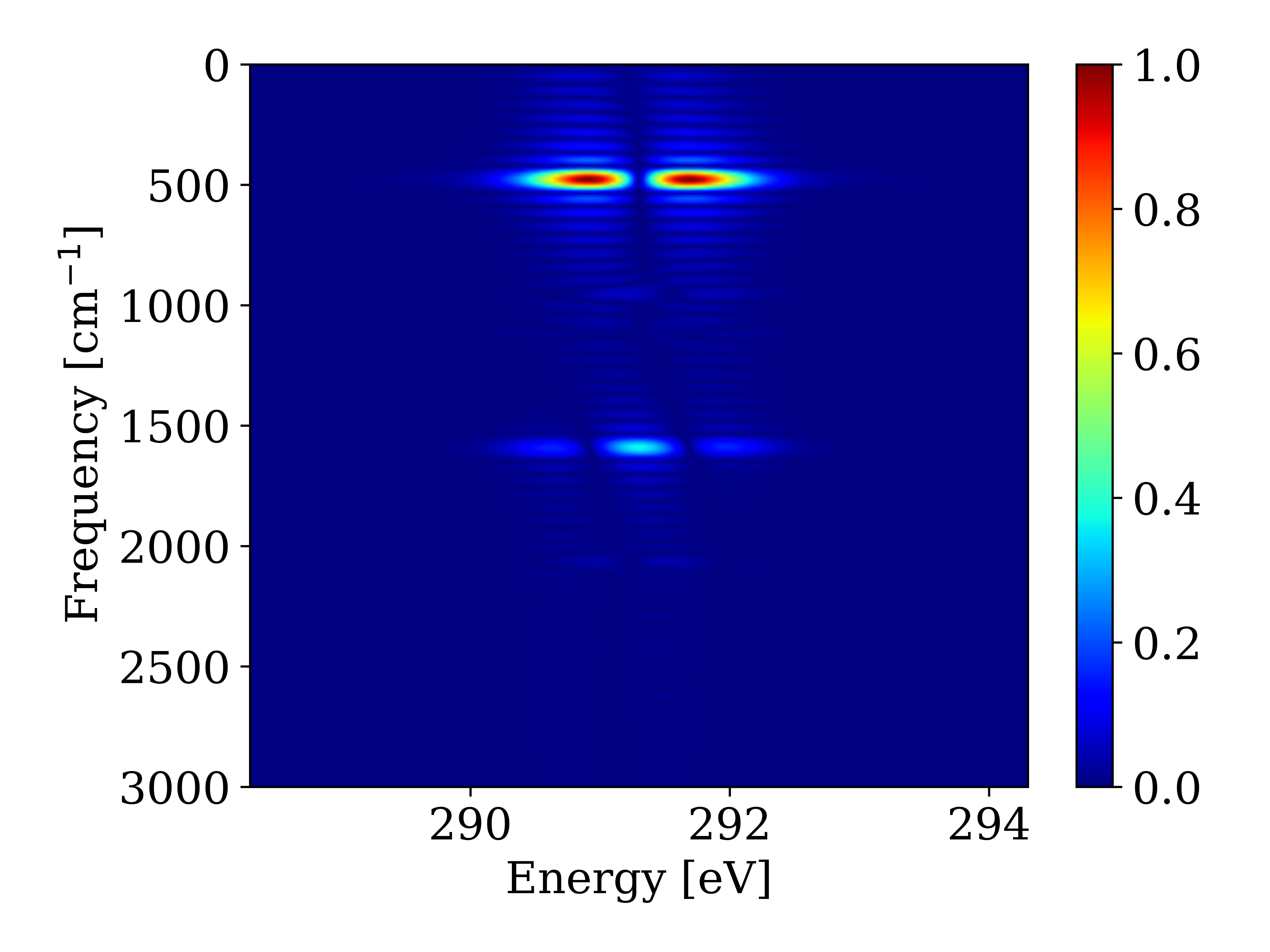}
\end{minipage}
\caption{Simulated transient C K-edge absorption (left) and the absolute values of its Fourier transform (right). The signals shown in both plots have been normalized such that the largest signal magnitude is 1.}
\label{fig:TDSE}
\end{figure}

A C K-edge transient absorption spectrum was computed using the $\left|\psi_i(x,t)\right|^2$ values from the simulations. A million normal mode displacements $\{q_i\}$ along each mode were sampled from the corresponding probability densities $\left|\psi_i(x,t)\right|^2$, for each timestep utilized for the time-dependent Schr{\"o}dinger equation simulations. For a given set of normal mode displacements $\{q_i\}$, the X-ray absorption energy was computed as $\Omega_C(\{q_i\})\approx\Omega_C(0)+ \displaystyle\sum\limits_i q_i\left(\dfrac{\partial E_C}{\partial q_i}\right)_{\{q_i\}=0}$, as discussed in the main text. The slopes $\left(\dfrac{\partial E_C}{\partial q_i}\right)_{\{q_i\}=0}$ for the three C 1s$\to 8t_2^*$ excitations were approximated by nonrelativistic ROKS gradient calculations at the equilibrium geometry (as shown in Table \ref{tab:forceprojectionC1s}). The vertical excitation energy at equilibrium geometry ($\Omega_C(0)$) was predicted to be 291.3 eV from relativistic (SF-X2C-1e) ROKS calculations (see Table II in main text). It was assumed that the oscillator strength does not undergo significant variation with normal mode displacement and was hence treated as a constant. Furthermore, any weak absorption from the formally dipole forbidden C 1s$\to 7a_1^*$ state from vibronic effects was not considered. The resulting C K-edge spectrum and its Fourier transform are shown in Fig. \ref{fig:TDSE}. Strong Fourier domain signal is only observed for the symmetric stretch ($\sim$ 479 cm$^{-1}$ from computation) in the  $<1000$ cm$^{-1}$ range. Some signal is also seen in the $\sim 1500$ cm$^{-1}$ range, from broadening effects at twice the fundamental frequency arising from the antisymmetric stretch breaking the degeneracy between the three C 1s$\to 8t_2^*$ excitations (see Fig. \ref{fig:psisq}). The experimental signal however does not have sufficient signal to noise in this regime for such behavior to be readily identifiable. There are some other weak Fourier domain signals which potentially arise from numerical error associated with the time-dependent Schr{\"o}dinger equation simulations and/or the underlying polarizability fits. 

\section{References}
\bibliography{biblio}